\documentclass[12pt,reqno]{amsart}

\usepackage{amsmath,amssymb,amsthm,
color, euscript, enumerate}

\usepackage[title]{appendix}
\usepackage{dsfont}                                                                          
\usepackage{longtable}
\usepackage{tikz}
\usepackage{caption}
\usepackage{multirow}
\usepackage{graphicx}

\newcommand{\RR}{{\mathbb R}}

\newcommand{\CC}{{\mathbb C}}
\newcommand{\be}{\begin{equation}}
\newcommand{\ee}{\end{equation}}
\newcommand{\ba}{\begin{array}}
\newcommand{\ea}{\end{array}}
\newcommand{\bea}{\begin{eqnarray}}
\newcommand{\eea}{\end{eqnarray}}

\usepackage{graphicx}
\usepackage{color}
\newcommand{\corr}[1]{{\color{black}{#1}}}
\usepackage{transparent}

\newtheorem{theorem}{Theorem} [section]

\newtheorem{lemma}{Lemma}[section]
\newtheorem{proposition}{Proposition}[section]
\newtheorem{definition}{Definition} [section]

\DeclareMathOperator{\sgn}{sgn}
\DeclareMathOperator{\res}{res}

\numberwithin{equation}{section}
\numberwithin{table}{section}

\allowdisplaybreaks

\begin{document}

\begin{center}
{\large   \bf  THE INVERSE SCATTERING THEORY OF KADOMTSEV PETVIASHVILI II EQUATIONS    }
 
\vskip 15pt

{\large  Derchyi Wu}

\vskip 5pt

{ Institute of Mathematics, Academia Sinica, 
Taipei, Taiwan}

\vskip 5pt
{\tt e-mail: {\tt mawudc@gate.sinica.edu.tw}}

\vskip 5 pt
{\today}
\end{center}

\vskip 10pt
\begin{center}{  \bf ABSTRACT}
\end{center}
\vskip 5pt

  An  overview of the inverse scattering theory of the Kadomtsev Petviashvili II  equation  with an emphasis on the inverse problem for perturbed KP multi line solitons is provided. It is shown that,  despite additional algebraic or analytic techniques are introduced due to new singular structures, there exists a consistency of the inverse scattering theories for different backgrounds  such as the vacuum, $1$-line solitons, and multi line solitons.

\tableofcontents

\section{Introduction}\label{S:introduction}

Originally derived as a model for small amplitude, long-wavelength, weakly two-dimensional waves in a weakly dispersive medium, the Kadomtsev Petviashvili II (KPII) equation
\be\label{E:KPII-intro}
(-4u_{x_3}+u_{x_1x_1x_1}+6uu_{x_1})_{x_1}+ 3u_{{x_2}{x_2}}=0     
\ee is a $(2+1)$-dimensional generalization of the Korteweg-de Vries (KdV) equation  and is related to various areas of mathematics and physics.  It has been shown to be integrable by the Lax pair
\begin{equation}\label{E:KPII-lax-1}
\left\{ 
{\begin{array}{l}
 (-\partial_{x_2}+\partial_{x_1}^2+u )\Phi(x,\lambda)=0,\\
 (-\partial_{x_3}+ \partial_{x_1}^3+\frac 32u\partial_{x_1}+\frac 34u_{x_1}+\frac 34\partial_{x_1}^{-1}u_{x_2}-\lambda^3   )\Phi (x,\lambda)=0. 
\end{array}}
\right.
\end{equation} since the beginning of the 1970's.

Thanks to the Lax pair, the initial value problem of the  KPII  equation can be solved by studying an inverse scattering theory (IST) and proving the scattering data evolves linearly. When the background is a vacuum, the debar method has been introduced to solve the IST  \cite{ABF83,Li86,GN88,Gr97,W87}. Around 2000, Boiti et al,   Villarroel  and Ablowitz started the investigation of the IST of the KPII equation when the background are $1$-line solitons  \cite{VA04,BP301}.   Boiti et al then integrated the Sato theory and set the foundation of the IST of the KPII equation when the background is a mult line soliton. Their achevements at least include:   deriving an explicit formula of the Green function,   $L^\infty$ estimates for the discrete part of the Green function, and  the $\mathcal D$-symmetry, the relation between values of the eigenfunction at multi value points \cite{P00,BP302,BP309,BP310,BP211,BP212,BP214}. Based on their work, we  complete a rigorous IST of the KPII equation for smooth perturbations of multi line solitons \cite{Wu20,Wu21,Wu22}. Our work is the first rigorous IST of a multi-dimensional integrable system when both continuous and discrete scattering data are present, and the support of continuous scattering data does not degenerate into contours in the complex plane.

In this paper, we give an  overview of the IST of the KPII  equation  with an emphasis on the inverse problem of the perturbed KP multi line solitons. The goal is to show that,  despite additional algebraic or analytic techniques are introduced due to new singular structures, there exists a consistency of the inverse scattering theories for different background potentials such as the vacuum, $1$-line solitons, and multi line solitons. That is, when discrete scattering data or continuous scattering data vanish, the forward and the inverse scattering transform constructed for perturbed line solitons degenerate into those transforms  for rapidly decaying potentials or for  multi line solitons.

The paper is organized as follows:  in Section \ref{S:vacuum}, we present an IST for the vacuum background. This theory, beautifully intertwining with the Fourier theory, outlines the approaches of IST's of the KPII equation for different backgrounds  and characterizes scattering properties when $\lambda\to\infty$ for multi line soliton background.

Without using the Sato theory, the IST for perturbed $1$-line solitons is firstly solved in Section \ref{S:1-soliton}. We define the forward scattering transform;  formulate the inverse problem as a system of a Cauchy integral equation  (CIE) and the $\mathcal D$-symmetry constraint;  solve the inverse problem by using H$\ddot{\mbox o}$lder interior estimates and deformation methods. We elucidate the connection between the forward and inverse problems and   highlight H$\ddot{\mbox o}$lder interior estimates and deformation methods in $\S$ \ref{SSS:I-3}, $\S$ \ref{SSS:I-5} since they are  major additional analytical tools. 

In Section \ref{S:N-line}, applying the Sato theory \cite{S81,SS82,S89a,S89b,BC07,BK03,KW13} and Boiti et al's  direct scattering theory of the KP  equation \cite{P00,BP302,BP309,BP310,BP211,BP212,BP214},   we  extend the approach of the IST for perturbed $1$-solitons to that for perturbed multi line  solitons.  We  state the complete theory but only  provide proofs of distinct features  which, additionally, shows  the TP (totally positive) condition is necessary and justifies differences between IST for perturbed $1$-solitons and  perturbed multi line solitons  are mainly algebraic.

\section{The IST for rapidly decaying potentials}\label{S:vacuum}

\medskip

\subsection{Statement of results}\label{SS:Lax}


 Taking a rapidly decaying potential $u_0(x_1,x_2)$ as the initial data,  one can solve the Cauchy problem of the KP equation using the inverse scattering theory (IST) \cite{ABF83,Li86,GN88,Gr97,W87}. Precisely, the direct scattering theory can be stated as
\begin{theorem} \cite{W87}\label{T:dp-vacuum}  
Given  
 $
      \sum_{|l|\le { d+7}} |\partial_{x_1}^{l_1}\partial_{x_2}^{l_2}u_0(x_1,x_2)|_  {L^1\cap L^2} \ll 1$, for $   d\ge 0$,
\begin{itemize}
     \item [(1)]    there exists  uniquly an eigenfunction  $\Phi(x_1,x_2,\lambda)=e^{ \lambda  x_1+ \lambda ^2x_2}m_0(x_1,x_2,\lambda)$ of the \sl{Lax equation} 
          \be\label{E:intro-Lax}
   \begin{gathered}
 (-\partial_{x_2}+\partial_{x_1}^2 +2\lambda\partial_{x_1}
+u_0(x_1,x_2))m_0(x_1,x_2 ,\lambda)=  0, \ \\
|\partial_x^l \left[m_0-1\right]|_{L^\infty}\le  C|\partial_x^lu_0|_{L^1\cap L^2}    
\end{gathered}\ee for $\forall\lambda\in\CC$.

       \item [(2)]   One can construct the {\sl forward scattering transform}
\be\label{E:intro-SD} 
\begin{split}
\mathcal S:u_0\mapsto &s _c(\lambda)= \frac{\sgn(\lambda_I)}{2\pi i} \left[{u_0(\cdot)m_0(\cdot,\lambda)}\right]^\wedge(\frac{\overline\lambda-\lambda}{2\pi i},\frac{\overline\lambda^2-\lambda^2}{2\pi i}), 
\end{split}
\ee 
satisfying
\be\label{E:debar-wick}\partial_{\overline\lambda}m_0 (x_1,x_2,\lambda) = {  {  s}_c(\lambda )}e^{(\overline\lambda-\lambda)x_1+(\overline\lambda^2-\lambda^2)x_2}   m(x_1,x_2,\overline\lambda),\ee  and the Cauchy integral equation
\be
\begin{split}
m_0(x_1,x_2,\lambda)= & 1+\mathcal C T_0m_0(x_1,x_2,\lambda). 
\end{split}
\label{E:intro-CIE-new}
\ee 
The algebraic and analytic constraints of the scattering data are
 \begin{gather}
  {\begin{array}{l} 
   |\left[|\overline\lambda-\lambda|^{l_1}   +| \overline\lambda^2-\lambda^2|^{l_2}\right] s_c (\lambda)|  _{
  L^\infty} 
  \le  {C \sum_{h=0}^{l} |\partial_{x}^{h}u_0|_{L^1\cap L^2}} , \end{array}}\label{E:intro-s-c-ana-c}\\
  s_c(\lambda)=  \overline{s_c( \overline\lambda)}. \label{E:intro-s-c-reality-c}
 \end{gather}
 

\end{itemize}
\end{theorem} Here $\lambda=\lambda_R+i\lambda_I,\,\overline\lambda=\lambda_R-i\lambda_I$, $\mathcal C$ is the Cauchy integral operator, $T_0$ is the continuous scattering operator $T$ at $x_3=0$, and $\widehat \phi$ is the Fourier transform defined by
\be\label{E:ct-operator}
\begin{split}
\mathcal C \phi  (x,\lambda)
\equiv & -\frac{1}{2\pi i}\iint_{\CC}\frac{\phi(x,\zeta)}{\zeta-\lambda}d\overline\zeta\wedge d\zeta, \quad x=(x_1,x_2,x_3),\\
 T  \phi  (x ,\lambda)
\equiv & {  s}_c(\lambda  )e^{(\overline\lambda-\lambda)x_1+(\overline\lambda^2-\lambda^2)x_2 +(\overline\lambda^3-\lambda^3)x_3 }\phi(x, \overline\lambda), \\
\widehat \phi(\xi,\lambda)\equiv& \iint_{\RR^2}e^{-2\pi i(x_1\xi_1+x_2\xi_2)}\phi(x_1,x_2,\lambda )dx_1  dx_2. 
\end{split}
\ee

A linearization theorem is
\begin{theorem} \label{T:l-vacuum}  
  If $\Phi(x,\lambda)= e^{ \lambda  x_1+ \lambda ^2x_2}    m(x, \lambda)$ satisfies the Lax pair \eqref{E:KPII-lax-1} and  
\[
\partial_{\overline\lambda}  m(x, \lambda)=  {  {  s}_c(\lambda,x_3)}e^{(\overline\lambda-\lambda)x_1+(\overline\lambda^2-\lambda^2)x_2}   m(x,\overline\lambda)  ,
\] then  
\be\label{E:linearization-D-evol-new}
\begin{gathered}
   {  s}_c(\lambda, x_3)=   {e^{ (\overline\lambda^3-{ \lambda}^3)x_3}}{  s}_c(\lambda ).  
   \end{gathered}\ee
\end{theorem}

The inverse scattering theory is stated as
\begin{theorem} \label{T:ip-vacuum} \cite{W87}
Given a small continuous scattering data $s_c(\lambda)$  decaying sufficiently rapidly in $(\overline\lambda-\lambda,\overline\lambda^2-\lambda^2)$,  one has
\begin{itemize}
\item [(1)]  the unique solvability of the  Cauchy integral equation   
\begin{gather}   
    {   m} (x , \lambda) =1 +\mathcal C  T 
     m (x , \lambda) ,  \label{E:intro-CIE-t}  \\
|\partial_x^l\left[m -1\right]|_{L^\infty}\le C|\left(1+|\xi|^{l_1}   +| \eta|^{l_2}\right)s_c|_{L^\infty\cap L^2(d\xi d\eta)},\label{E:intro-CIE-t-est} 
     \end{gather} where $2\pi i\xi= \overline\zeta-\zeta$, $ 2\pi i\eta=\overline\zeta^2-\zeta^2$;  
     \item [(2)] the Lax equation
  \begin{gather}
 \left(-\partial_{x_2}+\partial_{x_1}^2+2 \lambda\partial_{x_1}+ u (x) \right)  m (x ,\lambda)=0 ,\label{E:intro-Lax-u-new}
 \\
 u(x )= -\frac 1{\pi i}\partial_{x_1}\iint  T  m  \ d\overline\zeta\wedge d\zeta , \label{E:intro-Lax-u-new-1}
\\
   |(1+|\xi|^{k}+|\eta|^{k})\widehat u(x) |_{L^\infty}\le C |(1+|\xi|^{k+1}+|\eta|^{k})s_c|_{L^\infty\cap L^2(d\xi d\eta)},\label{E:intro-Lax-u-new-2}
\end{gather}   the inverse scattering transform is defined  by   
\be\label{E:inverse-transform-decay}
\mathcal S^{-1}(  s _c(\lambda) )\equiv -\frac 1{\pi i}\partial_{x_1}\iint  T  m  \ d\overline\zeta\wedge d\zeta ;\ee
\item[(3)] 
    the KPII equation is fulfilled
\be\label{E:intro-KP-ist-new}
(-4u_{x_3}+u_{x_1x_1x_1}+6uu_{x_1})_{x_1}+3u_{{x_2}{x_2}}=0 .
\ee

\end{itemize} 

\end{theorem}

\subsection{The strategy } \hfill \\
Detailed proof can be found in \cite{W87}. We highlight features of the proof.
\subsubsection{Proof of Theorem \ref{T:dp-vacuum}} 

\begin{itemize}
\item [$(1)$] {The Lax equation}   \eqref{E:intro-Lax} is proved by 
\begin{gather}
m_0(x_1,x_2,\lambda)= 1-\left[\frac{\widehat{u_0m_0}}{p_\lambda(\xi,\eta)}\right]^{\vee}=1-G_\lambda\ast u_0m_0, \label{E:c-integral-eq}\\
 p_\lambda(\xi,\eta)= (2\pi i\xi+\lambda)^2-(2\pi i\eta+\lambda^2),\quad 
 G_\lambda=\left[\frac{1}{p_\lambda}\right]^{\vee},\label{E:wick-infty-spec}\\
\left|\frac 1{p_\lambda}\right|_{L^1(\Omega_\lambda, d\xi d\eta)}\le  \frac C{(1+|\lambda_I|^2)^{1/2}},\ 
\left|\frac 1{p_\lambda}\right|_{L^2(\Omega_\lambda^c,d\xi d\eta)}\le \frac C{(1+|\lambda_I|^2)^{1/4}}, \label{E:wick-infty-spec-1}\\
\ \Omega_\lambda=\{(\xi,\eta)\in\RR^2\ :\ |p_\lambda(\xi,\eta)|<1\}.\nonumber 
\end{gather}

\item [$(2)$] {We prove  the forward scattering transform} is a $ \partial_{\bar\lambda}$-characterization of $m_0$. From \eqref{E:c-integral-eq}, \eqref{E:wick-infty-spec}, and 
\[
\begin{split}
\partial_{\bar\lambda}\left[ \frac{1}{p_\lambda}\right] 
=&\lim_{\epsilon\to 0}\partial_{\bar\lambda}\left[\frac{1}{p_\lambda} (1-e^{-\frac{1}{\epsilon^2}|p_\lambda|^2})\right] =\lim_{\epsilon\to 0}\left[\partial_{\bar\lambda}\overline{p_\lambda} \right]\frac{1}{\epsilon^2}e^{-\frac{1}{\epsilon^2}|p_\lambda|^2}\\
=&-\frac{\sgn(\lambda_I)}{2\pi i}\delta_{(\frac{\overline\lambda-\lambda}{2\pi i},\frac{\overline\lambda^2-\lambda^2}{2\pi i})},\\ p_\lambda(D)f=&e^{(\overline\lambda-\lambda)x_1+(\overline\lambda^2-\lambda^2)x_2  }p_{ \overline \lambda}(D)e^{-[(\overline\lambda-\lambda)x_1+(\overline\lambda^2-\lambda^2)x_2 ] }f,
\end{split}
\]one obtains
\begin{align*}
&\left[\partial_{\bar\lambda}G_\lambda\right]\ast u_0m_0= \left(\widehat{u_0m_0}\partial_{\bar\lambda}\left[\frac{1}{p_\lambda }\right]\right)^{\vee} \\
&\hskip1.05in=-\frac{\sgn(\lambda_I)}{2\pi i} e^{(\overline\lambda-\lambda)x_1+(\overline\lambda^2-\lambda^2)x_2  }\widehat{u_0m_0}(\frac{\overline\lambda-\lambda}{2\pi i},\frac{\overline\lambda^2-\lambda^2}{2\pi i},\lambda)\\
&\hskip1.05in=  -s_c(\lambda)e^{(\overline\lambda-\lambda)x_1+(\overline\lambda^2-\lambda^2)x_2  },\\
&G_\lambda\,  e^{(\overline\lambda-\lambda)x_1+(\overline\lambda^2-\lambda^2)x_2  }= e^{(\overline\lambda-\lambda)x_1+(\overline\lambda^2-\lambda^2)x_2  } \,   G_{ \overline\lambda}.
\end{align*}  
Consequently,
\begin{align}
 &\partial_{\bar\lambda}m_0(x_1,x_2, \lambda)= (1+G_\lambda\ast u_0)^{-1}1\label{E:sd-debar-m-0}\\
=&-(1+G_\lambda\ast u_0)^{-1}\partial_{\bar\lambda}G_\lambda\ast u_0m_0\nonumber\\
=&s_c(\lambda)(1+G_\lambda\ast u_0)^{-1}e^{(\overline\lambda-\lambda)x_1+(\overline\lambda^2-\lambda^2)x_2  }\nonumber\\
=&s_c(\lambda)e^{(\overline\lambda-\lambda)x_1+(\overline\lambda^2-\lambda^2)x_2  }(1+G_{\bar\lambda}\ast u_0)^{-1}1\nonumber\\
=&s_c(\lambda)e^{(\overline\lambda-\lambda)x_1+(\overline\lambda^2-\lambda^2)x_2  }m_0(x_1,x_2,\bar\lambda).\nonumber
\end{align}

Moreover, Liouville's theorem and the Lax equation imply that, there exists    $q(x_1,x_2)$   such that
\begin{gather}
m_0(x_1,x_2,\lambda)= q(x_1,x_2)+\mathcal CT_0m_0(x_1,x_2,\lambda),\label{E:CIE-initial}\\
u_0m_0= -2\lambda \partial_{x_1}q-\partial_{x_1}^2 q+ \partial_{x_2}q+\left(\partial_{x_2}-\partial_{x_1}^2 -2\lambda \partial_{x_1} \right)\mathcal CT_0m_0.\label{E:CIE-lax}
\end{gather}

Via a change of variables 
 \be\label{E:variables}
 \begin{split}
2\pi i\xi= \overline\zeta-\zeta,&\quad 2\pi i\eta=\overline\zeta^2-\zeta^2,\\
\zeta=-i\pi\xi+ \frac \eta{2\xi}, &\quad d\overline\zeta\wedge d\zeta=\frac{i\pi}{|\xi|}d\xi d\eta,
\end{split}
\ee 
and from \eqref{E:wick-infty-spec},     \eqref{E:wick-infty-spec-1},  we obtain,  
\begin{align}
&|\mathcal C T_0\phi|
\le   C|   \iint \frac {{   s_c(\zeta)e^{(\overline\zeta-\zeta)x_1+(\overline\zeta^2-\zeta^2)x_2}\phi}}{ \zeta-\lambda }d\overline\zeta\wedge d\zeta|\label{E:infty-spec-1}\\
\le &\ C|   \phi|_{L^\infty} \iint \frac {|{    s_c(\zeta(\xi,\eta))|}}{|(2\pi\xi)^ 2-4\pi i\xi\lambda+2\pi i \eta|}d\xi d\eta
\nonumber\\
\le &\ C|  \phi|_{L^\infty}  \{|  s_c(\zeta)| _{L^2(d\xi d\eta)}\left|\frac 1{p_\lambda}\right|_{L^2(\Omega^c_\lambda,d\xi d\eta)}   +{|  s_c(\zeta) |}_{ L^\infty(d\xi d\eta)}\left|\frac 1{p_\lambda}\right|_{L^1(\Omega_\lambda, d\xi d\eta)}\}, \nonumber
\end{align}and
\begin{eqnarray*}
&&\partial_{x_1}\mathcal CT_0  m=\mathcal C[ (\overline\lambda-\lambda)T_0  m+T_0(\partial_{x_1}   m)],\\
&&\partial_{x_1}^2\mathcal CT_0  m=\mathcal C[(\overline\lambda -\lambda  )^2T_0  m+2 (\overline\lambda-\lambda)T_0(\partial_{x_1}   m)+T_0(\partial_{x_1} ^2  m)],\\
&&\partial_{x_2}\mathcal CT_0  m=\mathcal C[ (\overline\lambda^2- \lambda^2) T _0 m +T_0(\partial_{x_2}  m)].
\end{eqnarray*}
Therefore, if  {$|\left(|\overline\lambda-\lambda|^{l_1}   +| \overline\lambda^2-\lambda^2|^{l_2}\right) s_c (\lambda)|  _{
  L^2\cap L^\infty(d\xi d\eta)}<\infty$, $l_1\le 2,\, l_2\le 1$,}   from \eqref{E:wick-infty-spec-1}, one has  $\left(\partial_{x_2}-\partial_{x_1}^2 -2\lambda \partial_{x_1} \right)\mathcal CT_0m_0\to 0$ as $\lambda_I\to\infty$. Together with \eqref{E:c-integral-eq}, \eqref{E:CIE-lax}, and letting $\lambda\gg 1$, yields
  \[
  q=q(x_2).\]Choosing $x_1\gg 1$, we obtains $q\equiv 1$. Hence the initial CIE \eqref{E:intro-CIE-new} is justified. \qed
\end{itemize}

\subsubsection{Proof of Theorem \ref{T:l-vacuum}} \hfill \\  
 Notice that
\[
 \partial_{\overline\lambda}\Phi(x, \lambda)=s_c(\lambda,x_3)\Phi(x,  \overline\lambda).
\]
Denote 
\[
\mathcal M_\lambda=- \partial_{x_3}+  \partial_{x_1}^3+\frac 32u\partial_{x_1}+\frac 34u_{x_1}+\frac 34\partial_{x_1}^{-1}u_{x_2}+\rho(\lambda) ,\ \
\rho(\lambda)= -\lambda ^3,
\] we have
\begin{align}
0=\,&\partial_{\overline\lambda}\left[\mathcal M_\lambda\Phi(x, \lambda)\right]=\mathcal M_\lambda\left[ \partial_{\overline\lambda}\Phi(x, \lambda)\right]=\mathcal M_\lambda\left[  s_c(\lambda,x_3)\Phi(x, \overline\lambda)\right]\label{E:linear}\\
=\,&\Phi(x, \overline\lambda)\left[ {-\partial_{x_3}}+\rho(\lambda)\right] s_c(\lambda,{x_3})+s_c(\lambda,{x_3})\left[ {\mathcal M_\lambda}-\rho(\lambda)\right] \Phi(x, \overline\lambda)\nonumber\\
=\,&\Phi(x, \overline\lambda)\left[-\partial_{x_3}+\rho(\lambda)\right] s_c(\lambda,{x_3})+ s_c(\lambda,{x_3})\left[{  \mathcal M_{ \overline \lambda}-\rho( \overline \lambda)}\right]\Phi(x, \overline\lambda)\nonumber\\
=\, &\Phi(x, \overline\lambda)\left[-\partial_{x_3}+\rho(\lambda)-\rho( \overline \lambda)\right] s_c(\lambda,{x_3}).\nonumber
\end{align} \qed

\subsubsection{Proof of Theorem  \ref{T:ip-vacuum}} 

\begin{itemize}

\item[$(1)$] 
{Unique solvability of the CIE \eqref{E:intro-CIE-t}} and the estimate  \eqref{E:intro-CIE-t-est}  follow  from 
\begin{multline}\label{E:infty-spec}
\hskip.6 in|\mathcal C T\phi| 
\le  \ C|  \phi|_{L^\infty}  \{|  s_c(x_3,\zeta)| _{L^2(d\xi d\eta)}\left|\frac 1{p_\lambda}\right|_{L^2(\Omega^c_\lambda,d\xi d\eta)}  \\  \hskip1 in +{|  s_c(x_3,\zeta) |}_{ L^\infty(d\xi d\eta)}\left|\frac 1{p_\lambda}\right|_{L^1(\Omega_\lambda, d\xi d\eta)}\}    \hskip.2 in
\end{multline}which is proved by the same argument as \eqref{E:infty-spec-1}.

\item[$(2)$] 
To derive the Lax equation \eqref{E:intro-Lax-u-new}, representation formula \eqref{E:intro-Lax-u-new-1}, and estimates \eqref{E:intro-Lax-u-new-2}  of the potential $u$, we  introduce the shorthand notation for the heat operator 
\[
\begin{gathered}
-\partial_{x_2}+\partial_{x_1}^2+2 \lambda\partial_{x_1}=-\nabla_2+\nabla_1^2 ,\\
\nabla_1=\partial_{x_1}+\lambda,\   \nabla_2=\partial_{x_2}+\lambda^2 ,\ \left[\nabla_j,\ T\right]=0.\end{gathered} 
\]  

Applying the heat operator to both sides of the CIE \eqref{E:intro-CIE-t},  
\be\label{E:cauchy-intro}
\begin{split}
  (-\nabla_2+\nabla_1^2  )m
=&   \left[-\nabla_2+\nabla_1^2,
  \mathcal CT  \right] m    +\mathcal CT(-\nabla_2+\nabla_1^2) m.
\end{split}
\ee
Formally,  
\begin{align}
& \left[-\nabla_2+\nabla_1^2 ,\mathcal CT \right]  m= \left[-\nabla_2+\nabla_1^2 ,\mathcal C \right] T m =2 \left[\lambda,\mathcal C \right] \partial_{x_1}\left(T  m\right)\label{E:rep-continuous-intro}\\
=&  { \frac 1{\pi i}\partial_{x_1}\iint  T m \ d\overline\zeta\wedge d\zeta\equiv-u(x) } ,  \nonumber \end{align}
along with the unique solvability of the CIE, yields the Lax equation
\begin{align}
\qquad\ \, &(-\nabla_2+\nabla_1^2 )m=-(1-\mathcal CT)^{-1}u(x)1=-u(x)(1-\mathcal CT)^{-1}1=-u(x)m(x,\lambda).\label{E:formal-derivation-intro}
\end{align}

Hence the key is to rigorously show boundedness of $(-\nabla_2+\nabla_1^2  )m$ and \eqref{E:rep-continuous-intro} by apriori estimates. To this aim, taking Fourier transform of the Lax equation \eqref{E:intro-Lax-u-new},
\[
\widehat u=-\widehat{m-1}\ast {\widehat u}-p_\lambda(\xi,\eta)\widehat{m-1}.
\]By proving
\begin{gather}
|\widehat{m-1}|_{L^1(d\xi d\eta)}\le C|s_c|_{L^2\cap L^\infty(d\xi d\eta)},\label{E:1}\\
|p_\lambda(\xi,\eta)\widehat{m-1}|_{L^2\cap L^\infty(d\xi d\eta)}\le C |s_c|_{L^2\cap L^\infty(d\xi d\eta)},\label{E:2-new}
\end{gather} one derives 
\begin{gather}
  |\widehat u(x) |_{L^2\cap L^\infty(d\xi d\eta)}\le C |s_c|_{L^2\cap L^\infty(d\xi d\eta)},\label{E:u-est-intro}\\
   (-\nabla_2+\nabla_1^2  )m\in L^\infty.\label{E:bdd-eigen}
\end{gather}  

We prove \eqref{E:1} by writing the CIE \eqref{E:intro-CIE-t} as
\begin{align}
\widehat{m-1}=&\left[\mathcal CT(m-1)   \right]^\wedge+ \left[\mathcal CT1   \right]^\wedge\label{E:m}\\
=&\left[\mathcal CT(m-1)   \right]^\wedge+\frac{2\pi is_c(\zeta(\xi,\eta))}{p_\lambda(\xi,\eta)},\nonumber\\
&\hskip1.05in {\color{blue}\left|\frac{2\pi is_c(\zeta(\xi,\eta))}{p_\lambda(\xi,\eta)}\right|_{L^1}\le C|s_c|_{L^2\cap L^\infty(d\xi d\eta)};} \nonumber\\
\left[\mathcal CTf   \right]^\wedge=&\left[-\frac 1{2\pi i}\iint_\CC\frac{s_c(\zeta)f(x,\bar\zeta)e^{(\overline\zeta-\zeta)x_1+(\overline\zeta^2-\zeta^2)x_2}}{\zeta-\lambda}d\overline\zeta\wedge d\zeta\right]^\wedge\nonumber\\
=& -\frac 1{2\pi i}\iint_\CC\frac{s_c(\zeta)\widehat f(\xi-\frac{\bar\zeta-\zeta }{2\pi i},\eta-\frac{\bar\zeta^2-\zeta ^2}{2\pi i},\bar\zeta)}{\zeta-\lambda}d\overline\zeta\wedge d\zeta\nonumber \\
=&R_{s_c}\widehat f,\hskip1in\textit{\color{blue}is a contraction of $\widehat f\in L^1(d\xi d\eta)$.}\nonumber
\end{align}
 
We prove \eqref{E:2-new} by writing \eqref{E:m} as
\begin{align*}
&p_\lambda \widehat{m-1}= p_\lambda R_{s_c}\widehat{m-1}+ {2\pi is_c(\zeta(\xi,\eta))}  
\equiv  M_{s_c}\left(p_\zeta \widehat{m-1} \right)+ {2\pi is_c(\zeta(\xi,\eta))},
\end{align*}with
\begin{gather*}
M_{s_c}f=\left(R_{s_c}  f\right)(\xi,\eta,\lambda)-\left(R_{s_c}f\right)(\xi,\eta, \frac{\eta }{2\xi}-i\pi\xi ),\\
 {|(1+|\xi|^k+|\eta|^k) M_{s_c}f|_{L^2 \cap L^\infty }\le C|(1+|\xi|^k+|\eta|^k)s_c|_{L^2\cap L^\infty }  | (1+|\xi|^k+|\eta|^k)f|_{L^2 \cap L^\infty }.}
\end{gather*}

\item[$(3)$] Finally, we justify {the KP equation} \eqref{E:intro-KP-ist-new} by verifying the Lax pairs. Namely, if  {$| (|\overline\lambda-\lambda|^{l_1}     +| \overline\lambda^2-\lambda^2|^{l_2} ) s_c (\lambda)|  _{
  L^2\cap L^\infty(d\xi d\eta)}<\infty$, $l_1\le 5,\, l_2\le 3$,} using the representation formula \eqref{E:intro-Lax-u-new-1}, we  define $\Phi(x,\lambda)= e^{ \lambda  x_1+ \lambda ^2x_2}    m(x, \lambda)$,
\be\label{E:v-evolution}
\mathcal M =- \partial_{x_3}+  \partial_{x_1}^3+\frac 32u\partial_{x_1}+\frac 34u_{x_1}+\frac 34\partial_{x_1}^{-1}u_{x_2} -\lambda ^3,
\ee and 
\[
\begin{split}
\mathcal M \Phi(x,\lambda)=&e^{ \lambda  x_1+ \lambda ^2x_2}\left(\mathcal M+3\lambda\partial_{x_1}^2+3\lambda^2\partial_{x_1}+\lambda^3+\frac 32u\lambda\right)m(x,\lambda)\\
\equiv&  e^{ \lambda  x_1+ \lambda ^2x_2}\mathfrak Mm.
\end{split}
\] By  reversing the procedure to prove \eqref{E:linear}, we obtain
\be\label{E:inverse-map-T-tilde}
\partial_{\overline\lambda}\left( \mathfrak Mm \right)(x,\lambda)=s_c(\lambda)e^{(\overline\lambda-\lambda)x_1+(\overline\lambda^2-\lambda^2)x_2 +(\overline\lambda^3-\lambda^3)x_3 }\left( \mathfrak Mm \right)(x, \overline\lambda).
\ee

Besides, as $ |\lambda|\to\infty$,  formally, letting $m(  x,\lambda)\sim\sum_{j=0}^\infty\frac{M_j(  x )}{\lambda^{j}}$, from the Lax equation \eqref{E:intro-Lax-u-new}, 
\begin{gather}
    2\partial_{x_1}M_{j+1}
=      (\partial_{x_2}-\partial_{x_1}^2-u)M_j,\label{E:v-evolution-KPII}\\
M_0=  1,\ 
M_1=  - \frac {1}2 \partial_{x_1}^{-1}u, \ 
M_2=   -\frac 14 \partial_{x_2} \partial_{x_1}^{-2}u+ \frac 14 u+\frac 14 \partial_{x_1}^{-1}\left(u \partial_{x_1}^{-1} u\right),\ \cdots\nonumber
\end{gather}  As a result, as $\lambda\to\infty$,  
\begin{align}
&\mathfrak Mm\label{E:v-evolution-asm}\\
\to&\frac34 u_{x_1}+\frac 34 \partial_{x_1}^{-1}u_{x_2} 
+3\lambda\partial_{x_1}^2(1+\frac{M_1}{\lambda}) +3\lambda^2\partial_{x_1}(1+\frac{M_1}{\lambda}+\frac{M_2}{\lambda^2})+\frac 32u\lambda \nonumber\\
=&\frac34 u_{x_1}+\frac 34 \partial_{x_1}^{-1}u_{x_2}+\left(
-\frac32u_{x_1}+3\partial_{x_1}[-\frac 14\partial_{x_2}\partial_{x_1}^{-2}u+\frac{u}{4}+\frac14\partial_{x_1}^{-1}(u\partial_{x_1}^{-1}u)] 
\right)\nonumber\\
+& \lambda\left( 3\partial_{x_1}M_1+\frac{3}{2}u\right) +\frac32u(-\frac 12\partial_{x_1}^{-1}u)\nonumber\\
=&0.\nonumber\end{align}
Using the unique solvability of the CIE, $\mathfrak M m(x,\lambda)=0$,  $\mathcal M \Phi(x,\lambda)=0$.  Hence   the Lax pair holds and the Cauchy problem of the KPII equation is solved.\end{itemize}\qed

\section{The IST for perturbed $1$ line solitons}\label{S:1-soliton}
\medskip

\subsection{$1$-line solitons}\label{SS:solitons}\hfill \\ 
The KPII equation \eqref{E:KPII-intro} possesses explicit interesting solutions, $   {\mathrm{Gr}(N,M)_{\ge 0}}$ KP  solitons,  which are regular in the entire $x_1x_2$-plane with peaks localized and non decaying along certain line segments and rays for each fixed time $x_3$. They can be constructed by the Sato theory \cite{S81,SS82,S89a,S89b,BC07,BK03,KW13}    
\be\label{E:line-tau}
u_s(x)= 2\partial^2_{x_1}\ln\tau(x) 
\ee     where the $\tau$-function is  the Wronskian determinant
\begin{align}
\tau(x)=&\left|
\left(
\begin{array}{cccc}
a_{11} &a_{12} & \cdots & a_{1M}\\
\vdots & \vdots &\ddots &\vdots\\
a_{N1} &a_{N2} & \cdots & a_{NM}
\end{array}
\right)
\left(
\begin{array}{ccc}
E_{1} & \cdots & \kappa_1^{N-1}E_1\\
E_{2} & \cdots & \kappa_2^{N-1}E_2\\
\vdots & \ddots &\vdots\\
E_{M} & \cdots & \kappa_M^{N-1}E_M\\
\end{array}
\right)
\right|\label{E:line-grassmannian} \\
=&\sum_{1\le j_1< \cdots< j_N\le M}\Delta_{j_1,\cdots,j_N}(A)E_{j_1,\cdots,j_N}(x),\nonumber
\end{align}
with $\kappa_1<\cdots<\kappa_M$,  $E_j(x)=\exp\theta_j(x)=\exp( \kappa_j x_1+\kappa_j^2 x_2+\kappa_j^3 x_3)$,
 $A=(a_{ij})\in {\mathrm{Gr}(N, M)_{\ge 0}}$ (TNN, full rank $N\times M$ real matrices with non negative minors),    $\Delta_{j_1,\cdots,j_N}(A)=\Delta_{J}(A)$  the $N\times N$ minor of the matrix $A$ whose columns are labelled by the   index set $J=\{j_1< \cdots< j_N\}\subset\{1,\cdots,M\}$, and $
 E_J=E_{j_1,\cdots,j_N}(x)=\Pi_{l<m}(\kappa_{j_m}- \kappa_{j_l})\exp ( \sum_{n=1}^N\theta_{j_n}(x)  )$. Moreover, $ \textrm{Gr}(N,M)_{> 0}$ KP solitons means all minors $\Delta_{j_1,\cdots,j_N}$ are positive,  {namely, fulfilling the totally positive (TP) condition}, form a dense subset of ${ \mathrm{Gr}(N,M)_{\ge 0}}$ KP solitons. In particular, the $   {\mathrm{Gr}(1,2)_{> 0}}$ KP  solitons
(or $1$-line solitons for simplicity)
\be \label{E:line-tau-oblique-1}
 u_s(x) 
=  \frac{(\kappa_1-\kappa_2)^2}2\textrm{sech}^2\frac{\theta_1(x)-\theta_2(x)-\ln a}2.
\ee are given by  $A=(1,a) $, $a>0$ in  \eqref{E:line-tau} - \eqref{E:line-grassmannian}.

A brief introduction on the Sato theory will be given in $\S$ \ref{SSS:N-solitons} and a formal IST which is applicable to a class  of functions at least containing  
multi line solitons has been shown in \cite{Wi00, Wi02, WS04}. Without involved with the Sato theory, we provide a rigorous IST  for perturbed $1$-line solitons in this section.  

\begin{lemma}[{\bf IST for $1$-line solitons}]\label{L:Sato-Lax} \cite[(2.12)]{BP214}, \cite[Theorem 6.3.8., (6.3.13) ]{D91}, \cite[Proposition 2.2, (2.21)]{K17}  
 Let $u_s(x)$ be a $   {\mathrm{Gr}(1,2)_{> 0}}$ KP  soliton.   The {Sato eigenfunction} $\varphi $ and the {Sato normalized eigenfunction} $\chi$, defined by  
\be\label{E:sato}
\begin{split}
\varphi(x,\lambda)&= e^{\lambda x_1+\lambda^2 x_2}\frac{(1-\frac{\kappa_1}\lambda)e^{\theta_1 (x )}+   (1-\frac{\kappa_2}\lambda)ae^{\theta_2(x )}}{e^{\theta_1(x )}+ a e^{\theta_2 (x )}}  \equiv e^{\lambda x_1+\lambda^2 x_2}\chi(x ,\lambda) 
\end{split}\ee  
 satisfy  the Lax equations  for $\lambda\in\CC\backslash\{0\}$, 
\be\label{E;spectral}
\begin{split}
  \left(-\partial_{x_2}+\partial^2_{x_1}+u_s(x )\right)\varphi(x ,\lambda)=&0,\\
  \left(-\partial_{x_2}+\partial^2_{x_1}+2\lambda\partial_{x_1}+u_s(x )\right)\chi(x ,\lambda)=&0;
\end{split}
\ee  
 and
\be\label{E:linearization-D-evol-new-1-s}
\begin{gathered}
\hskip1in{  { \chi}(x, \lambda) =1+ \frac{  \chi_{ 0, \res }(x   )}{\lambda  }  , } \qquad \textit{   $\lambda\in\CC\backslash\{0\}$, } \\
(e^{\kappa_1x_1+\kappa_1^2x_2+\kappa_1^3x_3}\chi(x,\kappa _1) ,e^{\kappa_2x_1+\kappa_2^2x_2+\kappa_1^3x_3}\chi(x,\kappa _2)){ \mathcal D ^\flat}=0,
   \end{gathered}
   \ee with
\be\label{E:intro-sym-N-D-new}
\begin{split}
& {\mathcal  D}^\flat = \textit{diag}\,(   
\kappa _1 , \kappa _2 )\, A^T=\left( \begin{array}{l}  \kappa _1\\ \kappa _2 a \end{array} \right).  
\end{split}\ee

  The {\sl forward scattering transform}  is defined by
\be\label{E:intro-SD-d} 
\begin{split}
\mathcal S:u_s(x_1,x_2,0)\mapsto &\{0,\kappa_1,\kappa_2, \mathcal D^\flat\},
\end{split}
\ee and the inverse scattering transform   by 
\be\label{E:intro-SD-dinv} 
\mathcal S^{-1}( \{0,\kappa_1, \kappa_2,\mathcal D^\flat\})= {-2  \partial_{x_1}\chi_{0,\res}(x)}.
\ee

\end{lemma}

\begin{proof}  
Writing 
\[
\ba{rl}
\tau(x )=&2e^{\frac 12(\theta_1(x) +\theta_2(x) +\ln a)}\cosh\frac 12(\theta_1(x) -\theta_2(x) -\ln a),\\
u_s(x)=&\frac{(\kappa_1-\kappa_2)^2}2\textrm{sech}^2\frac 12(\theta_1(x)-\theta_2(x)-\ln a) ,\\
\chi(x,\lambda)=&1-\frac{1}{\lambda}\partial_{x_1}\ln \tau(x), 
\ea\] one can  verify that the Sato normalized eigenfunction $\chi$ satisfies  the Lax equation \eqref{E;spectral}  by checking  $\left(-\partial_{x_2}+\partial^2_{x_1}+2\lambda\partial_{x_1}+u_s(x)\right)\chi(x,\lambda)$ is a holomorphic function tending to $0$ at $\infty$.

 Notice that \eqref{E:linearization-D-evol-new-1-s} can be viewed as a linear system of   $1+2$ variables $\chi_{0,\res}$,  $\chi(x,  \kappa_1)$,  and $\chi(x,  \kappa_2)$ by evaluating the first equation at $\kappa_1$ and $\kappa_2$. A direct computation yields $\chi_{0,\res}(x)=-\partial_{x_1}\ln \tau(x)$.

\end{proof}

\subsection{The direct problem for pertubed KP $1$-solitons}\label{SS:background} 
 
\subsubsection{Statement of results}\hfill \\  Building upon Boiti et al's work \cite{BP301,BP302,BP309,BP310,BP211,BP212,BP214}, rigorous direct scattering theory for perturbed $  {\mathrm{Gr}(1,2)_{> 0}}$ KP  solitons is carried out in  \cite{Wu20, Wu21}.   

\begin{theorem}  \label{T:Lax-1}\cite{Wu20, Wu21}
Given
\be\label{E:intro-ini-data-p}
  \begin{array}{c}
  u_0(x_1,x_2)=u_s(x_1,x_2,0)+v_0(x_1,x_2),\ \
  \textit{$ u_s(x)$  a ${\mathrm{Gr}(1,2)_{> 0}}$ KP soliton,  } 
  \\
  \textit{$   \sum_{|l|\le { d+8}} |\corr{(1+|x_1|+|x_2|)}\partial_x^lv _0|_  {L^1\cap L^\infty} \ll 1$,  $d\ge 0$,}
 \end{array}
 \ee one has
 \begin{itemize}   
 \item [(1)] for $\forall\lambda\in\CC\backslash\{0,\kappa_j\}$, the unique existence of the eigenfunction $\Phi(x_1,x_2,\lambda)=e^{ \lambda  x_1+ \lambda ^2x_2}m_0(x_1,x_2,\lambda)$ of the Lax equation 
   \begin{align}
(-\partial_{x_2}+\partial_{x_1}^2 +2\lambda\partial_{x_1}
+u_0(x_1,x_2))m_0(x_1,x_2 ,\lambda)= 0,   \label{E:intro-Lax-p}\\
\lim_{|x|\to\infty}m_0(x_1,x_2,\lambda)=   \chi(x_1,x_2,0,\lambda) . \label{E:intro-bdry-p} 
\end{align}

\item [(2)]  The {forward scattering transform} is defined by 
\be\label{E:intro-SD-p} 
\mathcal S(u_0 )=( 0, \kappa_1, \kappa_2,\mathcal  D,s _c(\lambda)) 
\ee which  satisfies the  Cauchy integral equation and the $\mathcal D$-symmetry 
\begin{gather}   
    {   m}_0(x_1,x_2, \lambda) =1+ \frac{   m_{0,  \res }(x_1,x_2  )}{\lambda  }  +\mathcal C  T_0
     m_0(x_1,x_2, \lambda) ,   \label{E:intro-CIE-0-p} \\
  (e^{\kappa_1x_1+\kappa_1^2x_2 }m_0(x_1,x_2,\kappa^+_1), e^{\kappa_2x_1+\kappa_2^2x_2 }m_0(x_1,x_2,\kappa^+_2))\mathcal D=0, \label{E:intro-CIE-0-p-d}\\
  m_0\in W_0, \quad\kappa_j^+=\kappa_j+0^+.  \label{E:intro-CIE-0-p-w}   \end{gather}      
      Here $\mathcal C$ is the Cauchy integral operator,      $T_0$ is the continuous scattering operator at $x_3=0$  defined by 
\eqref{E:ct-operator} with  $s_c(\lambda)$ being the continuous scattering data, 
\be \label{E:conti-sc-debar}
\begin{gathered}
\partial_{\overline\lambda}m_0(x_1,x_2, \lambda)
=   s_c(\lambda) e^{(\overline\lambda-\lambda)x_1+(\overline\lambda^2-\lambda^2)x_2  }m_0{(x_1,x_2, \overline\lambda)},\ \lambda\notin\RR,\\
  s_c(\lambda) 
  =   \frac {\sgn(\lambda_I)}{2\pi i} \left[\xi(\cdot,0,\bar\lambda)  v_0(\cdot) m_0 (\cdot,\lambda)\right]^\wedge(\frac{\overline\lambda-\lambda}{2\pi i},\frac{\overline\lambda^2-\lambda^2}{2\pi i}), 
\end{gathered}
\ee and  $\xi(x)= \xi(x,\lambda)
$ being the normalized Sato adjoint eigenfunction (see \eqref{E:sato-adjoint}, \eqref{E;spectral-new} for definition). Moreover, $\mathcal D$  can be computed by 
\begin{align}
&  \mathcal  D =  {\mathcal  D}^\sharp \times \left({\mathcal  D}_{11}^\sharp\right)^{-1} \times \kappa_1 = \left( \begin{array}{l}  \kappa _1\\ \mathcal D_{21} \end{array} \right),\label{E:intro-sym-N-D-0}\\
&\mathcal  D^\sharp=\left({ \ba{l}\mathcal D^\sharp_{1 1}\\
\mathcal D^\sharp_{2 1}
\ea}\right)= {\left({ \ba{l}\mathcal D^\flat_{11}+\dfrac{c_{11}\mathcal D^\flat_{11}}{1-c_{11}}+\dfrac{c_{12}\mathcal D^\flat_{21}}{1-c_{11}}\\
\mathcal D^\flat_{21}+\dfrac{c_{22}\mathcal D^\flat_{21}}{1-c_{22}}
\ea}\right)}, \label{E:intro-sym-N-D}\\ 
& \mathcal D^\flat= \textit{diag}\,(   
\kappa _1 , \kappa _2 )\, A^T=\left( \begin{array}{l}  \kappa _1\\ \kappa _2 a \end{array} \right),\label{E:intro-sym-N-D-1}
\end{align}with $c_{jl}=-\int\Psi _j(x_1,x_2,0 ) v_0(x_1,x_2)\varphi_l(x_1,x_2,0  )dx_1dx_2$, $\Psi_j(x)$, $\varphi_l(x)$   residues of the adjoint eigenfunction $\Psi(x,\lambda)$ (see \eqref{E:adjoint} for definition) at $\kappa_j$ and values of the Sato eigenfunction $\varphi(x,\lambda)$ at $\kappa_l$; ${ W_0 }=W_{(x_1,x_2,0)}$ is the eigenfunction space defined in Definition \ref{D:phase-0}.      
 
Finally, the scattering data  
 $\mathcal S(u_0)$ satisfies the algebraic and analytic constraints 
\begin{align}
& s_c(\lambda)=
\left\{
{\ba{ll}
 {\frac{ \frac {i}{ 2} \sgn(\lambda_I)}{\overline\lambda-\kappa_j}\frac{ \gamma_j}{1-\gamma _j|\alpha|}}+\sgn(\lambda_I)  h_j(\lambda),&\lambda\in   D^ \times_{ \kappa_j },\\
\sgn(\lambda_I) {  \hbar_0}(\lambda),&\lambda\in    D^\times _{ 0},
\ea}
\right.\label{E:intro-s-c-N}\\
&    \mathcal  D =  \left(    \kappa _1, \mathcal D_{21}   \right)^T,\nonumber
\end{align}
 and   
 \begin{align}
 &  {\begin{array}{l} 
 |(1-\sum_{j=1}^2E_{{\kappa_j}}  )  \sum_{|l|\le {d+8}}|\left(|\overline\lambda-\lambda|^{l_1}   +| \overline\lambda^2-\lambda^2|^{l_2}\right) s_c (\lambda)|  _{ 
  L^\infty} \\
   +   \sum_{j=1}^2(|\gamma_j|+|h_j|_{ \color{black}L^\infty(D_{\kappa_j})})+ |\hbar_0|_{  \color{black}C^1(D_{0})}+{|\mathcal D   -\mathcal D^\flat|_{L^\infty}} \\
  \le  {C\sum_{|l|\le  { d+8}} |{\color{black} (1+|x_1|+|x_2|) } \partial_{x }^{l }      v_0|_{L^1\cap L^\infty}} , \end{array}}\label{E:intro-s-c-ana}\\
  &s_c(\lambda)=  \overline{s_c( \overline\lambda)},
  h_j(\lambda)=-\overline{h_j( \overline\lambda)}, 
 \hbar_0(\lambda)=-\overline{\hbar_0( \overline\lambda)}. \label{E:intro-s-c-reality}
 \end{align} 
 Here $D_{z,a\delta}=\{   \lambda=z+re^{i\alpha}:0\le r\le {a \delta},|\alpha|\le\pi\}$, $D_{z,a\delta}^\times=D_{z,a\delta}\backslash\{z\} $, $1\ge \delta=\frac 12\inf \{|z-z'|:z, z'  \in  \{ 0,\, \kappa_1,\, \kappa_2 \},\ z\ne z'\}$,   $  E_{z,a\delta}(\lambda)\equiv 1$ on $D_{z,a\delta}$, $  E_{z,a\delta} (\lambda)\equiv 0$ elsewhere.  We suppress the $a\delta$-dependence for simplicity if $a=1$.

\end{itemize}
\end{theorem}  
\begin{definition}\label{D:phase-0} The {eigenfunction  space} ${ W_0 }=W_{x_1,x_2,0}$  consists of $\phi$ satisfying 
\begin{itemize}
\item [$(a)$] $\phi (x_1,x_2, \lambda)=\overline{ \phi (x_1,x_2, \overline\lambda)};$ 
\item [$(b)$] $(1-  E_{0} )\phi(x_1,x_2, \lambda)\in L^\infty;$ 
\item [$(c)$] for $\lambda \in D_{0}^\times$, 
$
\phi(x_1,x_2, \lambda)=\frac{ {\phi_{0,\res}(x_1,x_2)}}{\lambda  }  +\phi_{0,r}(x_1,x_2, \lambda)$, $ \phi_{0,\res}$,   $\phi_{0,r}  \in L^\infty( D_{0})$; 
\item[$(d)$] for $\lambda =\kappa_j+re^{i\alpha}\in D_{\kappa_j}^\times$, $\phi=\phi^\flat+\phi^\sharp $, 
$\phi^\flat =\sum_{l=0}^\infty \phi_l(X_1,X_2)(-\ln(1-\gamma_j|\alpha|))^l \in L^\infty(D_{\kappa_j}) $, $\phi^\sharp \in C_{\tilde\sigma}^{\mu } (D_{\kappa_j,\frac{1}{\tilde\sigma}})\cap  L^\infty (D_{\kappa_j})$, $\phi^\sharp(x_1,x_2,\kappa_j)=0$. 
\end{itemize}Here 
\be\label{E:sigma-parameter-0}
\tilde\sigma=   \max\{ 1, |X_1|,  \sqrt{|X_2|}\},
\ee is  the rescaling parameter with $X_k$ defined by  the coefficients of the phase function of   $T$, 
\be\label{E:phase-0}
\begin{split}
\wp(x_1,x_2,\lambda)=& i[(\overline\lambda- \lambda){x_1}+(\overline\lambda^2-\lambda^2){x_2}] = X_1r\sin\alpha  +X_2r^2 \sin2\alpha   
\equiv \wp(r,\alpha,X),\\
 X_1 = &2(x_1+2x_2z  ),\   
 X_2 =   2 x_2 , \ \lambda=z+re^{i\alpha} \in D_z.
 \end{split}
 \ee  
     Finally, $C_{\tilde\sigma}^{\mu } (D_{\kappa_j,\frac{1}{\tilde\sigma}})=C (D_{\kappa_j,\frac{1}{\tilde\sigma}})\cap H_{\tilde\sigma}^{\mu } (D_{\kappa_j,\frac{1}{\tilde\sigma}})$ and the rescaled H$\ddot{\mbox{o}}$lder space $H_{\tilde\sigma}^{\mu } (D_{z,\frac{1}{\tilde\sigma}})$  for $z\in\RR$,  consists of  functions $\phi(x,\lambda)\equiv \phi(r,\alpha,X_1,X_2)$,    satisfying
\be\label{E:rescaled-holder-0}
\begin{split}
&|\phi|_{ H_{\tilde\sigma}^\mu (D_{z,\frac{1}{\tilde\sigma}})} \equiv   {\color{black}\sup_{\textrm{\tiny ${\ba{c}  \tilde r_1,  \tilde r_2\le 1,
 |\alpha_1|,|\alpha_2|\le \pi\ea}$}} \frac{|\phi(\frac{ \tilde r_1}{\tilde\sigma}, \alpha_1,X_1,X_2)-\phi(\frac{ \tilde r_2}{\tilde\sigma},\alpha_2,X_1,X_2)|}{ |\tilde r_1e^{i\alpha_1}-\tilde r_2e^{i\alpha_2}|^\mu} } <\infty
 \end{split}
\ee for $
\lambda_j=z+r_j e^{i\alpha_j}=z+\frac{\tilde r_j}{\tilde\sigma} e^{i\alpha_j}\in D_{z,\frac{1}{\tilde\sigma}}$.

\end{definition}
 \begin{theorem}  \label{T:1-soliton-evolution} \cite{Wu20,Wu21} If $\Phi= e^{ \lambda  x_1+ \lambda ^2x_2}    m(x, \lambda)$ satisfies the Lax pair \eqref{E:KPII-lax-1} and  
\begin{gather}
\partial_{\overline\lambda}  m(x, \lambda)=  {  {  s}_c(\lambda,x_3)}e^{(\overline\lambda-\lambda)x_1+(\overline\lambda^2-\lambda^2)x_2}   m(x,\overline\lambda)  ,\label{E:debar-evol}\\
  (e^{\kappa_1x_1+\kappa_1^2x_2}m(x,\kappa^+_1),e^{\kappa_2x_1+\kappa_2^2x_2}m(x,\kappa^+_2)){ \mathcal D(x_3)}=0,\label{E:D-evol}
\end{gather} with $\mathcal D(x_3)=(\kappa _1, \mathcal D_{21}(x_3)  )^T$ then  
\be\label{E:linearization-D-evol}
\begin{gathered}
 {  s}_c(\lambda, x_3)=   {e^{ (\overline\lambda^3-{ \lambda}^3)x_3}}{  s}_c(\lambda ),\quad
   {\mathcal {\mathcal D}}_{mn}(x_3)=  { e^{ (\kappa_m^3-\kappa_n^3)x_3}}  {\mathcal D}_{mn}.  
   \end{gathered}\ee   	
 
\end{theorem}

 We make several remarks to conclude this subsection. 
   
For perturbed $1$-line solitons, besides the CIE, we have extra $\mathcal D$-symmetry which is defined by the values of the eigenfunction at $\kappa_j^+$. The decomposition $\phi^\sharp+\phi^\flat$ and the scaled topology $C^\mu_{\tilde\sigma}$ of the eigenfunction space $W_0$ are introduced to obtain the pointwise structures of the CIE at $\kappa_j$ for the inverse problem. To see this, 
for a perturbed $1$-soliton, away from $0$, $\kappa_j$, the continuous data $s_c$ and eigenfunction $m_0$ are as regular as those in case of rapidly decaying potentials. But 
\begin{itemize}
\item  At $0$, the boundary value $\chi$ has a simple pole; the eigenfunction $m_0$ has a "simple pole". Along with \eqref{E:intro-s-c-N}, near $0$, the Cauchy integral operator $\mathcal CT_0m_0$ is an  oscillatory singular operator.
\item At $\kappa_j$, the Green function $G$ is multi-valued; the eigenfunction $m_0$ is multi-valued; the continuous scattering data $s_c$ has a "simple pole  with a discontinuous residue" (see \eqref{E:intro-s-c-N}). Therefore, near $\kappa_j$, the Cauchy integral operator $\mathcal CT_0m_0$ is an oscillatory singular operator. 

\end{itemize}

Comparing \eqref{E:conti-sc-debar}-\eqref{E:intro-sym-N-D-1}, and  \eqref{E:sato-adjoint} with \eqref{E:intro-SD} and \eqref{E:intro-sym-N-D-new}, we prove  
when discrete scattering data or continuous scattering data vanish, the forward  scattering transform constructed for perturbed $1$ line solitons degenerate into those transforms  for rapidly decaying potentials or for $1$-line solitons.  

\subsubsection{The strategy of the proof of Theorem \ref{T:Lax-1}} \label{SS:proof-2.1} \hfill \\  
Throughout $\S$\ref{SS:proof-2.1}, $x=(x_1,x_2,0),\,x'=(x'_1,x'_2,0)$ for simplicity. 
\begin{itemize}
\item[$(1)$] The Lax equation  is proved by 
\begin{align}
m_0(x,\lambda)=& \chi(x , \lambda)-G \ast v_0m_0, \label{E:c-integral-eq-p}\\
  G(x ,x' , \lambda) =&G_c(x,x',\lambda)+G_d(x,x',\lambda),\label{E:wick-infty-spec-p}\\
G_c(x,x',\lambda)=&- \frac{\sgn(x_2-x_2')}{2\pi}e^{ \lambda (x_1'-x_1)+\lambda^2 (x_2'-x_2) } \int_{\RR} \theta((s^2-\lambda_I^2)(x_2 - x_2')) \,  \label{E:wick-infty-spec-1-p}\\
\times&  \varphi(x ,\lambda_R+is) \psi(x' ,\lambda_R+is) \, ds,\nonumber\\
G_d(x,x',\lambda)=&  -\theta(x_2' - x_2)e^{ \lambda (x_1'-x_1)+\lambda^2 (x_2'-x_2) } \label{E:wick-infty-spec-2-p}\\
\times&( \theta(\lambda_R - \kappa_1)\varphi_1(x ) \psi_1(x' )+  \theta(\lambda_
R-\kappa_2) \varphi_2(x) \psi_2(x')  ) ,\nonumber
\end{align}and
\begin{align}
G\ast f(x,\lambda)\equiv&\iint  G(x,x',\lambda)f(x' )dx',\nonumber\\
| G_c(x,x',\lambda)|\le& C (1+\frac1{\sqrt{|x_2-x_2'|}} ),\label{E:G-c-b}\\
| G_d(x,x',\lambda)|\le& C ,\label{E:G-d-b}\\
\lim_{|x|\to\infty}G(x,x',\lambda)&\ast f(x')  = 0.\nonumber
\end{align} Here  $\theta(s)$ is the Heaviside function, $\psi$, $\xi$  are the {Sato adjoint eigenfunction, Sato normalized adjoint eigenfunction} \cite[(2.12)]{BP214}, \cite[Theorem 6.3.8., (6.3.13) ]{D91} 
\begin{align}
 \psi(x_1,x_2,x_3,\lambda) 
 = & e^{-(\lambda x_1+\lambda^2 x_2)}\frac{\frac {e^{\theta_1(x_1,x_2,x_3) }}{(1-\frac{\kappa_1}\lambda)}+   \frac{ae^{\theta_2(x_1,x_2,x_3)}}{(1-\frac{\kappa_2}\lambda)}}{e^{\theta_1(x_1,x_2,x_3)}+ a e^{\theta_2(x_1,x_2,x_3) }} \label{E:sato-adjoint}\\
\equiv&  e^{-[\lambda x_1+\lambda^2x_2]} \xi(x_1,x_2,x_3,\lambda),\nonumber
\end{align}    satisfying 
\be\label{E;spectral-new}
\begin{split}
  &\left(\partial_{x_2}+\partial^2_{x_1}+u_s(x_1,x_2,x_3)\right)\psi(x_1,x_2,x_3,\lambda)=0,\\
   & \left(\partial_{x_2}+\partial^2_{x_1}{-}2\lambda\partial_{x_1}+u_s(x_1,x_2,x_3)\right)\xi(x_1,x_2,x_3,\lambda)=0.
\end{split}
\ee 
Finally,
\[
\begin{split}
 \varphi_j(x)\equiv&\varphi(x,\kappa _j)=e^{\kappa_j x_1+\kappa_j^2x_2} \chi_j(x),\ 
 \psi_j(x)\equiv \textrm{res}_{\lambda=\kappa_j}\psi(x,\lambda)=e^{-[\kappa_j x_1+\kappa_j^2x_2]}\xi_j(x).
 \end{split}
\]

We explain construction of the Green function   and estimates   in the following.
\begin{itemize}
\item [$\blacktriangleright$] {\bf Construction of the Green function \eqref{E:wick-infty-spec-p}-\eqref{E:wick-infty-spec-2-p}:} Using Fourier inversion theorem, the residue theorem, and the orthogonality
\be\label{E:orthogonality-1}
\sum_{j=1}^2\varphi_j (x)\psi_j(x')=0, 
\ee we first derive the orthogonality relation
\begin{multline}\label{E:ortho}
 \hskip.8in \delta({x-x'})= \delta({x_2-x'_2})  \{ \frac 1{2\pi}   {\int_{\mathbb R} } \varphi (x,\lambda_R+is)\psi (x',\lambda_R+is) ds\\
	 - {\sum_{j=1}^2 }\varphi_j(x)\psi_j(x')\theta(  \lambda_R-\kappa_j)\}.\hskip.2in
\end{multline}  

Therefore $G$ defined by \eqref{E:wick-infty-spec-p}-\eqref{E:wick-infty-spec-2-p} satisfies
\[
  \left(-\partial_{x_2}+\partial^2_{x_1}+2\lambda\partial_{x_1}+u_s(x)\right)  G(x,x', \lambda)= \delta(x-x')  
\] by applying \eqref{E:ortho} and 
\[
\begin{split}
\hskip.7in  \sgn (x_2-x_2')\theta((s^2-\lambda_I^2)(x_2-x_2'))=\theta(x_2-x'_2)\chi_{-}(s)-\theta(x_2'-x_2)\chi_{+}(s),
\end{split}\] where $\chi_{\pm}(s)$  the characteristic function for  $\{s|\,\textit{Re}(\left[\lambda+is\right]^2-\lambda^2)\gtrless 0\}$.

\item [$\blacktriangleright$] {\bf Estimates of the Green function \eqref{E:G-c-b}, \eqref{E:G-d-b}:} 
Firstly, note
\begin{align}
 &G_c(x,x', \lambda) \label{E:g-c}\\
=&-\frac{e^{i[\lambda_I(x_1'-x_1)+2\lambda_I\lambda_R(x'_2-x_2)]}}{2\pi}
\int_\RR ds\ \mbox{sgn}(x_2-x_2')\theta((s^2-\lambda_I^2)(x_2-x_2'))  \nonumber\\
 \times&  e^{(s^2-\lambda^2_I)(x_2'-x_2)- is[  (x_1'-x_1)+2 \lambda_R  (x_2'-x_2)] }\chi(x,\lambda_R+is)\xi(x',\lambda_R+is).\nonumber
\end{align} Hence  the following  proof for $G_c$ only requires totally non negative (TNN) condition. Precisely, if $\lambda\in D_{\kappa_1}^c\cap D_{\kappa_2}^c$, then a direct computation or special function properties yields
\[
\begin{split}
 |G_c (x,x', \lambda)| 
\le &\ C\theta(x_2-x_2')(\int^{-|\lambda_I|}_{-\infty}+\int^\infty_{|\lambda_I|})   e^{   (s^2-\lambda^2_I)(x_2'-x_2)}ds\\
&+C\theta(x'_2-x_2)\int_{-|\lambda_I|}^ {|\lambda_I| }e^{   (s^2-\lambda^2_I)(x_2'-x_2)}ds \\
\le & C (1+\frac 1{\sqrt{|x_2-x_2'|}} ).
\end{split}
\]

For $\lambda\in D_{\kappa_j}^\times$, decompose \eqref{E:g-c} as
\be\label{E:green-kappa-0}
G_c(x,x', \lambda)= -\frac{e^{i[\lambda_I(x_1'-x_1)+2\lambda_I\lambda_R(x'_2-x_2)]}}{2\pi}\left(I^{[1]}_j+I^{[2]}_j+I^{[3]}_j+I^{[4]}_j   \right), \ee
with
\begin{align}
I^{[1]}_j=:&\int_{- \delta}^ {\delta}\mbox{sgn}(x_2-x_2')\theta((s^2-\lambda_I^2)(x_2-x_2'))\chi(x, \lambda_R+is) \xi(x',\lambda_R+is)\label{E:green-kappa}\\
&\times  [e^{is[x_1-x_1'+2\lambda_R(x_2-x_2')]+(\lambda_I^2-s^2 )(x_2-x_2')}-1]  ds,\nonumber\\
I^{[2]}_j=:&\int_{- \delta}^ {\delta}\mbox{sgn}(x_2-x_2')\theta((s^2-\lambda_I^2)(x_2-x_2')) \nonumber\\
&\times[ \chi(x, \lambda_R+is) \xi(x',\lambda_R+is)
- \frac{\chi_j(x)\xi_j(x')} {\lambda_R+is-\kappa_j}
]ds, \nonumber\\
I^{[3]}_j=:&\int_{- \delta}^ {\delta}\mbox{sgn}(x_2-x_2')\theta((s^2-\lambda_I^2)(x_2-x_2'))\frac{\chi_j(x)\xi_j(x')} {\lambda_R+is-\kappa_j}ds,\nonumber\\
I^{[4]}_j=:& \left(\int_{-\infty}^{- \delta}+\int_\delta ^\infty\right)\mbox{sgn}(x_2-x_2')\theta((s^2-\lambda_I^2)(x_2-x_2')) \chi(x, \lambda_R+is)\nonumber\\
&\times \xi(x',\lambda_R+is)e^{(s^2-\lambda^2_I)(x_2'-x_2)- is[  (x_1'-x_1)+2 \lambda_R  (x_2'-x_2)] }  ds. \nonumber
\end{align} 

One can prove 
\[|I^{[1]}_j|,\,|I^{[2]}_j|,\,|I^{[3]}_j|<C,\quad |I^{[4]}_j| \le C (1+\frac1{\sqrt{|x_2-x_2'|}} ) , \] in particular, uniform estimates for $|I^{[1]}_j|$ are achieved by the 
 residue theorem, contour integrals, and changes of variables and  are involved a bit. 
 Hence  \eqref{E:G-c-b} is proved.

As for $G_d$, note $\varphi_j(x)$, $\psi_j(x')$ have $2$-cells and $0$-cells in nominators respectively, i.e.,
\[
\begin{split}
\varphi_1(x)= \frac{1-a\frac{\kappa_2}{\kappa_1}e^{\theta_1(x)+\theta_2(x)}}{e^{\theta_1(x)}+a e^{\theta_2(x)}},&\quad \varphi_2(x)=\frac{1- \frac{\kappa_1}{\kappa_2}e^{\theta_1(x)+\theta_2(x)}}{e^{\theta_1(x)}+a e^{\theta_2(x)}}, \\
\psi_1(x')=\frac{\kappa_1}{e^{\theta_1(x')}+a e^{\theta_2(x')}},&\quad \psi_2(x')=\frac{a\kappa_2}{e^{\theta_1(x')}+a e^{\theta_2(x')}} .
\end{split}
\] 
Following argument to  {permute and exchange} cells  \cite{BP211}, \cite{BP212},   
one obtains the decomposition
\be\label{E:discon-G-decomposition}
G_d(x,x',\lambda)=G_{d}^1(x,x',\lambda)+G_{d}^2(x,x',\lambda),
\ee
with
\begin{align}
&G_{d}^1(x,x',\lambda)\nonumber\\
=\,&(\kappa_2- \kappa_1 ) a\,e^{i\lambda_I[x'_1-x_1 +2\lambda_R(x_2'-x_2 )]}\,\theta(k_{12}(x_2'-x_2))\,\theta((\lambda_R-\kappa_1)(z_{12}-z'_{12})) \nonumber\\
&\times  {\sgn(z_{12}-z_{12}')}\dfrac{e^{-k_{12}(x'_2-x_2)+(\lambda_R-\kappa_1)(z'_{12}-z_{12})}\,e^{\theta_1(x')+\theta_2(x)}}{\tau(x)\tau(x')}  ,
\label{E:discon-G-decomposition-1}
\end{align}and
\begin{align}
&G_{d}^2(x,x',\lambda)\nonumber\\
= &(\kappa_1-\kappa_2 ) a\, e^{i\lambda_I[x'_1-x_1 +2\lambda_R(x_2'-x_2 )]}\theta(k_{12}(x_2'-x_2))\,\theta((\lambda_R-\kappa_2)(z_{12}-z'_{12})) \nonumber\\
&\times  {\sgn(z_{12}-z_{12}')}\dfrac{e^{-k_{12}(x'_2-x_2)+(\lambda_R-\kappa_2)(z'_{12}-z_{12})}\,e^{\theta_1(x )+\theta_2(x')}}{\tau(x)\tau(x')} ,
\label{E:discon-G-decomposition-2}
\end{align}where
\be\label{E:z-mn}
\begin{gathered}
z_{mn}=x_1+(\kappa_m+\kappa_n)x_2, \quad z'_{mn}=x'_1+(\kappa_m+\kappa_n)x'_2,\\
k_{mn}=\lambda_I^2-(\lambda_R-\kappa_m)(\lambda_R-\kappa_n),
\end{gathered}
\ee  for $m,\,n\in\{1,\,2\}$. Now all of the exponentials in the nominators of \eqref{E:discon-G-decomposition-1}, \eqref{E:discon-G-decomposition-2} are either bounded or dominated by the tau functions   in the denominators by the totally positive (TP) condition. So 
\eqref{E:G-d-b} is proved.
\end{itemize} 

\item[$(2)$] 
\begin{itemize}
\item [$\blacktriangleright$]The continuous scattering data $s_c$ will be extracted from $\partial_{\bar\lambda}m_0$. Thanks to \eqref{E:c-integral-eq-p},  we need to compute $ \partial_{\bar\lambda}  G(x,x' ,\lambda)$ first. To this aim, using  the holomorphic property,
\begin{gather*}
\dfrac 1{\pi}\partial_{\bar \lambda}\left(\frac 1{\lambda-a}\right)=\delta_{\lambda_R=a_R}\delta_{\lambda_I=a_I},\\
p_\lambda(D)f= e^{(\overline\lambda-\lambda)x_1+(\overline\lambda^2-\lambda^2)x_2  }p_{ \overline \lambda}(D)e^{-[(\overline\lambda-\lambda)x_1+(\overline\lambda^2-\lambda^2)x_2 ] }f,
\end{gather*}  we obtain
\be\label{E:p-debar-g}
\begin{gathered}
 \partial_{\bar\lambda}  G(x,x' ,\lambda)=  
 -\frac {\sgn (\lambda_I)}{2\pi i}e^{(\overline\lambda-\lambda)(x_1-x_1')+(\overline\lambda ^2-\lambda^2)(x_2-x_2')} \chi(x,\overline\lambda )\xi(x',\overline\lambda) ,\\
 G_\lambda\,  e^{(\overline\lambda-\lambda)x_1+(\overline\lambda^2-\lambda^2)x_2  }= e^{(\overline\lambda-\lambda)x_1+(\overline\lambda^2-\lambda^2)x_2  } \,   G_{ \overline\lambda}.
\end{gathered}\ee 

As a result, by setting $s_c(\lambda) 
  =   \frac {\sgn(\lambda_I)}{2\pi i} \left[\xi(\cdot,\bar\lambda)  v_0(\cdot) m_0 (\cdot,\lambda)\right]^\wedge(\frac{\overline\lambda-\lambda}{2\pi i},\frac{\overline\lambda^2-\lambda^2}{2\pi i})$,
\begin{align}
&\partial_{\overline\lambda}m_0 {(x,\lambda)}\label{E:p-debar-m0}\\
=\ & \partial_{\overline\lambda}\left[(1+G_\lambda\ast v_0)^{-1}\chi\right]\nonumber \\
=\ &- (1+ G_\lambda\ast v_0)^{-1}\left(\partial_{\bar\lambda}  G_\lambda\ast v_0\right)m _0{(x,\lambda)} \nonumber \\
=\ & 
-(1+ G_\lambda\ast v_0)^{-1}\frac {\sgn(\lambda_I)e^{(\overline\lambda-\lambda)(x_1-\corr{x_1'})+(\overline\lambda ^2-\lambda^2)(x_2-\corr{x_2'})}\chi(x, \overline\lambda)\corr{\xi(x', \overline\lambda)}}{-2\pi i} \ast \corr{v_0m_0}\nonumber  \\
=\ &  s_c(\lambda)(1+ G_\lambda\ast v_0)^{-1}e^{(\overline\lambda-\lambda) x_1 +(\overline\lambda ^2-\lambda^2) x_2 }\chi(x, \overline\lambda) \nonumber \\
=\ &  s_c(\lambda) e^{(\overline\lambda-\lambda) x_1 +(\overline\lambda ^2-\lambda^2) x_2 }(1+G_{ \overline\lambda}\ast v_0)^{-1}\chi(x, \overline\lambda)\nonumber  \\
=\ &  s_c(\lambda) e^{(\overline\lambda-\lambda)x_1+(\overline\lambda^2-\lambda^2)x_2  }m_0{(x, \overline\lambda)}.\nonumber 
\end{align} 

In the following, we investigate analytic constraints of the continuous data $s_c$ at $\infty,\,\kappa_j$, and $0$:

\noindent $\ast$ {\bf Hence 
 away from $\kappa_j$}, Fourier transform theory implies
\begin{align}
 \hskip.65in& |(1-\sum_{j=1}^2E_{ \kappa_j  }(\lambda))(|\overline\lambda-\lambda|^{l_1}+|\overline\lambda^2-\lambda^2|^{l_2} )  s_c(\lambda)|_{L^\infty} 
 \le    C \sum_{h=0}^{l} |\partial_{x}^{h}v_0| _{L^1\cap L^\infty}.\label{E:fourier}
  \end{align}

\noindent $\ast$ {\bf To characterize  $s_c$ at $\kappa_j$}, we need to investigate  $G(x,x',\lambda)$, $m_0(x,\lambda)$ for $\lambda=\kappa_j+re^{i\alpha}\in D_{ \kappa_j}^\times$. Refining estimates for $I^{[1]}_j$-$I^{[4]}_j$ in \eqref{E:green-kappa}, one obtains 
\[
\begin{split}
\hskip.5in -\frac{e^{i[\lambda_I(x_1'-x_1)+2\lambda_I\lambda_R(x'_2-x_2)]}}{2\pi} I^{[3]} _j 
\to\   [-1+\frac 12\theta(x_2-x_2')]\chi_j(x) \xi_j(x')\\
  +\theta(\lambda_R-\kappa_j) \theta(x_2'-x_2 )\chi_j(x) \xi_j(x')+\frac 1\pi \chi_j(x) \xi_j(x')|\alpha|,
  \end{split}
\] as $\lambda=\kappa_j+re^{i\alpha}\to\kappa_j$, $-\pi<\alpha<\pi$, which, together with $G_{d} $ \eqref{E:wick-infty-spec-2-p}, 
  yields discontinuity of $G$ at $\kappa_j$. 
\begin{gather}
 G(x,x', \lambda)={\mathfrak G}_j(x,x')+\frac 1\pi 
\chi_j(x)\xi_j(x')|\alpha|+\omega_j(x,x', \lambda),\label{E:g-asym-pm-i-prep-limit-sym}\\
  |{\mathfrak G}_j  |_{C(D_{ {\kappa}_j})} \le C (1+\frac1{\sqrt{|x_2-x_2'|}} ),\ \omega_j (x,x',\kappa_j)=0,\label{E:g-asym-pm-i-prep-limit-sym-1}\\ 
   |\omega_j  |_{L^\infty(D_{   \kappa _j})\cap C^\mu_{\tilde \sigma}(D_{ {\kappa}_j},\frac{1}{\tilde \sigma})  }\le C(1+\frac { 1+|x'| }{\sqrt{|x_2-x_2'|}} ). \label{E:g-asym-pm-i-prep-limit-sym-2}
 \end{gather}
 Here to derive \eqref{E:g-asym-pm-i-prep-limit-sym-2}, we have used
\begin{align*}
&  |e^{(\overline\lambda-\lambda) ( x_1-x_1')  +(\overline\lambda ^2-\lambda^2)  (x_2-x_2')  } |_{ H_{\tilde\sigma}^\mu (D_{z,\frac{1}{\tilde\sigma}})}\\
=&\sup_{\textrm{\tiny ${\ba{c}  \tilde r_1,  \tilde r_2 \le 1,
 |\alpha_1|,|\alpha_2|\le \pi\ea}$}} \\
 &\frac{| e^{(\overline\lambda_1-\lambda_1) ( x_1-x_1')  +(\overline\lambda _1^2-\lambda^2_1)  ( x_2-x_2') }-e^{(\overline\lambda_2-\lambda_2) ( x_1-x_1') +(\overline\lambda _2^2-\lambda^2_2)  ( x_2-x_2')  }|}{ |\tilde r_1e^{i\alpha_1}-\tilde r_2e^{i\alpha_2}|^\mu}\\
  =&\sup_{\textrm{\tiny ${\ba{c}  \tilde r_1,  \tilde r_2 \le 1,
 |\alpha_1|,|\alpha_2|\le \pi\ea}$}}\\
&   \frac{|e^{i([X_1-X_1']\frac{\tilde r_1}{\tilde\sigma}\sin\alpha_1+[X_2-X_2'](\frac{\tilde r_1}{\tilde\sigma})^2\sin2\alpha_1)}-e^{i([X_1-X_1']\frac{\tilde r_2}{\tilde\sigma}\sin\alpha_2+[X_2-X_2'](\frac{\tilde r_2}{\tilde\sigma})^2\sin2\alpha_2)}|}{ |\tilde r_1e^{i\alpha_1}-\tilde r_2e^{i\alpha_2}|^\mu} \\
 \le &\sup_{\textrm{\tiny ${\ba{c}  \tilde r_1,  \tilde r_2\le 1,
 |\alpha_1|,|\alpha_2|\le \pi\ea}$}}\\
&   \frac{|e^{i([X_1-X_1']\frac{\tilde r_1}{\tilde\sigma}\sin\alpha_1+[X_2-X_2'](\frac{\tilde r_1}{\tilde\sigma})^2\sin2\alpha_1)}-e^{i([X_1-X_1']\frac{\tilde r_2}{\tilde\sigma}\sin\alpha_2+[X_2-X_2'](\frac{\tilde r_2}{\tilde\sigma})^2\sin2\alpha_2)}|}{ |\tilde r_1e^{i\alpha_1}-\tilde r_2e^{i\alpha_2}| }\\
\le &C (1+|x'|),
\end{align*} for  $
\lambda_j=z+r_j e^{i\alpha_j}=z+\frac{\tilde r_j}{\tilde\sigma} e^{i\alpha_j}\in D_{z,\frac{1}{\tilde\sigma}}$, $z\in\{0,\kappa_1,\kappa_2\}$.

Plugging \eqref{E:g-asym-pm-i-prep-limit-sym} and \eqref{E:g-asym-pm-i-prep-limit-sym-1} into \eqref{E:c-integral-eq-p}, and applying
\be\label{E:asy-kappa}\begin{split}
\wp_j(x,x',\alpha)=  &1+[{\mathfrak G}_j+\frac 1\pi 
\chi_j(x)\xi_j(x')\cot^{-1}\frac{\lambda_R-\kappa_j}{|\lambda_I|}]\ast  v_0,\\
m(x,\lambda)=&(1+\wp_j^{-1}\omega_j  \ast  v_0)^{-1}\wp_j^{-1}\chi(x,\lambda)\\
=&\wp_j^{-1}\chi(x,\lambda)+\cdots+(-\wp_j^{-1}\omega_j  \ast  v_0)^k\wp_j^{-1}\chi(x,\lambda)+\cdots,
\end{split}\ee one has  
  \be\label{E:m-kappa1}
m_0 (x,\kappa_j+0^+e^{i\alpha})= \wp_j^{-1}\chi(x,\kappa_j)=  \frac{\Theta_j(x)}{1-\gamma_j |\alpha|},
\ee with
\be\label{E:leading-m-coeff}
\begin{split}
\Theta_j(x)=& [1+{\mathfrak G}_j(x,x')\ast  v_0(x')]^{-1}\chi_j(x'),\\
\gamma_j =& -\frac 1\pi\iint\xi_j(x )v_0(x )\Theta_j(x)dx.
\end{split}
\ee
Combining  \eqref{E:sato-adjoint}, \eqref{E:g-asym-pm-i-prep-limit-sym}, \corr{similar argument as that for deriving  \eqref{E:g-asym-pm-i-prep-limit-sym-2},}   and \eqref{E:m-kappa1} (see \eqref{E:m-0-dec} also),
\begin{align*}
 s_c(\lambda)  
  = &  \frac {\sgn(\lambda_I)}{2\pi i} \left[\xi(\cdot,\bar\lambda)  v_0(\cdot) m_0 (\cdot,\lambda)\right]^\wedge(\frac{\overline\lambda-\lambda}{2\pi i},\frac{\overline\lambda^2-\lambda^2}{2\pi i})\\
=&  \frac {\sgn(\lambda_I)}{2\pi i} \iint e^{(\overline\lambda-\lambda)x_1+(\overline\lambda^2-\lambda^2)x_2} \left(\frac{\xi_j(x)}{\bar\lambda-\kappa_j}+h.o.t.\right) \\
\times& v_0(x) \left(\frac{\Theta_j(x)}{1-\gamma_j |\alpha|}+h.o.t.\right)dx\\
=& \dfrac{\frac {i}{ 2} {\sgn}(\lambda_I)}{\overline\lambda-\kappa_j}\dfrac { \gamma_j}{1-\gamma_j |\alpha|}+\sgn(\lambda_I)h_j(\lambda)
\end{align*} with $|h_j|_{ \color{black}L^\infty(D_{\kappa_j})}<|  {\color{black}  (1+|x|) v_0}|_{L^1\cap L^\infty}$.
 
\noindent $\ast$ {\bf Similarly, for $\lambda\in D_{ 0}^\times$}, using $|m_{0,r}  |_{C^1(  D_0)}<C (1+|x|)\sum_{|k|=0}^1|\partial_x^k(1+|x|) v_0|_{L^1\cap L^\infty}$,
\[
\begin{split}
 s_c(\lambda)  
  = &  \frac {\sgn(\lambda_I)}{2\pi i} \left[\xi(\cdot,\bar\lambda)  v_0(\cdot) m_0 (\cdot,\lambda)\right]^\wedge(\frac{\overline\lambda-\lambda}{2\pi i},\frac{\overline\lambda^2-\lambda^2}{2\pi i})\\
=&  \frac {\sgn(\lambda_I)}{2\pi i} \iint e^{(\overline\lambda-\lambda)x_1+(\overline\lambda^2-\lambda^2)x_2} \left( 0\cdot\lambda +h.o.t.\right) \\
\times& v_0(x) \left(\frac{m_{0,\res}(x)}{\lambda}+h.o.t.\right)dx\\
=& {\sgn}(\lambda_I)\hbar_0(\lambda)
\end{split}
\]
with $|\hbar_0|_{ \color{black}C^1(D_{0})}<|{\color{black}(1+|x|)  v_0}|_{L^1\cap L^\infty}$.

\item[$\blacktriangleright$] Proof of $m_0\in W_0$ follows from :  
\begin{itemize}
\item [$(a)$] $m_0 (x, \lambda)=\overline{ m_0 (x, \overline\lambda)}$ follows from the reality of $u_0$.
\item[$(b)$] $(1-  E_{0} )m_0(x, \lambda)\in L^\infty$ follows from the integral equation \eqref{E:c-integral-eq-p} and $(1-  E_{0} )\chi\in L^\infty$.
 
\item[$(c)$] For $\lambda \in D_{0}^\times$ $
m_0(x_1,x_2, \lambda)=\frac{ {m_{0,\res}(x_1,x_2)}}{\lambda  }  +m_{0,r}(x_1,x_2, \lambda)$, with $ m_{0,\res}$,   $m_{0,r}  \in L^\infty( D_{0})$ follows from \eqref{E:c-integral-eq-p} and 
\be\label{E:chi}
\chi(x,\lambda)=1-\frac{1}{\lambda}\frac{e^{\kappa_1\theta_1(x)}+a\kappa_2 e^{\theta_2(x)}}{e^{\theta_1(x)}+a e^{\theta_2(x)}} .\ee

\item[$ (d)$]  
Applying \eqref{E:g-asym-pm-i-prep-limit-sym}-\eqref{E:m-kappa1}, and \eqref{E:chi},  
one has, for $\lambda =\kappa_j+re^{i\alpha}\in D_{\kappa_j}^\times$, 
 \be\label{E:m-0-dec}
\begin{gathered} 
\qquad\qquad\qquad m_0=m_0^\flat+m_0^\sharp ,\qquad
m_0^\flat =\wp_j^{-1}\chi(x,\kappa_j),\\
\qquad\qquad\qquad m_0^\sharp=\wp_j^{-1}\chi(x,\lambda)-\wp_j^{-1}\chi(x,\kappa_j)+\sum_{k=1}^\infty(-\wp_j^{-1}\omega_j  \ast  v_0)^k\wp_j^{-1}\chi(x,\lambda).
\end{gathered}
\ee 
Therefore,
\[
\begin{gathered} 
m_0^\flat =\sum_{l=0}^\infty m_{0,l}(X_1,X_2)(-\ln(1-\gamma_j|\alpha|))^l \in L^\infty(D_{\kappa_j}) ,\\
m_0^\sharp \in C_{\tilde\sigma}^{\mu } (D_{\kappa_j,\frac{1}{\tilde\sigma}})\cap L^\infty  (D_{\kappa_j}),\quad  m_0^\sharp(x_1,x_2,\kappa_j)=0.
\end{gathered}
\]
\end{itemize}

\item[$\blacktriangleright$]   To prove the initial CIE,   we shall derive the following Cauchy integral estimates for $m_0(x,\lambda)$, 
\begin{align}
&|\mathcal C   T m_0| _{L^\infty} \le C      |   v_0|_{L^1\cap L^\infty}
 ,\label{E:est-CIE-m-0}\\
 & \mathcal C T m_0 (x, \lambda)\to 0 ,\quad\textit{ as $|\lambda|\to\infty$, $\lambda_I\ne 0$ .} \label{E:asym-CIE-m-0}
\end{align}

Using \eqref{E:est-CIE-m-0} and \eqref{E:asym-CIE-m-0}, then applying Liouville's theorem, 
\be \label{E:unique-direct-problem-q-p}
  m_0(x, \lambda)=g(x )+\frac{  m_{ 0,\res }(x )}{\lambda }  +\mathcal CT _0 m_0(x ,\lambda),
\ee and
\be\label{E:unique-computation}
\begin{split} 
u(x )  m_0(x,  \lambda)=\left(\partial_{x_2}-\partial_{x_1}^2-2 \lambda\partial_{x_1}\right)
[g(x )+\frac    { m_{ 0,\res }(x )}{\lambda }    ]\\
+\left(\partial_{x_2}-\partial_{x_1}^2-2 \lambda\partial_{x_1}\right)\mathcal CT_0  m_0.
\end{split} 
\ee

Therefore, if  $\partial_{x }^kv_0 \in  {L^1\cap L^\infty}$, $ 0\le |k|\le 4 $,   letting   $|\lambda|\gg 1$,  $\lambda_I\ne 0$, one has first $g(x)\equiv g(x_2)$. Letting $x_1\gg 1$, $\lambda_R\ne \kappa_1,\,\kappa_2$,   we justify  $g\equiv 1$ and establish the CIE.

  To derive \eqref{E:est-CIE-m-0} and \eqref{E:asym-CIE-m-0}, we decompose 
\begin{align}
 |\mathcal C T_0m_0|
 \label{E:infty-spec-1-p-0} 
  \le &\iint_{D_0} \frac {{   s_c(\zeta)e^{(\overline\zeta-\zeta)x_1+(\overline\zeta^2-\zeta^2)x_2}m_0}}{ \zeta-\lambda }d\overline\zeta\wedge d\zeta|  \\
  +&\sum_{j=1,2}\iint_{D_{\kappa_j}} \frac {{   s_c(\zeta)e^{(\overline\zeta-\zeta)x_1+(\overline\zeta^2-\zeta^2)x_2}m_0}}{ \zeta-\lambda }d\overline\zeta\wedge d\zeta| \nonumber\\
  +&\iint_{\CC\backslash (D_0\cup D_{\kappa_1}\cup D_{\kappa_2})} \frac {{   s_c(\zeta)e^{(\overline\zeta-\zeta)x_1+(\overline\zeta^2-\zeta^2)x_2}m_0}}{ \zeta-\lambda }d\overline\zeta\wedge d\zeta|\nonumber\\
  \equiv &P_1+P_2+P_3 .\nonumber
\end{align}

Applying  Stokes' theorem,  the Sokhotski--Plemelj formula,   $|m_{0,r} |_{L^\infty(   D_0)}$,   $|m_0|_{L^\infty(  D_{\kappa_j})}<C |v_0|_{L^1\cap L^\infty}$, 
\begin{align}
&\qquad\qquad\qquad P_1= |m_{0,r}(x,\lambda)-\frac {1} {2\pi i} \oint_{|\zeta|={\delta}}\frac{  m_{0,r}(x,  \zeta)}{\zeta- \lambda}d\zeta|\le C |v_0|_{ L^1\cap L^\infty},\label{E:stokes-m-0}\\
&\qquad\qquad\qquad P_2\le \sum_{j=1}^2| m(x,\lambda)-\frac {1} {2\pi i}\ \oint_{|\zeta-\kappa_j|=\delta}\frac{  m(x,  \zeta)}{\zeta- \lambda} d\zeta|\le C|v_0|_{L^1\cap L^\infty}.\nonumber
\end{align}

Estimate $P_3$ is derived by the change of variables \eqref{E:variables} 
and from \eqref{E:wick-infty-spec},     \eqref{E:wick-infty-spec-1},  we obtain,  
\begin{align}
&\quad \qquad P_3
\le   C|   \iint_{\CC\backslash (D_0\cup D_{\kappa_1}\cup D_{\kappa_2})} \frac {{   s_c(\zeta)e^{(\overline\zeta-\zeta)x_1+(\overline\zeta^2-\zeta^2)x_2}m_0}}{ \zeta-\lambda }d\overline\zeta\wedge d\zeta|\label{E:infty-spec-1-p}\\
&\quad \qquad\le    C| m_0|_{L^\infty(D_0^c)} \iint_{\CC\backslash (D_0\cup D_{\kappa_1}\cup D_{\kappa_2})} \frac {|{    s_c(\zeta(\xi,\eta))|}}{|(2\pi\xi)^ 2-4\pi i\xi\lambda+2\pi i \eta|}d\xi d\eta
\nonumber\\
&\quad \qquad\le    C|m _0|_{L^\infty(D_0^c)}  \{|  s_c(\zeta)| _{L^2(D^c_{\kappa_1}\cap D^c_{\kappa_2},d\xi d\eta)}\left|\frac 1{p_\lambda}\right|_{L^2(\Omega^c_\lambda,d\xi d\eta)} \nonumber\\
 &\quad \qquad+{|  s_c(\zeta) |}_{ L^\infty(D^c_{\kappa_1}\cap D^c_{\kappa_2},d\xi d\eta)}\left|\frac 1{p_\lambda}\right|_{L^1(\Omega_\lambda, d\xi d\eta)}\}\le C|v|_{L^\infty\cap L^1} \nonumber
\end{align}

\item[$\blacktriangleright$]  
  \cite{BP214} To prove the $\mathcal D$-symmetry \eqref{E:intro-CIE-0-p-d},  introduce the total Green function $\mathcal K(x,x',\lambda)$ defined by 
\be \label{E:total-green}
\begin{split}
\mathcal K(x,x',\lambda)=& \mathcal G-\mathcal G\ast v_0\,\mathcal K,\\
\mathcal K(x,x',\lambda)=& \mathcal G-\mathcal K\ast v_0\,\mathcal G,\\
\mathcal G(x,x', \lambda) =&e^{ \lambda (x_1-x_1')+\lambda^2 (x_2-x_2') } G(x,x', \lambda).
\end{split}
\ee One has
\be \label{E:sym-0-K}
\begin{gathered}
\overrightarrow{\mathcal L_{v_0}}\mathcal K  =\mathcal K\overleftarrow{\mathcal L_{v_0}}=\delta(x-x') ,
\end{gathered}
\ee where 
\[
\mathcal L_{v_0}=\mathcal L+v_0,\quad \mathcal L=-\partial_{x_2}+\partial^2_{x_1}+u_s(x),
\] with $\overrightarrow {\mathcal L}$   the operator $\mathcal L$ applying to the $x$ variable of $\mathcal K$ and $\overleftarrow{\mathcal L}$   the operator  applying to the $x'$ variable of $\mathcal K$.

 Therefore the eigenfunction $\Phi $ and adjoint eigenfunction $\Psi $, defined by  
\be\label{E:adjoint}
\begin{gathered}
\mathcal L^\dagger\Psi (x,\lambda)\equiv  \left(\partial_{x_2}+\partial^2_{x_1}+u_s(x)\right)\Psi (x,\lambda)= -v_0(x)\Psi (x,\lambda)\\
\Psi (x,\lambda)\to  \ \psi(x,\lambda),
\end{gathered}\ee can be written as
\be \label{E:eigen-adj-eigen}
\begin{split}
\Phi(x,\lambda)=& \mathcal K(x,x',\lambda)\ast_{x'}\overleftarrow{\mathcal L}\varphi (x',\lambda)
\equiv  \mathcal K \ast \overleftarrow{\mathcal L}\varphi ,\\ 
\Psi(x' ,\lambda)=&\psi(x ,\lambda)\ast_{x }\overrightarrow{\mathcal L}\mathcal K (x,x',\lambda)
\equiv  \psi\ast\overrightarrow{\mathcal L}\mathcal K ,
\end{split}
\ee with  $\varphi$ and $\psi$  the Sato   eigenfunction and the Sato adjoint eigenfunction (see \eqref{E:sato}, \eqref{E;spectral}, \eqref {E:sato-adjoint}, \eqref{E;spectral-new}).

In view of \eqref{E:eigen-adj-eigen}, the $\mathcal D$-sysmmetry is reduced to study symmetry of
\begin{align}
&(\varphi_1(x),   \varphi_2  (x))\mathcal D^\flat=0,\label{E:eigen-sym}\\
&\mathcal G  _{j-1} =\mathcal G  _j +\varphi_j(x)\psi_j(x'),\label{E:g-sym}\\
&\mathcal K _{j-1} =\mathcal K _j +\frac{\Phi_j(x)\Psi_j(x')}{1-c_{jj}},\label{E:k-sym}
\end{align} which rely on successively using \eqref{E:eigen-adj-eigen} and \eqref{E:total-green}  \cite{BP214, Wu21}. Therefore,
\begin{gather}  
\sum_{j= 1}^{ 2}\frac{\Phi_j(x)\Psi_j(x')}{1-c_j}=0,\label{E:K-general-sum}\\
\mathcal K_{l}=\mathcal K_i+\sum_{j=l+1}^{i+2}\frac{\Phi_j(x)\Psi_j(x')}{1-c_j}.\label{E:K-general} 
\end{gather}  Here $
c_j= c_{jj}$ and the mod $2$-condition is adopted. 

Applying $  \overleftarrow{\mathcal L}\varphi_i$ to \eqref{E:K-general} from the right and using \eqref{E:eigen-adj-eigen}, we obtain
\be \label{E:K-phi-general} 
\mathcal K_l \ast \overleftarrow{\mathcal L}\varphi_i=\Phi_i+\sum_{j=l+1}^{i+2}\frac{\Phi_j(x)c_{ji}}{1-c_j}.
\ee   Summing \eqref{E:K-phi-general} up with the matrix $\mathcal D_{im}$ and using the symmetry \eqref{E:eigen-sym}, we derive
\be \label{E:higher-kdv-sym}
\sum_{i=1}^{2} \Phi_i\mathcal D^\flat_{im}+\sum_{i=1}^{2}\sum_{j=l+1}^{i+2}\frac{\Phi_j(x)c_{ji}\mathcal D^\flat_{im}}{1-c_j}=0.
\ee  Taking $l=2$ in \eqref{E:higher-kdv-sym} and using \eqref{E:K-general-sum},   we prove  
\be\label{E:d-sharp-1-line}
(e^{\kappa_1x_1+\kappa_1^2x_2 }m_0(x_1,x_2,\kappa^+_1), e^{\kappa_2x_1+\kappa_2^2x_2 }m_0(x_1,x_2,\kappa^+_2))\mathcal D^\sharp=0.
\ee Multiplying $\left(\mathcal D_{12}^\sharp\right)^{-1}\kappa_1$ from the right to both sides, we justify  
\[
(e^{\kappa_1x_1+\kappa_1^2x_2 }m_0(x_1,x_2,\kappa^+_1), e^{\kappa_2x_1+\kappa_2^2x_2 }m_0(x_1,x_2,\kappa^+_2))\mathcal D =0.
\]
\end{itemize} 
\end{itemize}

\subsubsection{The strategy of the proof of Theorem   \ref{T:1-soliton-evolution}}  \hfill \\  
 
  Notice that
\[
\begin{split}
 &\partial_{\overline\lambda}\Phi(x, \lambda)=s_c(\lambda,x_3)\Phi(x,  \overline\lambda),\\
 &-\kappa_1\Phi(x, \kappa_1)=  \mathcal D_{21}(x_3)\Phi(x, \kappa_2).
\end{split} 
\]
Denote 
\[
\mathcal M_\lambda=- \partial_{x_3}+  \partial_{x_1}^3+\frac 32u\partial_{x_1}+\frac 34u_{x_1}+\frac 34\partial_{x_1}^{-1}u_{x_2}+\rho(\lambda) ,\ \
\rho(\lambda)= -\lambda ^3,
\] we have
\begin{align}
0=\,&\partial_{\overline\lambda}\left[\mathcal M_\lambda\Phi(x, \lambda)\right]=\mathcal M_\lambda\left[ \partial_{\overline\lambda}\Phi(x, \lambda)\right]=\mathcal M_\lambda\left[  s_c(\lambda,x_3)\Phi(x, \overline\lambda)\right]\label{E:linear-p}\\
=\,&\Phi(x, \overline\lambda)\left[ {-\partial_{x_3}}+\rho(\lambda)\right] s_c(\lambda,{x_3})+s_c(\lambda,{x_3})\left[ {\mathcal M_\lambda}-\rho(\lambda)\right] \Phi(x, \overline\lambda)\nonumber\\
=\,&\Phi(x, \overline\lambda)\left[-\partial_{x_3}+\rho(\lambda)\right] s_c(\lambda,{x_3})+ s_c(\lambda,{x_3})\left[{  \mathcal M_{ \overline \lambda}-\rho( \overline \lambda)}\right]\Phi(x, \overline\lambda)\nonumber\\
=\, &\Phi(x, \overline\lambda)\left[-\partial_{x_3}+\rho(\lambda)-\rho( \overline \lambda)\right] s_c(\lambda,{x_3}),\nonumber
\end{align} and 
\begin{align}
0= &-\mathcal M_{\kappa_1}\left[\kappa_1\Phi(x, \kappa_1)\right]= \mathcal M_{\kappa_1}\left[   \mathcal D_{21}(x_3)\Phi(x,  \kappa_2)\right]\label{E:s-linear-evolution-p}\\
= &\Phi(x, \kappa_1)\left[ {-\partial_{x_3}}+\rho(\kappa_1)\right]  \mathcal D_{21}(x_3) +  \mathcal D_{21}(x_3)\left[ {\mathcal M_{\kappa_1}}-\rho(\kappa_1)\right]\Phi(x, \kappa_2)\nonumber\\
&\Phi(x, \kappa_1)\left[ {-\partial_{x_3}}+\rho(\kappa_1)\right]  \mathcal D_{21}(x_3) +  \mathcal D_{21}(x_3)\left[ {\mathcal M_{\kappa_2}}-\rho(\kappa_2)\right]\Phi(x, \kappa_2)\nonumber\\
=  &\Phi(x, \kappa_1)\left[-\partial_{x_3}+\rho(\kappa_1)-\rho( \kappa_2)\right]  \mathcal D_{21}(x_3).\nonumber
\end{align}

\subsection{The inverse problem for perturbed KP $1$-solitons}\label{SS:IP}  

\subsubsection{Statement of results}
\begin{definition}\label{E:generic-sd}
Let $0<\epsilon_0\ll 1$,  $d\ge 0$, and $u_s$ be a $  {\mathrm{Gr}(1, 2)_{> 0}}$ KP soliton defined by $\{\kappa_j\}, A=(1,a)$. A scattering data  $ {\mathcal S} =(\{0\},\{\kappa_1,\kappa_2\}, \mathcal D ,s _c(\lambda))$ is called    $d$-admissible   if  
\begin{align}
& s_c(\lambda)=
\left\{
{\ba{ll}
 {\frac{ \frac {i}{ 2} \sgn(\lambda_I)}{\overline\lambda-\kappa_j}\frac{\gamma_j}{1-\gamma _j|\alpha|}}+\sgn(\lambda_I)  h_j(\lambda),&\lambda\in   D^ \times_{ \kappa_j },\\
\sgn(\lambda_I) {  \hbar_0}(\lambda),&\lambda\in    D^\times _{ 0},
\ea}
\right.\label{E:intro-s-c-N-1}\\
&\mathcal D=\corr{(\kappa_1,\mathcal D   _{21})^T},\label{E:intro-s-d-N-1}
\end{align} and
  \be\label{E:epsilon-0-BPP}
\ba{l}
\epsilon_0\ge     (1-\sum_{j=1}^2E_{{\kappa_j}}  )  \sum_{|l|\le {d+8}}|\left(|\overline\lambda-\lambda|^{l_1} +| \overline\lambda^2-\lambda^2|^{l_2}\right) s_c (\lambda)|   _{    
 L^\infty}
 \\
  \qquad +  \sum_{j=1}^2(|\gamma_j|+|h_j|_{ \color{black}L^\infty(D_{\kappa_j})})+ |\hbar_0|_{ \color{black}C^1(D_{0})}+  |\mathcal D  - { \mathcal D}^\flat|_{L^\infty},\nonumber\\
  s_c(\lambda)=  \overline{s_c( \overline\lambda)},\,
  h_j(\lambda)=-\overline{h_j( \overline\lambda)}, \,
 \hbar_0(\lambda)=-\overline{\hbar_0( \overline\lambda)},\\
  \mathcal D^\flat= \textit{diag}\,(   
\kappa _1 , \kappa _2 )\, A^T= (    \kappa _1, \kappa _2 a  )^T. 
\ea
\ee
\end{definition} 

Define $T$ as the continuous scattering operator
\be\label{E:cauchy-operator-p}
T  \phi  (x ,\lambda)
\equiv  {  s}_c(\lambda  )e^{(\overline\lambda-\lambda)x_1+(\overline\lambda^2-\lambda^2)x_2 +(\overline\lambda^3-\lambda^3)x_3 }\phi(x, \overline\lambda).
\ee

\begin{definition}\label{D:phase} The {eigenfunction  space} ${ W }=W_x$  consists of $\phi$ satisfying 
\begin{itemize}
\item [$(a)$] $\phi (x, \lambda)=\overline{ \phi (x, \overline\lambda)};$ 
\item [$(b)$] $(1-   E_{0} )\phi(x, \lambda)\in L^\infty;$ 
\item [$(c)$] for $\lambda \in D_{0}^\times$, 
$
\phi(x, \lambda)=\frac{ {\phi_{0,\res}(x)}}{\lambda  }  +\phi_{0,r}(x, \lambda)$, $ \phi_{0,\res}$,   $\phi_{0,r}  \in L^\infty( D_{0})$; 
\item[$(d)$] for $\lambda =\kappa_j+re^{i\alpha}\in D_{\kappa_j}^\times$, $\phi=\phi^\flat+\phi^\sharp $, 
$\phi^\flat =\sum_{l=0}^\infty \phi_l(X)(-\ln(1-\gamma_j|\alpha|))^l \in L^\infty(D_{\kappa_j}) $, $\phi^\sharp \in C_{\tilde\sigma}^{\mu } (D_{\kappa_j,\frac{1}{\tilde\sigma}})\cap L^\infty (D_{\kappa_j})$, $\phi^\sharp(x,\kappa_j)=0$. 
\end{itemize}Here 
\be\label{E:sigma-parameter}
\tilde\sigma=   \max\{ 1, |X_1|,  {|X_2|},  {|X_3|}\},
\ee is the rescaling parameter with 
   $X_k$  the coefficients of the phase function of   $T$, 
\be\label{E:phase}
\begin{split}
\wp( x,\lambda)=& i[(\overline\lambda- \lambda){x_1}+(\overline\lambda^2-\lambda^2){x_2}+(\overline\lambda^3-\lambda^3){x_3}]\qquad \lambda=z+re^{i\alpha} \in D_z \\
=&  X_1r\sin\alpha  +X_2r^2 \sin2\alpha  +X_3r^3 \sin3\alpha  
\equiv \wp(r,\alpha,X),\\
 X_1( x,z)= &2(x_1+2x_2z +3x_3z^2),\   
 X_2(x,z)=   2(x_2 +3x_3z),\  
  X_3(x,z)= 2x_3.  
 \end{split}
 \ee  

Moreover, $C_{\tilde\sigma}^{\mu } (D_{\kappa_j,\frac{1}{\tilde\sigma}})=C (D_{\kappa_j,\frac{1}{\tilde\sigma}})\cap H_{\tilde\sigma}^{\mu } (D_{\kappa_j,\frac{1}{\tilde\sigma}})$ and the rescaled H$\ddot{\mbox{o}}$lder space $H_{\tilde\sigma}^{\mu } (D_{z,\frac{1}{\tilde\sigma}})$  for $z\in\RR$,  consists of  functions $\phi(x,\lambda)\equiv \phi(r,\alpha,X)$,  $X=(X_1,X_2,X_3)$,  satisfying
\be\label{E:rescaled-holder}
|\phi|_{ H_{\tilde\sigma}^\mu (D_{z,\frac{1}{\tilde\sigma}})} \equiv  {\color{black}\sup_{\textrm{\tiny ${\ba{c}  \tilde r_1,  \tilde r_2\le   1,
 |\alpha_1|,|\alpha_2|\le \pi\ea}$}} \frac{|\phi(\frac{ \tilde r_1}{\tilde\sigma}, \alpha_1,X)-\phi(\frac{ \tilde r_2}{\tilde\sigma},\alpha_2,X)|}{ |\tilde r_1e^{i\alpha_1}-\tilde r_2e^{i\alpha_2}|^\mu} }\quad<\infty
\ee for $
\lambda_j=z+r_j e^{i\alpha_j}=z+\frac{\tilde r_j}{\tilde\sigma} e^{i\alpha_j}\in D_{z,\frac{1}{\tilde\sigma}}$.  
 
  Finally, for   $\phi\in W$,   define 
\be\label{E:W} 
\begin{split}
 |\phi|_W \equiv & |(1-  E_{0})\phi|_{L^\infty}   + (| \phi_{0,\res}|_{L^\infty}+ | \phi_{0,r}|_{L^\infty(D_{0})}) \\
+& \sum_{j=1}^2 (| \phi^\flat     |_{ L^\infty(D_{\kappa_j})}+| \phi^\sharp     |_{ C_{\tilde\sigma}^{\mu } (D_{\kappa_j,\frac{1}{\tilde\sigma}})\cap L^\infty(D_{\kappa_j})})  . 
\end{split}\ee

\end{definition}

Our main results are
\begin{theorem}\label{T:1-line-inverse}\cite{Wu22} 
Let $u_s$ be  a $  {\mathrm{Gr}(1, 2)_{> 0}}$ KP soliton  defined by $\{\kappa_j\}, A=(1,a)$. There exists a positive constant $\epsilon_0\ll 1$, such that for any {$d$-admissible scattering data} $\mathcal S =(\{0\},\kappa_1,\kappa_2,\mathcal D,s_c(\lambda))$, 
\begin{itemize}
\item [(1)]
the system  of the Cauchy integral equation  and the {$\mathcal D$-symmetry},
\begin{gather}   
 {  {   m}(x, \lambda) =1+ \frac{   m_{0, \res }(x  )}{\lambda }  +\mathcal C  T
     m ,\ \lambda\neq 0,}  \label{E:intro-CIE} \\
{ (e^{\kappa_1x_1+\kappa_1^2x_2+\kappa_1^3x_3}m(x,\kappa^+_1), e^{\kappa_2x_1+\kappa_2^2x_2+\kappa_2^3x_3}m(x,\kappa^+_2))\mathcal D=0 },\label{E:intro-sym}
     \end{gather} is uniquely solved in $W$ satisfying
\be\label{E:intro-asymp}
  { \sum_{0\le l_1+2l_2+3l_3\le d+5}| \partial^l_{x}\left[m(x ,\lambda)-  \chi (x ,\lambda)\right]|_{W}\le C\epsilon_0}.
 \ee

\item [(2)] Moreover,      
\begin{gather}
 \left(-\partial_{x_2}+\partial_{x_1}^2+2 \lambda\partial_{x_1}+ u (x) \right)  m (x ,\lambda)=0 ,\label{E:intro-Lax-u}\\\
 u(x )\equiv - 2  \partial_{x_1}     m_{0,\res}(x  )-\frac 1{\pi i}\partial_{x_1}\iint  T  m  \ d\overline\zeta\wedge d\zeta ,\label{E:intro-u-rep}\\
\sum_{0\le l_1+2l_2+3l_3\le d+4}|\partial^l_x\left[u(x )-u_s(x )\right]|_{L^\infty }\le C \epsilon_0.\label{E:intro-Lax-u-asymp}
\end{gather}
We define the inverse scattering transform by
\be\label{E:inverse scattering-transform-1}
\mathcal S^{-1}( \{z_n,\kappa_j, \mathcal D,s _c(\lambda)\})=-\frac 1{\pi i}\partial_{x_1}\iint  T  m \ d\overline\zeta\wedge d\zeta  -2  \partial_{x_1}  m_{z_0,\res}(x) ;
\ee

\item[(3)]  $u :\mathbb R\times\mathbb R\times \mathbb R^+\to \mathbb R$ solves the KPII equation
\be\label{E:intro-KP-ist}
(-4u_{x_3}+u_{x_1x_1x_1}+6uu_{x_1})_{x_1}+3u_{{x_2}{x_2}}=0.
\ee
\end{itemize} 
\end{theorem}

\subsubsection{The strategy of the proof of Theorem \ref{T:1-line-inverse}}\label{SS:strategy} 

\begin{itemize}
\item[$(1)$] The system of the CIE and $\mathcal D$-symmetry, \eqref{E:intro-CIE} and \eqref{E:intro-sym}, represents both analytic and algebraic aspects of the IST toward an understanding of the KPII equation. We will show existence, uniqueness, and estimates of this system by
taking the limit in $W$ of the iteration sequence 
\begin{align}
& \phi^{(k)}(x ,\lambda)  
=   1+ \frac{\color{black}\phi^{(k)}_{0,\res} (x   )}{\lambda  }  +\mathcal CT \phi^{(k-1)}(x, \lambda)  ,\ \ k>0,\label{E:recursion-iteration-introduction}\\
&(e^{\kappa_1x_1+\kappa_1^2x_2+\kappa_1^3x_3}\phi^{(k)}(x,\kappa^+_1),  e^{\kappa_2x_1+\kappa_2^2x_2+\kappa_2^3x_3}\phi^{(k)}(x,\kappa^+_2))\mathcal D=0,\label{E:recursion-iteration-D-introduction}\\
&\phi^{(0)}(x ,\lambda)=  \chi(x,\lambda) .\label{E:recursion-iteration-bdry-introduction}
 \end{align}
 
 Evaluating the CIE \eqref{E:recursion-iteration-introduction} at $\kappa_1^+$, $\kappa_2^+$, and using  the   $\mathcal D$-symmetry  \eqref{E:recursion-iteration-D-introduction}, one obtains a linear system of $2+1$ variables $\phi^{(k)}(x,\kappa^+_1)$, $\phi^{(k)}(x,\kappa^+_2)$, and $\phi^{(k)}_{0,\res} (x   )$. Hence the iteration turns into
 \begin{align}
  \phi^{(k)}(x ,\lambda)  
= &  1+ \frac{\phi^{(k)}_{0,\res} (x   )}{\lambda  }  +\mathcal CT \phi^{(k-1)}(x, \lambda)  ,\ \ k>0,\label{E:recursion-iteration-introduction-new}\\
 \phi^{(k)}_{0,\res} (x   )=&-\frac{\kappa_1 e^{\kappa_1x_1+\kappa_1^2x_2+\kappa_1^3x_3}+a \kappa_2e^{\kappa_2x_1+\kappa_2^2x_2+\kappa_2^3x_3}}{e^{\kappa_1x_1+\kappa_1^2x_2+\kappa_1^3x_3}+a e^{\kappa_2x_1+\kappa_2^2x_2+\kappa_2^3x_3}}\label{E:recursion-iteration-D-introduction-new}\\
 &-\frac{\kappa_1 e^{\kappa_1x_1+\kappa_1^2x_2+\kappa_1^3x_3} }{e^{\kappa_1x_1+\kappa_1^2x_2+\kappa_1^3x_3}+a e^{\kappa_2x_1+\kappa_2^2x_2+\kappa_2^3x_3}}\mathcal CT \phi^{(k-1)}(x, \kappa^+_1)\nonumber\\
 &-\frac{ a \kappa_2e^{\kappa_2x_1+\kappa_2^2x_2+\kappa_2^3x_3}}{e^{\kappa_1x_1+\kappa_1^2x_2+\kappa_1^3x_3}+a e^{\kappa_2x_1+\kappa_2^2x_2+\kappa_2^3x_3}}\mathcal CT \phi^{(k-1)}(x, \kappa^+_2),\nonumber\\
 \phi^{(0)}(x ,\lambda)= & \chi(x,\lambda) .\label{E:recursion-iteration-bdry-introduction-new}
 \end{align}

By the totally positive (TP) condition, it reduces to deriving uniform estimates of $CT\phi^{(k)}$ on $D_\infty$, $D_{\kappa_j}$, $D_{0}$.
 \begin{itemize}
 \item [$\blacktriangleright$] {\bf Estimates on $D_\infty$:} Away from $0$, $\kappa_j$, the continuous data $s_c$ and eigenfunction $m_0$ are  regular as in the case of rapidly decaying potentials. Hence 
via the coordinates change \eqref{E:variables} and applying \eqref{E:wick-infty-spec-1},  
 we have
 \[
\ba{rl}
 &|\mathcal CT(1- E_{0}-\sum_{j=1}^2E_{\kappa_j})\phi|_{L^\infty(D_{\kappa_j}) }  \\
\le&C|(1- E_{0}-\sum_{j=1}^2E_{\kappa_j})\phi|_{L^\infty} |\iint \dfrac{(1- E_{0}-\sum_{j=1}^2E_{\kappa_j})|s_c(\zeta)| }{|p_\lambda (\xi,\eta)|}d\xi  d\eta|_{L^\infty(D_{\kappa_j})} \\
\le&C|(1- E_{0}-\sum_{j=1}^2E_{\kappa_j})\phi|_{L^\infty}
  \{ |(1- E_{0}-\sum_{j=1}^2E_{\kappa_j})s_c|_{L^2(d\xi d\eta)}\left|\dfrac 1{p_\lambda}\right|_{L^2(\Omega_\lambda^c,d\xi d\eta)}\nonumber\\
+&|(1- E_{0}-\sum_{j=1}^2E_{\kappa_j})s_c|_{L^\infty(d\xi d\eta)}\left|\dfrac 1{p_\lambda}\right|_{L^1(\Omega_\lambda ,d\xi d\eta)}\} \\
\le& C\epsilon_0|(1- E_{0}-\sum_{j=1}^2E_{\kappa_j})\phi|_{L^\infty}
\ea\]

 \item [$\blacktriangleright$] {\bf Estimates on $D_{\kappa_j}$:} To solve the inverse problem, we shall overcome the discontinuous leading singularity of $s_c$   by noting  
\be\label{E:stokes-introduction}
\frac{ \frac {i}{ 2} \sgn(\lambda_I)}{\overline\lambda-\kappa_j}\frac{ \gamma_j}{1-\gamma _j|\alpha|}=-\partial_{\bar\lambda}\ln(1-\gamma_j|\alpha|)\equiv \widetilde\gamma_j (\lambda)
\ee and applying Stokes' theorem. To achieve an $L^\infty$-estimates for the CIO  at $\kappa_j$, an oscillatory kernel, which is required to defined the $  {\mathcal D}$-symmetry, we shall use the scaled topology $C^\mu_{\tilde\sigma}(D_{\kappa_j,\frac{1}{\tilde\sigma}})$ introduced in Definition \ref{D:phase-0}. 
Precisely,   taking advantage of \eqref{E:stokes-introduction}, the two dimensional property of   $\mathcal C$,  $|h_j|_{ \color{black}L^\infty(D_{\kappa_j})}<|  {\color{black}  (1+|x|) v_0}|_{L^1\cap L^\infty}$ (assured by the $d$-admissible condition), and scaling invariant properties,  
 \begin{align}
 &|\mathcal CE_z\phi|_{L^\infty} \le   C|\phi|_{L^p(D_z)},\quad
|\mathcal CE_z\phi|_{H^\nu(D_z)} \le   C|\phi|_{L^p(D_z)},\label{E:vekua}\\
&\mathcal C\widetilde\gamma_jE_{\kappa_j}[-{ \ln (1-\gamma_j|\beta|)}]^l =\frac{[- { \ln (1-\gamma_j|\alpha|)}]^{l+1}}{l+1}\label{E:stokes} \\
 &\hskip.9in -\frac {1}{2\pi i}\oint_{|\zeta-\kappa_j|= \delta}\frac{\frac 1{l+1}[-\ln (1-\gamma_j|\beta|)]^{l+1}} { \zeta-\lambda }d\zeta ,\nonumber\\
 &|\mathcal C_\lambda\widetilde\gamma_j e^{-i\wp(x,\zeta)} E_{\kappa_j}  f(\kappa_j+se^{i\beta}) |_{C_{\tilde\sigma}^{\mu } (D_{\kappa_j,\frac 1{\tilde\sigma}})\cap L^\infty(D_{\kappa_j }) }\label{E:scale-invariant}\\
&\hskip.9in= |\mathcal C_{\tilde\lambda}\widetilde\gamma_je^{-i\wp(\frac{\tilde s}{\tilde\sigma},\beta, X)}E_{\kappa_j,\tilde\sigma\delta}   f(\kappa_j+\frac{\tilde s}{\tilde\sigma} e^{i\beta})|_{C ^{\mu } (D_{\kappa_j,1})\cap L^\infty(D_{\kappa_j,\tilde\sigma\delta})},\nonumber
\end{align}    where the  dilating   polar coordinates near $z=\kappa_j$ is defined by
\[
\begin{split}
 \lambda=z+ re^{i\alpha}=z+ \frac{\tilde r}{{\tilde\sigma}}e^{i\alpha},&\quad \zeta=z+  se^{i\beta}=z+ \frac{\tilde s}{{\tilde\sigma}}e^{i\beta},  \\
  \tilde\lambda=z+ \tilde r e^{i\alpha},&\quad \tilde\zeta=z+\tilde se^{i\beta},\\
 r,\,s\le\delta,\quad
 \tilde r,\,{\tilde s} \le {\tilde\sigma}\delta, &\quad |\alpha|,|\beta|\le \pi,  
 \end{split}
 \] 
 it then reduces to estimating $\mathcal C_\lambda\widetilde\gamma_je^{-i\wp }E_{\kappa_j} f$ for $f\in W$. To this aim,  decompose
 \begin{align}
&  \mathcal C_\lambda\widetilde\gamma_je^{-i\wp(x,\zeta)}E_{\kappa_j} f\equiv I_1+I_2+I_3+I_4+I_5,\label{E:scale-mu}
\end{align}
with
\begin{align}
&\hskip.6inI_1= -\frac {\theta(1-\tilde r)}{2\pi i}\iint_{ \tilde s<2} \frac{  \widetilde \gamma_j(\tilde s, \beta)  f^{ \flat}(\frac {\tilde s}{ {\tilde\sigma} },-\beta,X) }{\tilde\zeta-\tilde \lambda}d\overline{\tilde\zeta} \wedge d\tilde\zeta  ,\label{E:scal-1}\\
&\hskip.6inI_2= -\frac {\theta(1-\tilde r)}{2\pi i}\iint_{ \tilde s<2} \frac{  \widetilde \gamma_j(\tilde s, \beta)[ e^{-i\wp(\frac {\tilde s}{ {\tilde\sigma} },\beta,X)}-1]f^{\flat}(\frac {\tilde s}{ {\tilde\sigma} },-\beta,X) }{\tilde\zeta-\tilde \lambda}d\overline{\tilde\zeta} \wedge d\tilde\zeta ,\label{E:scal-2}\\
&\hskip.6inI_3= -\frac {\theta(1-\tilde r)}{2\pi i}\iint_{ \tilde s<2} \frac{  \widetilde \gamma_j(\tilde s, \beta) e^{-i\wp(\frac {\tilde s}{ {\tilde\sigma} },\beta,X)}f^{ \sharp}(\frac {\tilde s}{ {\tilde\sigma} },-\beta,X) }{{\tilde\zeta}-\tilde \lambda}d\overline{\tilde\zeta} \wedge d{\tilde\zeta} ,\label{E:scal-3}\\
&\hskip.6inI_4= -\frac {\theta(1-\tilde r)}{2\pi i}\iint_{2< \tilde s< {\tilde\sigma} \delta} \frac{ \widetilde \gamma_j(\tilde s, \beta) e^{-i\wp(\frac {\tilde s}{{\tilde\sigma}},\beta,X)}f(\frac {\tilde s}{{\tilde\sigma}},-\beta,X)  }{{\tilde\zeta}-\tilde \lambda}d\overline{\tilde\zeta} \wedge d{\tilde\zeta},\label{E:scal-4}\\
&\hskip.6inI_5= -\frac {\theta(\tilde r-1)}{2\pi i}\iint_{  \tilde s< {\tilde\sigma} \delta} \frac{  \widetilde\gamma_j(\tilde s, \beta) e^{-i\wp(\frac {\tilde s}{{\tilde\sigma}},\beta,X)}f(\frac {\tilde s}{ {\tilde\sigma} },-\beta,X) }{{\tilde\zeta}-\tilde\lambda}d\overline{\tilde\zeta} \wedge d{\tilde\zeta}.\label{E:scal-5}
\end{align}

   We shall  apply Stokes' theorem \eqref{E:stokes} and H$\ddot{\mbox{o}}$lder interior estimates  \cite{Ga66} to derive 
   \[
   |I^\flat_1|_{L^\infty},\ |I^\sharp_1|_{C^\mu_{\tilde\sigma}(D_{\kappa_j,\frac{1}{\tilde\sigma}})},\ |I_2 |_{C^\mu_{\tilde\sigma}(D_{\kappa_j,\frac{1}{\tilde\sigma}})},\ |I_3 |_{C^\mu_{\tilde\sigma}(D_{\kappa_j,\frac{1}{\tilde\sigma}})}.
   \] 
For the estimates of 
\[
|I_4|_{C^\mu_{\tilde\sigma}(D_{\kappa_j,\frac{1}{\tilde\sigma}}) },\qquad |I_5|_{L^\infty(D_{\kappa_j})}
\] which are integrals with {slow decaying kernels} on non uniformly $\tilde s$-compact domains.   Taking advantage of the $2$-dimensional property, we write them as iterated integrals in polar coordinates
\begin{align*}
&\hskip.6in I_4= -\frac {\theta(1-\tilde r)}{2\pi i}\int_{-\pi}^\pi  d\beta[\partial_ {\beta}   \ln (1-\gamma_j |\beta| )] 
       \int_{2< \tilde s< {\tilde\sigma} \delta} \frac{  e^{-i\wp(\frac {\tilde s}{ {\tilde\sigma} },\beta,X)}f(\frac {\tilde s}{{\tilde\sigma}},-\beta,X)}{\tilde s -  \tilde r e^{ i (\alpha-\beta) }} d\tilde s    , \\
&\hskip.6in I_5= -\frac { \theta(\tilde r-1)}{2\pi i}  \int_{-\pi}^\pi  d\beta[\partial_ {\beta}   \ln (1-\gamma_j |\beta| )] 
       \int_{0 }^{\widetilde\sigma \delta} \frac{  e^{-i\wp(\frac {\tilde s}{ {\tilde\sigma} },\beta,X)}f(\frac {\tilde s}{{\tilde\sigma}},-\beta,X)}{\tilde s -  \tilde r e^{ i (\alpha-\beta) }} d\tilde s.      
\end{align*}Then estimates will be derived by applying meromorphic  properties in $\tilde s$, the deformation method, and stationary point analysis of $\wp$. Main ideas are, near stationary points, we deform 
\be\label{E:deform-intro}
\tilde s\in\RR\quad\longrightarrow\quad \tilde se^{i\tau}\in\CC
\ee such that the segment of $ \tilde s $   deforms into a union  of line segments $\Gamma$'s and arcs $S$'s, satisfying
 \begin{align}
 (a)\ &\ \textit{on $\Gamma$'s,    $\mathfrak {Re} ({-i\wp(\frac {\tilde se^{i\tau}}{ {\tilde\sigma} },\beta,X)})\le -\frac 1C |\sin (k\beta)|\tilde s^k $ } \nonumber \\
 (b)\ &\  \textit{on $\Gamma$'s,  ${|{\tilde\zeta}-\tilde\lambda}|\ge \frac 1C\max\{\tilde s,\tilde r\}$,}\label{E:deform-intro-1}\\
(c)\ &\ \textit{on $S$'s,  $\mathfrak {Re} ({-i\wp(\frac {\tilde se^{i\tau}}{  \tilde\sigma},\beta,X)})\le 0$. }\nonumber
\end{align} 
  Therefore, 
   estimates can be done if $|{{\tilde\zeta}-\tilde\lambda}|\ge \frac 1C$.  When $|{{\tilde\zeta}-\tilde\lambda}|\le \frac 1C$, we shall prove  $I_5$ near the stationary is no longer a singular integral and estimates can be obtained  by means of \eqref{E:vekua} (see Proposition \ref{P:non-homogeneous-F3} for $\tilde\sigma=\sqrt[k]{|X_k|}$,   $k=3,2 $).

Estimates for  $\tilde\sigma=|X_1|$ are more involved because $e^{\mathfrak {Re} ({-i\wp(\frac {\tilde se^{i\tau}}{ {\tilde\sigma} },\beta,X)})}   $   either decays too slowly or creates a $\frac 1{|\sin\beta|}$ singularity. To overcome difficulties, we shall either take advantage of the scaling invariant properties of the Hilbert transform or   a finer decomposition  (see Proposition \ref{P:non-homogeneous-F1} and remarks before Definition \ref{D:renormal}).  

Details to prove the key estimates for $I_3$, $I_4$ and $I_5$ will be provided in $\S$ \ref{SSS:I-3}, and $\S$ \ref{SSS:I-5}. Since they are major additional analytic features of the inverse problem for multi-soliton backgrounds.

 \item [$\blacktriangleright$] {\bf Estimates on $D_0$ :} Decompose
\begin{align*}
&\qquad  \mathcal C  TE_{0}\phi    \\
&\quad =   -\frac { 1}{ 2\pi i}\iint_{ D_{0, {\tilde\sigma}\delta}} \frac{  \sgn(\beta) \hbar_0( \frac{\tilde s}{\tilde\sigma},\beta) e^{-i\wp(\frac {\tilde s}{ {\tilde\sigma} },\beta,X)}\phi_{0,\res}(x) }{ (\tilde\zeta-\tilde \lambda) \overline{\tilde\zeta}   }d\overline{\tilde\zeta} \wedge d\tilde\zeta +\mathcal C_\lambda TE_{0}\phi_{0,r} \\
&\quad \equiv   II_1+II_2+II_3+II_4+II_5, 
\end{align*}
where
\begin{align*}
&\qquad II_1= -\frac {\theta(1-\tilde r)}{2\pi i}\iint_{ \tilde s<2} \frac{  \sgn(\beta) \hbar_0(\frac{\tilde s}{\tilde\sigma}, \beta)  \phi_{0,\res}(x) }{(\tilde\zeta-\tilde \lambda) \overline{\tilde\zeta} }d\overline{\tilde\zeta} \wedge d\tilde\zeta  ,\\
&\qquad II_2= -\frac {\theta(1-\tilde r)}{2\pi i}\iint_{ \tilde s<2} \frac{  \sgn(\beta) \hbar_0(\frac{\tilde s}{\tilde\sigma}, \beta)[ e^{-i\wp(\frac {\tilde s}{ {\tilde\sigma} },\beta,X)}-1]\phi_{0,\res}( x) }{(\tilde\zeta-\tilde \lambda)  \overline{\tilde\zeta}  }d\overline{\tilde\zeta} \wedge d\tilde\zeta ,\\
&\qquad II_3= \ \mathcal C_\lambda E_{0}T\phi_{0,r} ,\\
&\qquad II_4= -\frac {\theta(1-\tilde r)}{2\pi i}\iint_{2< \tilde s< {\tilde\sigma} \delta} \frac{  \sgn(\beta) \hbar_0(\frac{\tilde s}{\tilde\sigma}, \beta)  e^{-i\wp(\frac {\tilde s}{ {\tilde\sigma} },\beta,X)} \phi_{0,\res}(x) }{(\tilde\zeta-\tilde \lambda)  \overline{\tilde\zeta}  }d\overline{\tilde\zeta} \wedge d{\tilde\zeta},\\
&\qquad II_5= -\frac {\theta(\tilde r-1)}{2\pi i}\iint_{  \tilde s< {\tilde\sigma} \delta} \frac{  \sgn(\beta) \hbar_0(\frac{\tilde s}{\tilde\sigma}, \beta)  e^{-i\wp(\frac {\tilde s}{ {\tilde\sigma} },\beta,X)} \phi_{0,\res}(x) }{(\tilde\zeta-\tilde \lambda)  \overline{\tilde\zeta}  }d\overline{\tilde\zeta} \wedge d{\tilde\zeta}. 
\end{align*}

From $|\hbar_0|_{C^1(D_0)}<|v_0|_{L^1\cap L^\infty}$  (assured by the $d$-admissible condition), the mean value theorem,  and the Hilbert transform theory \cite{Ga66},   
\[
\begin{split}
|II_1|_{L^\infty(D_{0})},\,|II_2|_{L^\infty(D_{0})}\le & C\epsilon_0 |\phi_{0,\res}|_{L^\infty} .
\end{split}
\] 

Writing $\hbar_0(\zeta)=\hbar_0(0)+ [\hbar_0(\zeta)-\hbar_0(0)]$ and decompose $II_4=II_{41}+II_{42}$, $II_5=II_{51}+II_{52}$ with
\begin{align*}
II_{41}=&-\frac {\theta(1-\tilde r)}{2\pi i}\iint_{2< \tilde s< {\tilde\sigma} \delta} \frac{  \sgn(\beta) \hbar_0(0)  e^{-i\wp(\frac {\tilde s}{ {\tilde\sigma} },\beta,X)} \phi_{0,\res}(x) }{(\tilde\zeta-\tilde \lambda)  \overline{\tilde\zeta}  }d\overline{\tilde\zeta} \wedge d{\tilde\zeta},\\
II_{51}=&-\frac {\theta(\tilde r-1)}{2\pi i}\iint_{  \tilde s< {\tilde\sigma} \delta} \frac{  \sgn(\beta) \hbar_0(0)  e^{-i\wp(\frac {\tilde s}{ {\tilde\sigma} },\beta,X)} \phi_{0,\res}(x) }{(\tilde\zeta-\tilde \lambda)  \overline{\tilde\zeta}  }d\overline{\tilde\zeta} \wedge d{\tilde\zeta}.
\end{align*}
{Thanks to $\tilde s$-meromorphic properties  and adapting argument for estimating $I_4$, $I_5$ (see $\S$ \ref{SSS:I-3})} for $ II_{41} $, $ II_{51} $, we obtain
\[
\begin{split}
|II_{41}|_{L^\infty(D_{0})},\,|II_{51}|_{ L^\infty(D_{0})}\le & C\epsilon_0|\phi_{0,\res}|_{L^\infty}.
\end{split}
\]

For the remaining terms, by \eqref{E:vekua} and $|\hbar_0|_{C^1(D_0)}<|v_0|_{L^1\cap L^\infty}$  (assured by the $d$-admissible condition),  
\[
\begin{gathered}
|II_{42}|_{L^\infty(D_{0})},\,|II_{52}|_{ L^\infty(D_{0})}\le   C\epsilon_0|\phi_{0,\res}|_{L^\infty},\\
|II_3|_{L^\infty(D_{0})} \le    C\epsilon_0 |\phi_{0,r}|_{L^\infty(D_{0})}.
\end{gathered}
\] 
\end{itemize}

\item[(2)]   To derive the Lax equation and the representation formula, firstly, introduce the shorthand notation for the heat operator 
\[
\begin{gathered}
-\partial_{x_2}+\partial_{x_1}^2+2 \lambda\partial_{x_1}=-\nabla_2+\nabla_1^2 ,\\
\nabla_1=\partial_{x_1}+\lambda,\   \nabla_2=\partial_{x_2}+\lambda^2 ,\ \
J  \, \phi   =  \frac{\phi _{0,\res}(x)}{\lambda },\end{gathered} 
\]
formally, 
\begin{gather} 
 (-\nabla_2+\nabla_1^2  )m 
=  \left[-\nabla_2+\nabla_1^2,
  J  +
  \mathcal CT  \right] m 
+      
  (J    +\mathcal CT)(-\nabla_2+\nabla_1^2) m,  
  \label{E:inverse-lax}\\
  (e^{\kappa_1x_1+\kappa_1^2x_2+\kappa_1^3x_3}\left[(-\nabla_2+\nabla_1^2  )m(x,\kappa_1^+)\right]  ,\hskip1.6in \label{E:linearization-D-evol-new-1-s-J-d-p}\\
  \hskip1.6in e^{\kappa_2x_1+\kappa_2^2x_2+\kappa_2^3x_3}\left[(-\nabla_2+\nabla_1^2  )m(x,\kappa_2^+)\right] ){ \mathcal D  }=0,\nonumber\\
 -u(x)\equiv     
 \left[-\nabla_2+\nabla_1^2 ,
  J +\mathcal CT\right]  m
=   {+\frac 1{\pi i}\partial_{x_1}\iint  T  m \ d\overline\zeta\wedge d\zeta}  +2  \partial_{x_1}  m_{0,\res}(x). \label{E:lax-t-cauchy}
\end{gather}

If we can prove rigorously  
\be\label{E:u-est}
\begin{gathered}
(-\nabla_2+\nabla_1^2  )m\in W,\\
\sum_{0\le l_1+2l_2+3l_3\le d+4}\ |\partial_x^l\left[u(x)-u_s(x)\right]\,|_{L^\infty}\le C\epsilon_0,
\end{gathered}
\ee
together with \eqref{E:linearization-D-evol-new-1-s-J-d-p}, then the unique of the CIE and the $\mathcal D$ symmetry constraint imply
\begin{align}
 (-\nabla_2+\nabla_1^2 )m= -(1-J-\mathcal CT)^{-1}u(x)1=&-u(x)(1-J-\mathcal CT)^{-1}1\label{E:formal-derivation}\\
 =&-u(x)m(x,\lambda).\nonumber
\end{align}Therefore, \eqref{E:intro-Lax-u}-\eqref{E:intro-Lax-u-asymp} are verified.

Rigorous argument is carried out by introducing
\be\label{E:CIE-J}
\phi^{(k)}=  1+J\phi^{(k)}+\mathcal CT\phi^{(k-1)}, \qquad
    J  \, \phi^{(k)}  =  \frac{\phi^{(k)}_{0,\res}(x)}{\lambda}, 
\ee   proving 
\begin{align*}
&\textit{$\left[-\nabla_2+\nabla_1^2 ,  J  \right]\phi^{(k)}$ is  independent of $\lambda$,}\\
&\textit{$\left[-\nabla_2+\nabla_1^2 ,  \mathcal CT   \right]\phi^{(k-1)}$ is  independent of $\lambda$,}
\end{align*}  
\begin{align}
  \left[-\nabla_2+\nabla_1^2 ,  J  \right]\phi^{(k)}\to \ &\left[-\nabla_2+\nabla_1^2 ,  J  \right]m=2\partial_{x_1} m_{0,\res}(x),\nonumber\\
\left[-\nabla_2+\nabla_1^2 ,  \mathcal CT  \right]\phi^{(k-1)}\to &\  \left[-\nabla_2+\nabla_1^2 ,  \mathcal CT \right]m= \frac 1{\pi i}\partial_{x_1}\iint  T m  \, d\overline\zeta\wedge d\zeta  , \nonumber\\
 & \nonumber\\
 J   (-\nabla_2+\nabla_1^2  ) \phi^{(k)}\to \ & J   (-\nabla_2+\nabla_1^2  ) m\quad \textit{in $W$},\label{E:lax-limit}\\
 { \mathcal CT  (-\nabla_2+\nabla_1^2  ) \phi^{(k)}\to  }\ &  {\mathcal CT (-\nabla_2+\nabla_1^2  ) m\quad \textit{in $W$},} \nonumber\\
 & \nonumber\\
 \sum_{0\le l_1+2l_2+3l_3\le d+4} |\partial_x^l (  [ - \nabla_2& +\nabla_1^2 ,  J   ]m+u_s ) |_{L^\infty}\le C\epsilon_0,  \nonumber\\
|\left[-\nabla_2+\nabla_1^2 ,  \mathcal CT \right]\phi^{(k-1)}&-\left[-\nabla_2+\nabla_1^2 ,  \mathcal CT \right] \chi|_{L^\infty}\le(C\epsilon_0)^{k } .\nonumber 
\end{align} 
 
They are proved by using the  {estimates of the CIO's} and the iteration method. 

\item[(3)] {\bf The KP equation:} The KP equation will be derived by justifying the Lax pair. By the representation formula \eqref{E:intro-Lax-u-asymp} and $\Phi(x, \lambda)= e^{ \lambda  x_1+ \lambda ^2x_2}  m(x, \lambda)$, we define the evolution operators 
\[\begin{split}
\mathcal M =&- \partial_{x_3}+  \partial_{x_1}^3+\frac 32u\partial_{x_1}+\frac 34u_{x_1}+\frac 34\partial_{x_1}^{-1}u_{x_2} -\lambda ^3, \\
\mathcal M \Phi(x,\lambda)=&e^{ \lambda  x_1+ \lambda ^2x_2}\left(\mathcal M+3\lambda\partial_{x_1}^2+3\lambda^2\partial_{x_1}+\lambda^3+\frac 32u\lambda\right)m(x,\lambda) \\
\equiv  & e^{ \lambda  x_1+ \lambda ^2x_2}\left(\mathfrak Mm\right)(x,\lambda), 
\end{split}\] One can reverse the procedure, \eqref{E:linear-p}, \eqref{E:s-linear-evolution-p},  in the linearization theorem to prove 
\begin{gather}
\partial_{\overline\lambda}\left( \mathfrak Mm \right)(x,\lambda)=s_c(\lambda)e^{(\overline\lambda-\lambda)x_1+(\overline\lambda^2-\lambda^2)x_2 +(\overline\lambda^3-\lambda^3)x_3 }\left( \mathfrak Mm \right)(x, \overline\lambda),\label{E:inverse-map-T-tilde-p}\\
(e^{\kappa_1x_1+\kappa_1^2x_2+\kappa_1^3x_3}\mathfrak Mm(x,\kappa _1^+) ,e^{\kappa_2x_1+\kappa_2^2x_2+\kappa_2^3x_3}\mathfrak Mm(x,\kappa _2^+)){ \mathcal D  }=0. \label{E:soliton-debar-p}
\end{gather}  
 As $ |\lambda|\to\infty$, letting
\begin{align}
m \sim&\sum_{j=0}^\infty\frac{M_j(  x )}{\lambda^{j}} , \, \textit{}   \ \
\mathfrak Mm\sim  q_2(x)\lambda^2+q_1(x)\lambda+q_0(x)+\frac {q_{0,\res}}\lambda+\cdots ,\label{E:soliton-debar-0-p}
 \end{align} from the Lax equation \eqref{E:intro-Lax-u},  
\begin{gather}
\qquad    2\partial_{x_1}M_{j+1}
=      (\partial_{x_2}-\partial_{x_1}^2-u)M_j,\nonumber\\
\qquad M_0=  1,\ 
M_1=  - \frac {1}2 \partial_{x_1}^{-1}u,\  
M_2=   -\frac 14 \partial_{x_2} \partial_{x_1}^{-1}u+ \frac 14 u+\frac 14 \partial_{x_1}^{-1}\left(u \partial_{x_1}^{-1} u\right),\cdots\label{E:v-evolution-KPII-new-p} 
\end{gather}  

As a result, as $\lambda\to\infty$,  
\begin{align}
 &\mathfrak Mm\label{E:v-evolution-asm-s-p} \\
\to&\frac34 u_{x_1}+\frac 34 \partial_{x_1}^{-1}u_{x_2} 
+3\lambda\partial_{x_1}^2(1+\frac{M_1}{\lambda}) +3\lambda^2\partial_{x_1}(1+\frac{M_1}{\lambda}+\frac{M_2}{\lambda^2})+\frac 32u\lambda \\
=&\frac34 u_{x_1}+\frac 34 \partial_{x_1}^{-1}u_{x_2}+\left(
-\frac32u_{x_1}+3\partial_{x_1} (-\frac 14\partial_{x_2}\partial_{x_1}^{-2}u+\frac{u}{4}+\frac14\partial_{x_1}^{-1}[u\partial_{x_1}^{-1}u])
\right) \nonumber\\
+&\lambda\left( 3\partial_{x_1}M_1+\frac{3}{2}u\right)+\frac32u(-\frac 12\partial_{x_1}^{-1}u)=0.\nonumber\end{align}
Therefore, $q_2=q_1=q_0\equiv 0$ and then $\mathfrak M m\in W$. Together with \eqref{E:inverse-map-T-tilde-p}, \eqref{E:soliton-debar-p}, and  {unique solvability of the system of the CIE and $\mathcal D$ symmetry},  yields $\mathfrak M m(x,\lambda)=0$ and  the Lax pair.

\end{itemize} 

\subsubsection{Highlight: estimates for $I_3$}\label{SSS:I-3}\hfill \\

The scaled H$\ddot{\mbox{o}}$lder estimate  for 
\[
I_3= -\frac {\theta(1-\tilde r)}{2\pi i}\iint_{ \tilde s<2} \frac{  \widetilde \gamma_j(\tilde s, \beta) e^{-i\wp(\frac {\tilde s}{ {\tilde\sigma} },\beta,X)}f^{ \sharp}(\frac {\tilde s}{ {\tilde\sigma} },-\beta,X) }{{\tilde\zeta}-\tilde \lambda}d\overline{\tilde\zeta} \wedge d{\tilde\zeta}  \] is a reminiscence of the H$\ddot{\mbox{o}}$lder estimate of the Beltrami's equation \cite[Theorem 1.32]{V62} which involves mainly estimates of 
\[
I_3'g=\iint_{ D} \frac{   g(\zeta) }{( \zeta - \lambda)^2}d\overline{ \zeta} \wedge d{ \zeta} ,\qquad |g|_{C^\mu}<\infty.
\]Both leading singular terms of $I_3$ and $I_3'$ can be  integrated by Stokes' theorem.
\begin{proposition}\label{P:i-3}
\be \label{E:mean-vekua-3}
 | I_3|_{C^\mu_{\tilde\sigma}(D_{\kappa_1,\frac{1}{\tilde\sigma}})}\le C\epsilon_0    | f^\sharp|_{C^\mu_{\tilde\sigma}(D_{\kappa_1,\frac{1}{\tilde\sigma} })} .
 \ee
\end{proposition}

\begin{proof}

From $  f^\sharp\in C^{\mu}_{\tilde\sigma}(D_{\kappa_1,\frac{1}{\tilde\sigma}})$ and  $  f^\sharp(x,\kappa_1)=0$, 
\[
| \widetilde\gamma_1(\tilde s,\beta)f ^\sharp(\frac {\tilde s}{ \tilde\sigma},\beta,X)|_{L^\infty(D_{\kappa_1 })}\le C \epsilon_0  |f^\sharp|_{H^\mu_{\tilde\sigma}(D_{\kappa_1,\frac{1}{\tilde\sigma}})}\tilde s^{\mu-1}.
\] Therefore, an improper integral yields
\be \label{E:mean-vekua}
 | I_3|_{L^\infty(D_{\kappa_1})}\le C\epsilon_0  |f^\sharp|_{H^\mu_{\tilde\sigma}(D_{\kappa_1,\frac{1}{\tilde\sigma}})} .
 \ee
 
 To derive the $H^\mu_{\tilde\sigma}$-estimate of $I_3$, let $\tilde \lambda_j=\kappa_1+\tilde r_je^{i\alpha_j}$, $ \tilde r_j \le 1$, $j=1,2$, define
 \be\label{E:phi-sharp}
 \varphi_{f^\sharp}(x,\zeta)=e^{-i\wp(\frac {\tilde s}{ \tilde\sigma },\beta,X) }{f^\sharp}(x, \overline\zeta),
 \ee and decompose  
\begin{align} 
&  I_3(x,\lambda_1)-I_3(x, \lambda_2)\label{E:linear-norm-dec}\\
=& -\frac {\tilde\lambda_1-\tilde\lambda_2}{4\pi i} \iint_{\tilde s< 2}\widetilde\gamma_1({\tilde\zeta})\frac{ \varphi_{f^\sharp}(\frac {\tilde s}{  \tilde\sigma},\beta,X)
-\varphi_{f^\sharp}(\frac {\tilde r_1}{ \tilde\sigma },\alpha_1,X)
 }{({\tilde\zeta}-\tilde\lambda_1)({\tilde\zeta}-\tilde\lambda_2)} d\bar{\tilde\zeta}\wedge d{\tilde\zeta}\nonumber\\
- &  \frac {\tilde\lambda_1-\tilde\lambda_2}{4\pi i} \iint_{\tilde s< 2}\widetilde\gamma_1({\tilde\zeta})\frac{ \varphi_{f^\sharp}(\frac {\tilde s}{  \tilde\sigma},\beta,X)
-\varphi_{f^\sharp}(\frac {\tilde r_2}{ \tilde\sigma},\alpha_2,X)
 }{({\tilde\zeta}-\tilde\lambda_1)({\tilde\zeta}-\tilde\lambda_2)} d\bar{\tilde\zeta}\wedge d{\tilde\zeta}\nonumber\\
+&\frac {\varphi_{f^\sharp}(\frac {\tilde r_1}{ \tilde\sigma },\alpha_1,X)}{4\pi i} \iint_{\tilde s< 2}\widetilde\gamma_1({\tilde\zeta})[\frac{ 1
}{{\tilde\zeta}-\tilde\lambda_2}-\frac{ 1
}{{\tilde\zeta}-\tilde\lambda_1}] d\bar{\tilde\zeta}\wedge d{\tilde\zeta}\nonumber\\
+&\frac {\varphi_{f^\sharp}(\frac {\tilde r_2}{ \tilde\sigma },\alpha_2,X)}{4\pi i} \iint_{\tilde s< 2}\widetilde\gamma_1({\tilde\zeta})[\frac{ 1
}{{\tilde\zeta}-\tilde\lambda_2}-\frac{ 1
}{{\tilde\zeta}-\tilde\lambda_1}] d\bar{\tilde\zeta}\wedge d{\tilde\zeta}.\nonumber
\end{align}

In view of  $  f^\sharp\in C^{\mu }_{\tilde\sigma}(D_{\kappa_1,\frac{1}{\tilde\sigma}})$ and  $  f^\sharp(x,\kappa_1)=0$, we have
\be\label{E:mu-radial}
|\varphi_{f^\sharp}(\frac {\tilde r}{ \tilde\sigma },\alpha,X)|_{L^\infty(D_{\kappa_1 })}\le C |f^\sharp|_{H^\mu_{\tilde\sigma}(D_{\kappa_1,\frac{1}{\tilde\sigma} })}\tilde r^\mu .
\ee Along with \eqref{E:stokes-introduction}, yields
\begin{align*}
|\frac {\varphi_{f^\sharp}(\frac {\tilde r_1}{ \tilde\sigma},\alpha_1,X)}{4\pi i}  \iint_{\tilde s\le 2}\widetilde\gamma_1({\tilde\zeta})[\frac{ 1
}{{\tilde\zeta}-\tilde\lambda_2}-\frac{ 1
}{{\tilde\zeta}-\tilde\lambda_1}] d\bar{\tilde\zeta}\wedge d{\tilde\zeta}| 
\le & C\epsilon_0 |f^\sharp|_{H^\mu_{\tilde\sigma}(D_{\kappa_1,\frac{1}{\tilde\sigma} })}  |\tilde\lambda_1-\tilde\lambda_2|^\mu,   \tilde r_1=\tilde r_2,\\
|\frac {\varphi_{f^\sharp}(\frac {\tilde r_1}{ \tilde\sigma},\alpha_1,X)}{4\pi i} \iint_{\tilde s\le 2}\widetilde\gamma_1({\tilde\zeta})[\frac{ 1
}{{\tilde\zeta}-\tilde\lambda_2}-\frac{ 1
}{{\tilde\zeta}-\tilde\lambda_1}] d\bar{\tilde\zeta}\wedge d{\tilde\zeta}| 
\le & 0,\qquad  \alpha_1=\alpha_2
\end{align*}respectively. Therefore, 
\be \label{E:I_1-mu-3}
|\frac {\varphi _{f^\sharp}(\frac {\tilde r_1}{ \tilde\sigma},\alpha_1,X)}{4\pi i} \iint_{\tilde s\le 2}\widetilde\gamma_1({\tilde\zeta})[\frac{ 1
}{{\tilde\zeta}-\tilde\lambda_2}-\frac{ 1
}{{\tilde\zeta}-\tilde\lambda_1}] d\bar{\tilde\zeta}\wedge d{\tilde\zeta}|
\le  C\epsilon_0 |f^\sharp|_{C^{\mu }_{\tilde\sigma}(D_{\kappa_1,\frac{1}{\tilde\sigma}})}  |\tilde\lambda_1-\tilde\lambda_2|^\mu.
\ee

In an entirely similar way, 
\be \label{E:I_1-mu-4}
|\frac {\varphi_{f^\sharp}(\frac {\tilde r_2}{ \tilde\sigma},\alpha_2,X)}{4\pi i} \iint_{\tilde s\le 2}\widetilde\gamma_1({\tilde\zeta})[\frac{ 1
}{{\tilde\zeta}-\tilde\lambda_2}-\frac{ 1
}{{\tilde\zeta}-\tilde\lambda_1}] d\bar{\tilde\zeta}\wedge d{\tilde\zeta}| 
\le  C\epsilon_0 |f^\sharp |_{C^{\mu }_{\tilde\sigma}(D_{\kappa_1,\frac{1}{\tilde\sigma}})}  |\tilde\lambda_1-\tilde\lambda_2|^\mu.
\ee

Let us now investigate the first term on the right hand side of \eqref{E:linear-norm-dec}.  Applying \eqref{E:vekua}, it suffices to derive the estimate  for all $\lambda_1$, $\lambda_2$ with $\tilde D\subset \{\tilde s\le 2\}  $ being a disk centred at $\tilde\lambda_1$ with radius $l$ and $l=2|\tilde\lambda_2-\tilde\lambda_1|$. Write
\begin{align}
&-\frac {\tilde\lambda_1-\tilde\lambda_2}{4\pi i}\iint_{\tilde s\le 2}\widetilde\gamma_1({\tilde\zeta})\frac{ \varphi _{f^\sharp}(\frac {\tilde s}{  \tilde\sigma},\beta,X)
-\varphi _{f^\sharp}(\frac {\tilde r_1}{ \tilde\sigma},\alpha_1,X)
 }{({\tilde\zeta}-\tilde\lambda_1)({\tilde\zeta}-\tilde\lambda_2)} d\bar{\tilde\zeta}\wedge d{\tilde\zeta}\label{E:holder-ext}\\
 =&-\frac {\tilde\lambda_1-\tilde\lambda_2}{4\pi i}  \iint_{\tilde D}\widetilde\gamma_1({\tilde\zeta})\frac{ \varphi _{f^\sharp}(\frac {\tilde s}{  \tilde\sigma },\beta,X)
-\varphi _{f^\sharp}(\frac {\tilde r_1}{ \tilde\sigma},\alpha_1,X)
 }{({\tilde\zeta}-\tilde\lambda_1)({\tilde\zeta}-\tilde\lambda_2)} d\bar{\tilde\zeta}\wedge d{\tilde\zeta}\nonumber\\
 -& \frac {\tilde\lambda_1-\tilde\lambda_2}{4\pi i}\iint_{\{\tilde s\le 2\}/\tilde D}\widetilde\gamma_1({\tilde\zeta})\frac{ \varphi _{f^\sharp}(\frac {\tilde s}{ \tilde\sigma},\beta,X)
-\varphi _{f^\sharp}(\frac {\tilde r_1}{ \tilde\sigma },\alpha_1,X)
 }{({\tilde\zeta}-\tilde\lambda_1)({\tilde\zeta}-\tilde\lambda_2)} d\bar{\tilde\zeta}\wedge d{\tilde\zeta}.\nonumber
\end{align}

Let $\tilde D_0=\{\zeta: |\zeta-\tilde \lambda_1|<\frac{3l}2\} $.  
 \begin{itemize}
 \item [$\blacktriangleright$] If ${\tilde\zeta}\in \{\tilde s\le 2\}/\tilde D$ and   $\kappa_1\in \tilde D_0$, then
 \[
  \frac 1C\le|\frac{{\tilde\zeta}-\tilde\lambda_1}{{\tilde\zeta}-\tilde\lambda_2}|,|\frac{{\tilde\zeta}-\kappa_1}{{\tilde\zeta}-\tilde\lambda_1}|,|\frac{{\tilde\zeta}-\kappa_1}{{\tilde\zeta}-\tilde\lambda_2}|\le C.
 \] In this case, using $  f^\sharp\in C^{\mu }_{\tilde\sigma}(D_{\kappa_1,\frac{1}{\tilde\sigma}})$  and  {\cite[Chapter 1,\S6.1]{V62}}, 
 \begin{align}
&|-\frac {\tilde\lambda_1-\tilde\lambda_2}{4\pi i}\iint_{\{\tilde s\le 2\}/\tilde D}\widetilde\gamma_1({\tilde\zeta})\frac{ \varphi _{f^\sharp}(\frac {\tilde s}{  \tilde\sigma },\beta,X)
-\varphi _{f^\sharp}(\frac {\tilde r_1}{  \tilde\sigma },\alpha_1,X)
 }{({\tilde\zeta}-\tilde\lambda_1)({\tilde\zeta}-\tilde\lambda_2)} d\bar{\tilde\zeta}\wedge d{\tilde\zeta}|\label{E:holder-ext-1}\\
 \le &C\epsilon_0|f^\sharp|_{C^{\mu }_{\tilde\sigma}(D_{\kappa_1,\frac{1}{\tilde\sigma}})}|\tilde\lambda_1-\tilde\lambda_2|\iint_{\{\tilde s\le 2\}/\tilde D} \frac{1}{|\tilde\zeta-\lambda_2| |{\tilde\zeta}-\tilde\lambda_1|^{2-\mu} } d\bar{\tilde\zeta}\wedge d{\tilde\zeta} \nonumber\\
  \le & C\epsilon_0|f^\sharp|_{C^{\mu }_{\tilde\sigma}(D_{\kappa_1,\frac{1}{\tilde\sigma}})}|\tilde\lambda_1-\tilde\lambda_2|^\mu.\nonumber
\end{align}
 \item [$\blacktriangleright$] If ${\tilde\zeta}\in \{\tilde s\le 2\}/\tilde D$ and $\kappa_1\notin \tilde D_0$  then
 \[
 \begin{gathered}
 \frac 1C\le|\frac{{\tilde\zeta}-\tilde\lambda_1}{{\tilde\zeta}-\tilde\lambda_2}|\le C,\quad
 |\tilde\lambda_1 -\tilde\lambda_2 |\le \frac 1C\min\{| {\tilde\lambda_1}-\kappa_1 |,|{\tilde\lambda_2}-\kappa_1 |\}. 
 \end{gathered}
 \]
In this case, using $  f^\sharp\in C^{\mu }_{\tilde\sigma}(D_{\kappa_1,\frac{1}{\tilde\sigma}})$  and  {\cite[Chapter 1,\S6.1]{V62}},  
\begin{align}
&|-\frac {\tilde\lambda_1-\tilde\lambda_2}{4\pi i}\iint_{\{\tilde s\le 2\}/\tilde D}\widetilde\gamma_1({\tilde\zeta})\frac{ \varphi _{f^\sharp}(\frac {\tilde s}{  \tilde\sigma },\beta,X)
-\varphi _{f^\sharp}(\frac {\tilde r_1}{  \tilde\sigma },\alpha_1,X)
 }{({\tilde\zeta}-\tilde\lambda_1)({\tilde\zeta}-\tilde\lambda_2)} d\bar{\tilde\zeta}\wedge d{\tilde\zeta}|\label{E:holder-ext-2}\\
 \le & C\epsilon_0|f^\sharp|_{C^{\mu }_{\tilde\sigma}(D_{\kappa_1,\frac{1}{\tilde\sigma}})}|\tilde\lambda_1-\tilde\lambda_2|\iint_{\{\tilde s\le 2\}/\tilde D} \frac{ 1
 }{|\tilde\zeta-\kappa_1| |{\tilde\zeta}-\tilde\lambda_1|^{2-\mu} } d\bar{\tilde\zeta}\wedge d{\tilde\zeta} \nonumber\\
  \le & C\epsilon_0|f^\sharp|_{C^{\mu }_{\tilde\sigma}(D_{\kappa_1,\frac{1}{\tilde\sigma}})}|\tilde\lambda_1-\tilde\lambda_2|^\mu.\nonumber
\end{align}
 \end{itemize}
Therefore the second term on the RHS of \eqref{E:holder-ext} is done.

Let $\tilde L(\zeta)=0$ be the line perpendicular to $\overline{\lambda_1\lambda_2}$ and passing through $\frac 12(\lambda_1+\lambda_2)$.  Set
\[
\begin{split}
\tilde D_{\tilde\lambda_1,\pm}=\tilde D\cap\{ \zeta: L(\zeta)L(\lambda_1)\gtrless 0\}.  
\end{split}
\] Therefore, thanks to $ f^\sharp \in C^\mu_{\tilde\sigma}(D_{\kappa_1,\frac{1}{\tilde\sigma}})$,   setting $\eta=\frac{\tilde\zeta-\tilde\lambda_1}{|\tilde\lambda_1-\lambda_2|}$,  $\frac{\tilde\zeta-\kappa_1}{|\tilde\lambda_1-\lambda_2|}=\eta-r_0e^{i\alpha_0}$,  and using   {\cite[Chapter 1,\S6.1]{V62}},   
\begin{align}
&|\frac {\tilde\lambda_1-\tilde\lambda_2}{4\pi i} \iint_{\tilde D_{\tilde\lambda_1,+}}\widetilde\gamma_1({\tilde\zeta})\frac{ \varphi _{f^\sharp}(\frac {\tilde s}{  \tilde\sigma},\beta,X)
-\varphi _{f^\sharp}(\frac {\tilde r_1}{ \tilde\sigma},\alpha_1,X)
 }{({\tilde\zeta}-\tilde\lambda_1)({\tilde\zeta}-\tilde\lambda_2)} d\bar{\tilde\zeta}\wedge d{\tilde\zeta}|\label{E:lambda-1-p}\\
 \le &C\epsilon_0|\tilde\lambda_1-\tilde\lambda_2||f^\sharp|_{C^\mu_{\tilde\sigma}(D_{\kappa_1,\frac{1}{\tilde\sigma}})}  |\iint_{\tilde D_{\tilde\lambda_1,+}}  \frac 1{|\tilde\zeta-\kappa_1||{\tilde\zeta}-\tilde\lambda_1|^{1-\mu}|{\tilde\zeta}-\tilde\lambda_2|}d\bar{\tilde\zeta}\wedge d{\tilde\zeta}|\nonumber\\
 \le &C\epsilon_0|\tilde\lambda_1-\tilde\lambda_2|^\mu|f^\sharp|_{C^\mu_{\tilde\sigma}(D_{\kappa_1,\frac{1}{\tilde\sigma}})}  |\iint_{\{|\eta|\le 2\}\cap \tilde D_{\tilde\lambda_1,+}}  \frac 1{|\eta-r_0e^{i\alpha_0}||\eta|^{1-\mu} |\eta-e^{i\alpha'}| } d \eta_Rd\eta_I|\nonumber\\
 \le &C\epsilon_0|\tilde\lambda_1-\tilde\lambda_2|^\mu|f^\sharp|_{C^\mu_{\tilde\sigma}(D_{\kappa_1,\frac{1}{\tilde\sigma}})}  |\iint_{\{|\eta|\le 2\}\cap \tilde D_{\tilde\lambda_1,+}}  \frac 1{|\eta-r_0e^{i\alpha_0}||\eta|^{1-\mu}  } d \eta_Rd\eta_I|\nonumber\\
\le &C\epsilon_0|f^\sharp|_{C^\mu_{\tilde\sigma}(D_{\kappa_1,\frac{1}{\tilde\sigma}})}|\tilde\lambda_1-\tilde\lambda_2|^\mu. \nonumber
\end{align} 

By analogy,
\begin{align*}
&|\frac {\tilde\lambda_1-\tilde\lambda_2}{4\pi i} \iint_{\tilde D_{\tilde\lambda_1,-}}\widetilde\gamma_1({\tilde\zeta})\frac{ \varphi _{f^\sharp}(\frac {\tilde s}{  \tilde\sigma},\beta,X)
-\varphi _{f^\sharp}(\frac {\tilde r_1}{ \tilde\sigma},\alpha_1,X)
 }{({\tilde\zeta}-\tilde\lambda_1)({\tilde\zeta}-\tilde\lambda_2)} d\bar{\tilde\zeta}\wedge d{\tilde\zeta}|\\
 \le &|\frac {\tilde\lambda_1-\tilde\lambda_2}{4\pi i} \iint_{\tilde D_{\tilde\lambda_1,-}}\widetilde\gamma_1({\tilde\zeta})\frac{ \varphi _{f^\sharp}(\frac {\tilde s}{  \tilde\sigma},\beta,X)
-\varphi _{f^\sharp}(\frac {\tilde r_2}{ \tilde\sigma},\alpha_2,X)
 }{({\tilde\zeta}-\tilde\lambda_1)({\tilde\zeta}-\tilde\lambda_2)} d\bar{\tilde\zeta}\wedge d{\tilde\zeta}|\\
 +&|\frac {\tilde\lambda_1-\tilde\lambda_2}{4\pi i} \iint_{\tilde D_{\tilde\lambda_1,-}}\widetilde\gamma_1({\tilde\zeta})\frac{ \varphi _{f^\sharp}(\frac {\tilde r_2}{  \tilde\sigma},\alpha_2,X)
-\varphi _{f^\sharp}(\frac {\tilde r_1}{ \tilde\sigma},\alpha_1,X)
 }{({\tilde\zeta}-\tilde\lambda_1)({\tilde\zeta}-\tilde\lambda_2)} d\bar{\tilde\zeta}\wedge d{\tilde\zeta}|\\
  \le &C\epsilon_0|f^\sharp|_{C^\mu_{\tilde\sigma}(D_{\kappa_1,\frac{1}{\tilde\sigma}})}|\tilde\lambda_1-\tilde\lambda_2|^\mu\\
 +&|\frac {\tilde\lambda_1-\tilde\lambda_2}{4\pi i} \iint_{\tilde D_{\tilde\lambda_1,-}}\widetilde\gamma_1({\tilde\zeta})\frac{ \varphi _{f^\sharp}(\frac {\tilde r_2}{  \tilde\sigma},\alpha_2,X)
-\varphi _{f^\sharp}(\frac {\tilde r_1}{ \tilde\sigma},\alpha_1,X)
 }{({\tilde\zeta}-\tilde\lambda_1)({\tilde\zeta}-\tilde\lambda_2)} d\bar{\tilde\zeta}\wedge d{\tilde\zeta}|.
\end{align*}
Applying $ f^\sharp \in C^\mu_{\tilde\sigma}(D_{\kappa_1,\frac{1}{\tilde\sigma}})$,  Stokes' theorem,   and  
 $| {\tilde\zeta-\tilde\lambda_1}| $, $ | {\tilde\zeta-\tilde\lambda_2}|\sim| {\tilde\lambda_1}-\tilde\lambda_2|$ on the boundary of $\tilde D_{\tilde\lambda_1,-}$ (assured by $|\tilde\lambda|\le 1$, $\tilde D\subset\{\tilde s<2\}$), 
\begin{align*}
&|\frac {\tilde\lambda_1-\tilde\lambda_2}{4\pi i} \iint_{\tilde D_{\tilde\lambda_1,-}}\widetilde\gamma_1({\tilde\zeta})\frac{ \varphi _{f^\sharp}(\frac {\tilde r_2}{  \tilde\sigma},\alpha_2,X)
-\varphi _{f^\sharp}(\frac {\tilde r_1}{ \tilde\sigma},\alpha_1,X)
 }{({\tilde\zeta}-\tilde\lambda_1)({\tilde\zeta}-\tilde\lambda_2)} d\bar{\tilde\zeta}\wedge d{\tilde\zeta}|\\
 \le &C|f^\sharp|_{C^\mu_{\tilde\sigma}(D_{\kappa_1,\frac{1}{\tilde\sigma}})}|\tilde\lambda_1-\tilde\lambda_2|^{1+\mu} \iint_{\tilde D_{\tilde\lambda_1,-}}\widetilde\gamma_1({\tilde\zeta})\frac{ 1 }{({\tilde\zeta}-\tilde\lambda_1)({\tilde\zeta}-\tilde\lambda_2)} d\bar{\tilde\zeta}\wedge d{\tilde\zeta}|\\
 =&C|f^\sharp|_{C^\mu_{\tilde\sigma}(D_{\kappa_1,\frac{1}{\tilde\sigma}})}|\tilde\lambda_1-\tilde\lambda_2|^{1+\mu} \iint_{\tilde D_{\tilde\lambda_1,-}} \frac{ \partial_{\overline\zeta}\left[\ln(1-\gamma|\beta|) \frac 1{ {\tilde\zeta}-\tilde\lambda_1}\right]}{ {\tilde\zeta}-\tilde\lambda_2  } d\bar{\tilde\zeta}\wedge d{\tilde\zeta}|\\
\le &C\epsilon_0|f^\sharp|_{C^\mu_{\tilde\sigma}(D_{\kappa_1,\frac{1}{\tilde\sigma}})}|\tilde\lambda_1-\tilde\lambda_2|^\mu.
\end{align*}  
Therefore the first term on the RHS of \eqref{E:holder-ext} is done. Thus  
\begin{multline}\label{E:I_1-mu-1}
|\frac {\tilde\lambda_1-\tilde\lambda_2}{4\pi i}\iint_{\tilde s\le 2}\widetilde\gamma_1({\tilde\zeta})\frac{ \varphi _{f^\sharp}(\frac {\tilde s}{  \tilde\sigma },\beta,X)
-\varphi _{f^\sharp}(\frac {\tilde r_1}{ \tilde\sigma },\alpha_1,X)
 }{({\tilde\zeta}-\tilde\lambda_1)({\tilde\zeta}-\tilde\lambda_2)} d\bar{\tilde\zeta}\wedge d{\tilde\zeta}|\\
 \le C\epsilon_0|f^\sharp|_{C^\mu_{\tilde\sigma}(D_{\kappa_1,\frac{1}{\tilde\sigma}})}|\tilde\lambda_1-\tilde\lambda_2|^\mu,\quad\textit{for $|\tilde\lambda_j-\kappa_1|\le 1,\ j=1,2$. }
\end{multline}

In an entirely similar way, 
\begin{multline}\label{E:I_1-mu-2}
|\frac {\tilde\lambda_1-\tilde\lambda_2}{4\pi i}\iint_{\tilde s\le 2}\widetilde\gamma_1({\tilde\zeta})\frac{ \varphi _{f^\sharp}(\frac {\tilde s}{  \tilde\sigma },\beta,X)
-\varphi _{f^\sharp}(\frac {\tilde r_2}{ \tilde\sigma },\alpha_2,X)
 }{({\tilde\zeta}-\tilde\lambda_1)({\tilde\zeta}-\tilde\lambda_2)} d\bar{\tilde\zeta}\wedge d{\tilde\zeta}| 
 \\
 \le C\epsilon_0|f^\sharp|_{C^\mu_{\tilde\sigma}(D_{\kappa_1,\frac{1}{\tilde\sigma}})}|\tilde\lambda_1-\tilde\lambda_2|^\mu,\quad\textit{for $|\tilde\lambda_j-\kappa_1|\le 1,\ j=1,2$. }
\end{multline}  

Plugging \eqref{E:I_1-mu-3}, \eqref{E:I_1-mu-4}, \eqref{E:I_1-mu-1}, and \eqref{E:I_1-mu-2} into \eqref{E:linear-norm-dec}, we obtain
\be 
\label{E:linear-norm-est}
|I_3(x, \lambda_1 )-I_3(x, \lambda_2)|\le C\epsilon_0 |f^\sharp|_{C^\mu_{\tilde\sigma}(D_{\kappa_1,\frac{1}{\tilde\sigma}})} |\tilde\lambda_1-\tilde\lambda_2|^\mu 
\ee for $|\tilde\lambda_j-\kappa_1|\le 1,\ j=1,2$. 
Hence \eqref{E:mean-vekua-3} follows.

\end{proof}

\subsubsection{Highlight: estimates for $I_4,I_5$}\label{SSS:I-5}\hfill \\

Without loss of generality and for simplicity, we   assume   
\be\label{E:assumption}
\begin{gathered}
\kappa_j=\kappa_1,\quad |\lambda-\kappa_1|\le \frac \delta 2,\quad X_1> 0,\quad X_2,X_3\ge 0 , 
\end{gathered}
\ee in this section. 

For $ \widehat \sigma\in\{X_1,\sqrt{X_2},\sqrt[3]{X_3}\}$,  consider the  $\widehat\sigma$-scaled coordinates
\[
\zeta =\kappa_1+se^{i\beta}=\kappa_1+\frac{\widehat s }{\widehat \sigma}e^{i\beta}\quad\in D_{\kappa_1},
\] and the deformation 
\[
\widehat s\in\RR \longrightarrow \widehat se^{i\tau}\in\CC. 
\]
   Observe that
\[
\begin{split}
&\mathfrak {Re} ({-i\wp(\frac {\widehat se^{i\tau}}{\widehat \sigma},\beta,X)})=\frac{X_3}{\widehat \sigma ^3} \sin3\tau\sin3\beta\widehat s^3+  \frac{X_2}{\widehat \sigma^2} \sin2\tau\sin2\beta\widehat s^2 +\frac{X_1}{\widehat \sigma} \sin \tau\sin \beta\widehat s,\\
&\partial_{\widehat s}\wp(\frac{\widehat s}{\widehat \sigma},\beta,X)=3\frac{X_3}{\widehat \sigma^3}  \sin3\beta\widehat s^2+ 2 \frac{X_2}{\widehat \sigma^2}  \sin2\beta\widehat s  +\frac{X_1}{\widehat \sigma } \sin \beta.  
\end{split}
\] Hence as $|\tau|\ll 1$,   zero roots of $\partial_{\widehat s}\wp(\frac{\widehat s}{\widehat \sigma},\beta,X) $, the major obstruction of $(a)$ in \eqref{E:deform-intro-1}, are approximated  by those of  $\mathfrak {Re} ({-i\wp(\frac {\widehat se^{i\tau_\dagger}}{\widehat \sigma},\beta,X)})/ \widehat{s}$.

\begin{definition}\label{E:stationary} The stationary points are defined to be 
\be\label{E:determinant}
\widehat s_\pm = \frac{    - 1\pm\sqrt{1-\Delta}\, }{3\frac{X_3 }{\widehat\sigma X_2}\frac{\sin3\beta}{\sin2\beta}},\quad \Delta= 3\frac{ {X_1}{X_ 3}}{X_2^2}\frac{\sin\beta\sin3\beta}{\sin^22\beta},
\ee
which satisfy 
\[
\partial_{\widehat s}\wp(\frac{\widehat s_\pm}{\widehat \sigma},\beta,X)=3\frac{X_3}{\widehat \sigma^2} \sin3\beta \widehat s^2_\pm+2\frac{X_2}{\widehat \sigma^{2 }}\sin2\beta\widehat s_\pm+\frac{X_1}{\widehat \sigma }\sin\beta=0.
\]

Denote 
\be\label{E:Omega}
\ba{lll}
\Omega_1=\{0\le|\beta|\le\frac\pi 3 \},& \Omega_2=\{\frac {\pi} 3\le|\beta|\le\frac {\pi }2\},& \Omega_3=\{\frac {\pi} 2\le|\beta|\le\frac {2\pi }3\},\\
\Omega_4=\{\frac {2\pi} 3\le|\beta|\le\pi\},&  &
\ea
\ee one has Figure \ref{Fg:signature} for signatures of $sin (k\beta)$ on $\Omega_j$.  Moreover, according to the determinant $\Delta$, we have Table \ref{Tb:dynamic-FNH} for properties  of roots $\widehat s_\pm$. Notice that
\be\label{E:roots-distance}
  |\widehat s_ +-\widehat s_ -  
|=    |\frac{\sqrt{1-\Delta}}{3{ \frac{X_3}{\widehat\sigma X_2 } \frac{ \sin3\beta  }{  \sin2\beta} }}|.
\ee

{\begin{center}
\begin{figure} 
\begin{tikzpicture}[xscale=1.2 , yscale=.6]
\draw [  thick, ->](0,0) -- (3,0);
\draw [ thick, ->](0,0) -- (1.5,2.59) node [right]{$\frac{ \pi} 3$};
\draw [ thick, ->](0,0) -- (1.5,-2.59) node [right]{$-\frac{ \pi} 3$};
\draw [ thick,    ->](0,0) -- (0,3) ;
\draw [ thick,    ->](0,0) -- (0,-3) ;
\draw [ thick, ->](0,0) -- (-1.5,2.59) node [left]{$\frac{2\pi} 3$};
\draw [ thick, ->](0,0) -- (-1.5,-2.59) node [left]{$-\frac{2\pi} 3$};
\draw [ thick,   ->](0,0) -- (-3,0);
\node   at (3.5,1.6) {$(+,+,+)$};
\node   at (3.5,-1.6) {$(-,-,-)$};
\node  at (1,3.5) {$(+,+,-)$};\node  at (-1,3.5) {$(+,-,-)$};
\node  at (1,-3.5) {$(-,-,+)$};\node  at (-1,-3.5) {$(-,+,+)$};
\node   at (-3.5,1.6) {$(+,-,+)$};\node   at (-3.5,-1.6) {$(-,+,-)$};
\node [above] at (3.5,1.6) {$\Omega_1$};
\node [above] at (3.5,-1.6) {$\Omega_1$};
\node [above] at (1,3.5) {$\Omega_2$};
\node [above] at (1,-3.5) {$\Omega_2$};
\node [above] at (-1, 3.5) {$\Omega_3$};
\node [above] at (-1,-3.5) {$\Omega_3$};
\node [above] at (-3.5,1.6) {$\Omega_4$};
\node [above] at (-3.5,-1.6) {$\Omega_4$};
\end{tikzpicture}
\caption[Signatures]{\small Signatures of  $
( \sin\beta,\sin2\beta,\sin3\beta)$ for $X_1,X_2,X_3>0$}\label{Fg:signature}
\end{figure}
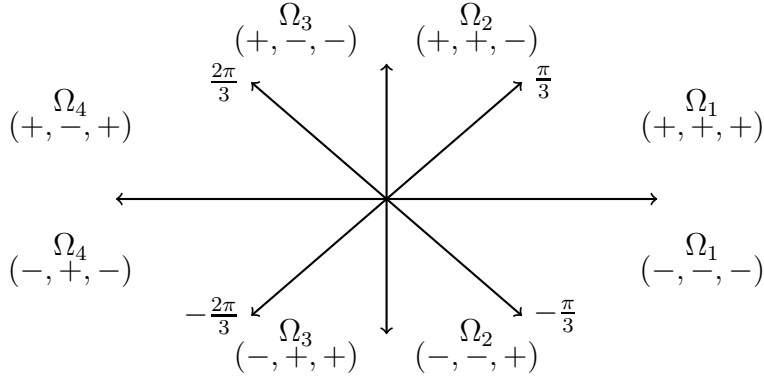
\end{center}}

\begin{table}
{\small\begin{center}
\vskip.1in
\begin{tabular}{|c|c|c|l|l|}
 \hline
 \bf{Type}&\bf{Subtype}&\bf{Range of $\Delta$}  &\bf{Properties of $\widehat s_\pm$}   &\bf{$\beta$-domain}\\
& &  && ({\bf order of } $\widehat s_\pm$)\\
   \hline\hline
 \multirow{2}*{$\mathfrak A(\beta,X)$}  & $\mathfrak A'$  &  {$  (2,\infty)$}   &\multirow{2}* {$\widehat s_\pm$ complex roots} & $\Omega_1(\frac{\widehat s_++\widehat s_-}{2}\le 0)$       \\
\cline{2-2}\cline{3-3}   
 & $\mathfrak A''$ & {$  (1,2)$}   &   & $ \Omega_4(\frac{\widehat s_++\widehat s_-}{2}\ge 0)$ \\
 \hline
 \multirow{2}*{$\mathfrak B(\beta,X)$}  & $\mathfrak B'$   &\multirow{2}* {$ { (\frac 12,1)}$}    &  {adjacent real roots w.  } & $ {\Omega_1 (\widehat s_-\le\widehat s_ +\le 0) }$       \\
\cline{2-2}\cline{5-5}  
 & $\mathfrak B''$ &   &   & $ {\Omega_4 (\widehat s_-\ge\widehat s_ +\ge 0)}$ \\
 \hline
 \multirow{2}*{$\mathfrak C(\beta,X)$}  & $\mathfrak C'$   &\multirow{2}* { $ (0, \frac { 1}2)$}    & {$\widehat s_ +=\frac{-\Delta}{6\frac{X _3}{\widehat \sigma X_2}\frac{\sin 3\beta}{\sin2\beta} } +\mbox{l.o.t.}, $} & $ {\Omega_1  (\widehat s_-\le\widehat s_+\le 0)}$    \\
\cline{2-2}\cline{5-5}  
 & $\mathfrak C''$ &   &$ {\widehat s_ -}= \frac{ -2 }{ 3\frac{X_3}{\widehat \sigma X_2}\frac{\sin3\beta}{\sin 2 \beta}  } +\mbox{l.o.t.}  $   & $ {\Omega_4   {(\widehat s_-\ge\widehat s_+\ge 0)}}$ \\
 \hline
 \multirow{2}*{$\mathfrak D(\beta,X)$}  &     &\multirow{2}* {$ (- \frac { 1}2,0)$}    &  {$\widehat s_ +=\frac{-\Delta}{6\frac{X _3}{\widehat \sigma X_2}\frac{\sin 3\beta}{\sin2\beta} } +\mbox{l.o.t.}, $   } & $ \Omega_2  (\widehat s_-\ge 0\ge\widehat s_ + )$     \\ 
 &   &   &$ {\widehat s_ -}= \frac{ -2 }{ 3\frac{X_3}{\widehat \sigma X_2}\frac{\sin3\beta}{\sin 2 \beta}  } +\mbox{l.o.t.}  $   & $ {\Omega_3   {(\widehat s_+\ge 0\ge \widehat s_ - )}}$ \\
 \hline
 \multirow{2}*{$\mathfrak E(\beta,X)$}  &     &\multirow{2}* {$  (-\infty,- \frac { 1}2)$}    &\multirow{2}* {$\widehat s_\pm$ real roots} & $ \Omega_2  (\widehat s_-\ge 0\ge\widehat s_ + )$      \\  
 &   &   &   & $ {\Omega_3  (\widehat s_+\ge 0\ge \widehat s_ - )}$ \\
 \hline
\end{tabular}
\end{center} }

\caption{\small Properties of $\widehat s_\pm$ and $\Delta$ for  Type $\mathfrak A,\cdots,\mathfrak E$ when $X_1>0,X_2,X_3\ge 0$} 
\label{Tb:dynamic-FNH}
\end{table}

\end{definition}
In the following, we will  define essential stationary points $\widehat s_{j,\ast}$ and decompose  $[0,\widehat\sigma]$ into intervals $\mho_j$ around $\widehat s_{j,\ast}$. The deformation will be defined on $\mho_j $.  
\begin{definition}\label{D:deformation}   We define 
the essential stationary points $\widehat s_{j,\ast}=\widehat s_{j,\ast}(\widehat\sigma,\beta, X)$ by
\begin{align}
\widehat s_{0,\ast} = &  0, \nonumber\\
\widehat s_{1,\ast}= &  \left\{
{\ba{ll}  
 { \frac{\widehat s_{+ }+\widehat s_{- }}2\gtrless 0}  , &       Type\,\mathfrak A' \wedge(\widehat{\sigma}\in\{\sqrt{X_2},\sqrt[3]{X_3}\}), \,\mathfrak A'' , \\
  {  \inf\widehat s_{\pm }}>0 , &     Type\, \mathfrak B'',\mathfrak C'',\\
      \sup\widehat s_{\pm }>0 ,&    Type\,  \mathfrak D, \mathfrak  E,  \\
     - ,&  Type\,  \mathfrak A '\wedge(\widehat{\sigma}=X_1),\, \mathfrak B' ,\mathfrak C' ,  
 \ea}\right.  \ &\nonumber\\ 
  \widehat s_{2,\ast}= &  \left\{
{\ba{ll}  
  { \sup\widehat s_{\pm }}, &   \hskip.3in     Type\, \mathfrak B'',\mathfrak C'' ,\\
 -  ,&  \hskip.3in    \textit{others, }
 \ea}\right.\label{E:F3-ast} 
\end{align}   where $-$  means no definition.    Given $0<\epsilon_1<\frac{\pi}{2k}\ll 1$, define neighborhood $   \mho_j(\widehat\sigma,\beta,X )$ of essential critical points $\widehat s_{j,\ast}$ by 
\begin{align}
\mho_0 
=&\left\{
{\ba{l }  
{  [0, \frac 12]},    
\hskip3.15 in        Type\, \mathfrak A'\wedge(\widehat{\sigma}\in\{\sqrt{X_2},\sqrt[3]{X_3}\}),\mathfrak A'', \\
  {  [0, \frac 1{ 2\cos\epsilon_1 } \widehat s _{1,\ast} ],}   \hskip2.65 in   Type\, \mathfrak B'',\mathfrak C'',   \mathfrak D ,  \mathfrak E,\\
    {[0, \widehat\sigma\delta]=[0,\frac 12]\cup[\frac 12,\widehat\sigma\delta]=\mho_{0,<}\cup\mho_{0,>}} ,  
\hskip.9 in   Type\,\mathfrak A '\wedge(\widehat{\sigma}=X_1), \, \mathfrak B',\mathfrak C', 
 \ea}\right.\nonumber\\
 \mho_1  
=&
\left\{
{\ba{l}  
{  [\frac 12,\widehat s_{1,\ast}]\cup[\widehat s_{1,\ast}, \widehat\sigma\delta] }\equiv\mho_{1,<}\cup\mho_{1,>}, 
\hskip .1in Type\, \{[\mathfrak A'\wedge(\widehat{\sigma}\in      \{\sqrt{X_2},\sqrt[3]{X_3}\})]\vee\mathfrak A''\}\wedge (\widehat{s}_{1,\ast}>0),\\
{  [\frac 12,  \widehat\sigma\delta] }\equiv \mho_{1,>}, 
\hskip 1.4in  Type\, \{[\mathfrak A'\wedge(\widehat{\sigma}\in\{\sqrt{X_2},      \sqrt[3]{X_3}\})]\vee \mathfrak A''\}\wedge(\widehat{s}_{1,\ast}<0),\\
  { [ (1-  \frac 1{2\cos\epsilon_1 } ) \widehat s_{1,\ast}, \widehat s_{1,\ast} ]\cup [ \widehat s_{1,\ast}, \widehat s_{1,\ast}+ \frac {\widehat s_{2,\ast}-\widehat s_{1,\ast}  }{2\cos\epsilon_1  }  ]\equiv\mho_{1,<}\cup\mho_{1,>},}      
 \hskip  1in  Type\, \mathfrak B'',\mathfrak C'',\\
 { [ (1-  \frac 1{2\cos\epsilon_1 } )\widehat s_{1,\ast}, \widehat s_{1,\ast} ]\cup [  \widehat s_{1,\ast},\widehat\sigma\delta   ] \equiv\mho_{1,<}\cup\mho_{1,>},}      
\hskip 1.8in  Type\ \,  \mathfrak D ,\  \mathfrak E,\\
  {\phi,}      
\hskip 3.9 in  Type\,\mathfrak A '\wedge(\widehat{\sigma}=X_1),\,  \mathfrak B',\mathfrak C', 
 \ea}\right.\nonumber\\
 \mho_2  
=&
\left\{
{\ba{l} 
  { [  \widehat s_{2,\ast}-\frac {\widehat s_{2,\ast}-\widehat s_{2,\ast}  }{2 \cos\epsilon_1 } , \widehat s_{2,\ast} ]\cup [  \widehat s_{2,\ast},\widehat\sigma\delta ] \equiv\mho_{2,<}\cup\mho_{2,>},}   \hskip.2in Type\, \mathfrak B'',\mathfrak C'',\\
 \phi      \hskip3.49 in   \textit{otherwise},
 \ea}\right.
 \label{E:f-beta-interval}
\end{align}

Write
\be\label{E:FNH-lambda-sigma}
\begin{gathered}
\lambda=\kappa_1+\frac{\widehat re^{i\alpha} }{\widehat \sigma}=\kappa_1+\frac{ \widehat s_{j,\ast}e^{i\beta}+\widehat r_je^{i\alpha_j}}{\widehat \sigma}  ,\\
 \widehat r_j=\widehat r_j(\widehat\sigma,\beta,X,\lambda),\ \alpha_j=\alpha_j(\widehat\sigma,\beta,X,\lambda),\,j=0,1,2.
\end{gathered}
\ee 
We define the deformation   defined by
\be \label{E:f3-deform-s}
\begin{gathered}
\zeta=\kappa_1+se^{i\beta}=\kappa_1+\frac{\widehat s e^{i\beta}}{\widehat\sigma}\\
 \widehat  s \mapsto \xi_j\equiv  \widehat s_{j,\ast} +\widehat s_je^{i\tau_j}  ,   \ 
  \widehat s\equiv\widehat s_{j,\ast}\pm\widehat s_j\in \mho_j ,\,  |\tau_j|\lessgtr\frac\pi 2,\,\widehat s_j\ge 0,\, j=0,1,2, 
\end{gathered} 
\ee
with 
\begin{center}
\begin{tabular}{lll}
\multirow{8}{*}{$   
 \left\{
{\ba{l}  
  {\tau_0\equiv 0  ,} \\
  { \tau_0\equiv 0  ,} \\
   { \pm\epsilon_1\gtrless\tau_0\gtrless 0  ,} \\
  {\pm\frac{\epsilon_1 }4\gtrless\tau_0\gtrless 0  ,} \\
  { \mp\epsilon_1\lessgtr\tau_0\lessgtr 0  ,} \\
  {\mp\frac{\epsilon_1 }4\lessgtr\tau_0\lessgtr 0  ,}\\
  { \mp\epsilon_1\lessgtr\tau_0\lessgtr 0  ,} \\
  {\mp\frac{\epsilon_1 }4\lessgtr\tau_0\lessgtr 0  ,} 
  \ea}\right.$}&$\textit{for }\sin 3\beta\gtrless 0,\ \ |\alpha_0-\beta|\le \frac{\epsilon_1}2$,&$\widehat s\in  [0,\frac 12]\subset \mho_0$, $Type\ \mathfrak A,\mathfrak B',\mathfrak C'$,\\
    &$ \textit{for }\sin 3\beta\gtrless 0,\ \ |\alpha_0-\beta|\ge \frac{\epsilon_1}2$,&$\widehat s\in  [0,\frac 12]\subset \mho_0$, $Type\ \mathfrak A,\mathfrak B',\mathfrak C'$,\\
    &$\textit{for }\sin 3\beta\gtrless 0,\ \ |\alpha_0-\beta|\le \frac{\epsilon_1}2$,&$  \widehat s\in\mho_0$, $Type\ \mathfrak D,\mathfrak E$,\\
    &$ \textit{for }\sin 3\beta\gtrless 0,\ \ |\alpha_0-\beta|\ge \frac{\epsilon_1}2$,&$\widehat s\in\mho_0 $,  $Type\ \mathfrak D, \mathfrak E$,\\
    &$\textit{for }\sin 3\beta\gtrless 0,\ \ |\alpha_0-\beta|\le \frac{\epsilon_1}2$,&$\widehat s\in   \mho_{0,>}$, $Type\ \mathfrak A',\mathfrak B',\mathfrak C'$,\\
    &$ \textit{for }\sin 3\beta\gtrless 0,\ \ |\alpha_0-\beta|\ge \frac{\epsilon_1}2$,&$\widehat s\in   \mho_{0,>}$, $Type\ \mathfrak A',\mathfrak B',\mathfrak C'$,\\
    &$\textit{for }\sin 3\beta\gtrless 0,\ \ |\alpha_0-\beta|\le \frac{\epsilon_1}2$,&$  \widehat s\in\mho_0$, $Type\, \mathfrak B'',\mathfrak C''$,\\
    &$ \textit{for }\sin 3\beta\gtrless 0,\ \ |\alpha_0-\beta|\ge \frac{\epsilon_1}2$,&$\widehat s\in\mho_0$, $Type\, \mathfrak B'',\mathfrak C''$,\\
 \multirow{4}{*}{$   
 \left\{
{\ba{l}  
  {    \pm\pi \gtrless\tau_2\gtrless \pm\pi\mp\epsilon_1,}   \\
   {  \mp\epsilon_1\lessgtr\tau_2\lessgtr 0  , }  \\
  {    \pm\pi \gtrless\tau_2\gtrless \pm\pi\mp\frac{\epsilon_1 }4 ,}   \\
  { \mp\frac{\epsilon_1 }4\lessgtr\tau_2\lessgtr 0 ,}  
 \ea}\right.$}&$\textit{for }\sin 3\beta\gtrless 0,\ \ ||\alpha_2-\beta|- \pi|\le \frac{\epsilon_1}2$,&$  \widehat s\in\mho_{2<}$, $Type\,  \mathfrak B'', \mathfrak  C''$,\\
    &$\textit{for } \sin 3\beta\gtrless 0,\ \ |\alpha_2-\beta|\le \frac{\epsilon_1}2$,&$ \widehat s\in\mho_{2>},$ $Type\,  \mathfrak B'', \mathfrak  C''$,\\
    &$\textit{for }\sin 3\beta\gtrless 0,\ \ ||\alpha_2-\beta|- \pi|\ge \frac{\epsilon_1}2$,&$ \widehat s\in\mho_{2<} $, $Type\,  \mathfrak B'', \mathfrak  C''$,\\
    &$\textit{for }\sin 3\beta\gtrless 0,\ \ |\alpha_2-\beta|\ge \frac{\epsilon_1}2$,&$\widehat s\in\mho_{2>} $, $Type\,  \mathfrak B'', \mathfrak  C''$,
    \end{tabular}
\end{center}
    \begin{center}
\begin{tabular}{lll}

    \multirow{12}{*}{$   
 \left\{
{\ba{l}  
{  \mp\pi\lessgtr\tau_1\lessgtr \mp\pi\pm\epsilon_1 ,} \\
{ \mp\epsilon_1\lessgtr \tau_1\lessgtr 0  ,} \\
  {   \mp\pi\lessgtr \tau_1\lessgtr \mp\pi\pm\epsilon_1 ,} \\
   {   \pm\epsilon_1\gtrless\tau_1\gtrless 0 , }\\
   {  \pm\pi\gtrless\tau_1\gtrless\pm\pi\mp\epsilon_1,}\\
   {  \mp\epsilon_1\lessgtr\tau_1\lessgtr 0,}\\   
   { \mp\pi\lessgtr\tau_1\lessgtr \mp\pi\pm\frac{\epsilon_1 }4 ,} \\
   {  \mp\frac{\epsilon_1}4\lessgtr\tau_1\lessgtr 0 ,} \\
 {  \mp\pi\lessgtr \tau_1\lessgtr \mp\pi\pm\frac{\epsilon_1 }4,} \\
  {      \pm\frac{\epsilon_1 }4 \gtrless\tau_1\gtrless 0  ,} \\
   { \pm\pi\gtrless\tau_1\gtrless \pm\pi\mp\frac{\epsilon_1}4,}\\
   { \mp\frac{\epsilon_1}4\lessgtr\tau_1\lessgtr 0,}
 \ea}\right.$}
    &$\textit{for }\sin 3\beta\gtrless 0,\ \ ||\alpha_1-\beta|- \pi|\le \frac{\epsilon_1}2$,&$ \widehat s\in\mho_{1<}\ne\phi,\,  Type\, \mathfrak A ,$\\
    &$\textit{for }\sin 3\beta\gtrless0,\ \ |\alpha_1-\beta|\le \frac{\epsilon_1}2$,&$\widehat s\in\mho_{1>}\ne\phi, \,  Type\, \mathfrak A ,$\\ 
    &$\textit{for }\sin 3\beta\gtrless0,\ \ ||\alpha_1-\beta|- \pi|\le \frac{\epsilon_1}2$,&$\widehat s\in\mho_{1<}, $ $Type\, \mathfrak B'',\mathfrak C''$,\\
    &$\textit{for }\sin 3\beta\gtrless0,\ \ |\alpha_1-\beta|\le \frac{\epsilon_1}2$,&$ \widehat s\in\mho_{1>} ,$ $Type\, \mathfrak B'',\mathfrak C''$,\\
    &$\textit{for }\sin 3\beta\gtrless0,\ \ ||\alpha_1-\beta|- \pi|\le \frac{\epsilon_1}2$,&$ \widehat s\in\mho_{1<} ,$ $Type\, \mathfrak D,\mathfrak E$,\\
     &$\textit{for }\sin 3\beta\gtrless 0,\ \ |\alpha_1-\beta|\le \frac{\epsilon_1}2$,&$ \widehat s\in\mho_{1>} ,$ $Type\, \mathfrak D,\mathfrak E,\mathfrak B',\mathfrak C'$,\\
    &$ \textit{for }\sin 3\beta\gtrless 0,\ \ ||\alpha_1-\beta|- \pi|\ge \frac{\epsilon_1}2$,&$ \widehat s\in\mho_{1<}\ne\phi,\,  Type\, \mathfrak A ,$ \\
    &$ \textit{for }\sin 3\beta\gtrless 0,\ \ |\alpha_1-\beta|\ge \frac{\epsilon_1}2$,&$\widehat s\in\mho_{1>}\ne\phi,\,  Type\, \mathfrak A ,$ \\
    &$\textit{for }\sin 3\beta\gtrless 0,\ \ ||\alpha_1-\beta|- \pi|\ge \frac{\epsilon_1}2$,&$ \widehat s\in\mho_{1<}, $ $Type\, \mathfrak B'',\mathfrak C''$, \\
    &$\textit{for }\sin 3\beta\gtrless 0,\ \ |\alpha_1-\beta|\ge \frac{\epsilon_1}2$,&$\widehat s\in\mho_{1>}, $ $Type\, \mathfrak B'',\mathfrak C''$, \\
    &$\textit{for }\sin 3\beta\gtrless 0,\ \ ||\alpha_1-\beta|- \pi|\ge \frac{\epsilon_1}2$,&$ \widehat s\in\mho_{1<} ,$ $Type\, \mathfrak D,\mathfrak E$,\\
     &$\textit{for }\sin 3\beta\gtrless 0,\ \ |\alpha_1-\beta|\ge \frac{\epsilon_1}2$,&$ \widehat s\in\mho_{1>} ,$ $Type\, \mathfrak D, \mathfrak E,\mathfrak B',\mathfrak C'$,
\end{tabular}
\end{center}
 and
 \begin{center}
\begin{tabular}{lll} 
\multirow{8}{*}{$ \tau_{0,\dagger}=  
 \left\{
{\ba{l}  
  { 0  ,} \\
  {0  ,} \\
  { \pm\epsilon_1   ,} \\
  {\pm\frac{\epsilon_1 }4   ,}\\
  { \mp\epsilon_1   ,} \\
  {\mp\frac{\epsilon_1 }4   ,}\\
  { \mp\epsilon_1   ,} \\
  {\mp\frac{\epsilon_1 }4   ,} 
  \ea}\right.$}&$\textit{for }\sin 3\beta\gtrless 0,\ \ |\alpha_0-\beta|\le \frac{\epsilon_1}2$,&$\widehat s\in  [0,\frac 12]\subset \mho_{0 }$, $Type\ \mathfrak A ,\mathfrak B',\mathfrak C'$,\\
    &$ \textit{for }\sin 3\beta\gtrless 0,\ \ |\alpha_0-\beta|\ge \frac{\epsilon_1}2$,&$\widehat s\in  [0,\frac 12]\subset \mho_{0 }$, $Type\ \mathfrak A ,\mathfrak B',\mathfrak C'$,\\
    &$\textit{for }\sin 3\beta\gtrless 0,\ \ |\alpha_0-\beta|\le \frac{\epsilon_1}2$,&$  \widehat s\in\mho_0,\,Type\, \mathfrak D, \mathfrak E $, \\
    &$ \textit{for }\sin 3\beta\gtrless 0,\ \ |\alpha_0-\beta|\ge \frac{\epsilon_1}2$,&$\widehat s\in\mho_0 ,\,Type\, \mathfrak D, \mathfrak E$,\\
    &$\textit{for }\sin 3\beta\gtrless 0,\ \ |\alpha_0-\beta|\le \frac{\epsilon_1}2$,&$\widehat s\in    \mho_{0,> }$, $Type\ \mathfrak A' ,\mathfrak B',\mathfrak C'$,\\
    &$ \textit{for }\sin 3\beta\gtrless 0,\ \ |\alpha_0-\beta|\ge \frac{\epsilon_1}2$,&$\widehat s\in    \mho_{0,> }$, $Type\ \mathfrak A' ,\mathfrak B',\mathfrak C'$,\\
    &$\textit{for }\sin 3\beta\gtrless 0,\ \ |\alpha_0-\beta|\le \frac{\epsilon_1}2$,&$  \widehat s\in\mho_0 $,  $Type\, \mathfrak B'',\mathfrak C''$,\\
    &$ \textit{for }\sin 3\beta\gtrless 0,\ \ |\alpha_0-\beta|\ge \frac{\epsilon_1}2$,&$\widehat s\in\mho_0 $,  $Type\, \mathfrak B'',\mathfrak C''$,
    \\
 \multirow{12}{*}{$  \tau_{1,\dagger}=   
 \left\{
{\ba{l}  
{    \mp\pi\pm\epsilon_1 ,} \\
{ \mp\epsilon_1   ,} \\
  {    \mp\pi\pm\epsilon_1 ,} \\
   {   \pm\epsilon_1 , }\\
   {\pm\pi\mp\epsilon_1 ,}\\
   {  \mp\epsilon_1 ,}\\   
   {   \mp\pi\pm\frac{\epsilon_1 }4 ,} \\
   {  \mp\frac{\epsilon_1}4 ,} \\
 {  \mp\pi\pm\frac{\epsilon_1 }4,} \\
  { \pm\frac{\epsilon_1 }4   ,} \\
   {\pm\pi\mp\frac{\epsilon_1}4 ,}\\
   { \mp\frac{\epsilon_1}4 ,}
 \ea}\right.$}
    &$\textit{for }\sin 3\beta\gtrless 0,\ \ ||\alpha_1-\beta|- \pi|\le \frac{\epsilon_1}2$,&$ \widehat s\in\mho_{1<}\ne\phi,\,  Type\, \mathfrak A ,$\\
    &$\textit{for }\sin 3\beta\gtrless0,\ \ |\alpha_1-\beta|\le \frac{\epsilon_1}2$,&$\widehat s\in\mho_{1>}\ne\phi, \,  Type\, \mathfrak A ,$\\
    &$\textit{for }\sin 3\beta\gtrless0,\ \ ||\alpha_1-\beta|- \pi|\le \frac{\epsilon_1}2$,&$\widehat s\in\mho_{1<}, $ $Type\, \mathfrak B'', \mathfrak C''$,\\
    &$\textit{for }\sin 3\beta\gtrless0,\ \ |\alpha_1-\beta|\le \frac{\epsilon_1}2$,&$ \widehat s\in\mho_{1>} ,$ $Type\, \mathfrak B'', \mathfrak C''$,\\
    &$\textit{for }\sin 3\beta\gtrless0,\ \ ||\alpha_1-\beta|- \pi|\le \frac{\epsilon_1}2$,&$ \widehat s\in\mho_{1<} ,$ $Type\, \mathfrak D, \mathfrak E$,\\
     &$\textit{for }\sin 3\beta\gtrless 0,\ \ |\alpha_1-\beta|\le \frac{\epsilon_1}2$,&$ \widehat s\in\mho_{1>} ,$ $Type\, \mathfrak D, \mathfrak E$,\\
    &$ \textit{for }\sin 3\beta\gtrless 0,\ \ ||\alpha_1-\beta|- \pi|\ge \frac{\epsilon_1}2$,&$ \widehat s\in\mho_{1<},\  Type\, \mathfrak A ,$ \\
    &$ \textit{for }\sin 3\beta\gtrless 0,\ \ |\alpha_1-\beta|\ge \frac{\epsilon_1}2$,&$\widehat s\in\mho_{1>},\  Type\, \mathfrak A ,$ \\
    &$\textit{for }\sin 3\beta\gtrless 0,\ \ ||\alpha_1-\beta|- \pi|\ge \frac{\epsilon_1}2$,&$ \widehat s\in\mho_{1<}, $ $Type\, \mathfrak B'', \mathfrak C''$, \\
    &$\textit{for }\sin 3\beta\gtrless 0,\ \ |\alpha_1-\beta|\ge \frac{\epsilon_1}2$,&$\widehat s\in\mho_{1>}, $ $Type\, \mathfrak B'', \mathfrak C''$, \\
    &$\textit{for }\sin 3\beta\gtrless 0,\ \ ||\alpha_1-\beta|- \pi|\ge \frac{\epsilon_1}2$,&$ \widehat s\in\mho_{1<} ,$ $Type\, \mathfrak D, \mathfrak E$,\\
     &$\textit{for }\sin 3\beta\gtrless 0,\ \ |\alpha_1-\beta|\ge \frac{\epsilon_1}2$,&$ \widehat s\in\mho_{1>} ,$ $Type\, \mathfrak D, \mathfrak E$,  
\end{tabular}
\end{center}
    \begin{center}
\begin{tabular}{lll}     
 \multirow{4}{*}{$ \tau_{2,\dagger}=    
 \left\{
{\ba{l}  
  {\pm\pi\mp\epsilon_1   ,}   \\
   {\mp\epsilon_1  , }  \\
  {\pm\pi\mp\frac{\epsilon_1 }4    ,}   \\
  {\mp\frac{\epsilon_1 }4  ,}  
 \ea}\right.$}&$\textit{for }\sin 3\beta\gtrless 0,\ \ ||\alpha_2-\beta|- \pi|\le \frac{\epsilon_1}2$,&$  \widehat s\in\mho_{2<},$ $Type\, \mathfrak B'', \mathfrak C''$,\\
    &$\textit{for } \sin 3\beta\gtrless 0,\ \ |\alpha_2-\beta|\le \frac{\epsilon_1}2$,&$ \widehat s\in\mho_{2>},$ $Type\, \mathfrak B'', \mathfrak C''$,\\
    &$\textit{for }\sin 3\beta\gtrless 0,\ \ ||\alpha_2-\beta|- \pi|\ge \frac{\epsilon_1}2$,&$ \widehat s\in\mho_{2<},$ $Type\, \mathfrak B'', \mathfrak C''$,\\
    &$\textit{for }\sin 3\beta\gtrless 0,\ \ |\alpha_2-\beta|\ge \frac{\epsilon_1}2$,&$\widehat s\in\mho_{2>}, $ $Type\, \mathfrak B'', \mathfrak C''$,    
    \end{tabular}
\end{center}

\end{definition}

The following lemmas  show   that the deformation fulfills goals of \eqref{E:deform-intro-1} partially. 
\begin{table}
{\begin{center}
\vskip.1in
\begin{tabular}{|l|l|}
   \hline
\bf{Case}& \bf{Type $\mathfrak A$} \cr
   \hline\hline
   $\widehat s \in\mho_0$&$\blacktriangleright \mathfrak{Re}(-i\wp(\frac{\widehat s e^{i\tau }}{{\widehat\sigma}},\beta,X)) $\\
   &$\color{blue}=\frac{X_3}{\widehat\sigma^3} \sin3\tau\sin3\beta\widehat s(\widehat s-  \frac{3\sin2\tau }{ 2\sin3\tau }\widehat s_{1,\ast})^2 +\frac{X_1}{\widehat\sigma}\sin\tau\sin\beta\widehat s(1- \frac{3\sin^22\tau}{4\sin\tau\sin3\tau}\frac{1}{\Delta })$ \\
 \hline
 $\widehat s \in\mho_1$&$\blacktriangleright \mathfrak{Re}(-i\wp(\frac{\widehat s_1 e^{i \tau_1 } +\widehat s_{1,\ast}}{{\widehat\sigma}},\beta,X)) $\\
 &$\color{blue}=  \frac{X_3}{\widehat\sigma^3}\sin3\tau_1\sin3\beta \widehat s_1^3 +\frac{X_1}{\widehat\sigma}\sin\tau_1\sin\beta\widehat s_1(1-\frac{1}{\Delta })  $   \\
   \hline
 \hline
\bf{Case} &\bf{Type $\mathfrak B$,   $ \mathfrak C$} \cr
\hline\hline  
$\widehat s\in\mho_0$& $\blacktriangleright\mathfrak{Re}(-i\wp(\frac{\widehat s e^{i\tau}}{{\widehat\sigma}},\beta,X))$\\  
& $\color{blue}=\frac{X_3}{\widehat\sigma^3}\sin3\tau \sin3\beta \widehat s   (\widehat s-\frac{    - 1+\sqrt{1-\frac 43 \frac{\sin\tau \sin3\tau  }{\sin^22\tau }\Delta}  }{ 3\frac{X_3 }{ \widehat\sigma X_2}\frac{ \sin3\beta}{ \sin2\beta}\frac{2\sin3\tau}{3\sin2\tau}})    (\widehat s- \frac{    - 1-\sqrt{1-\frac 43 \frac{\sin\tau \sin3\tau }{\sin^22\tau }\Delta}  }{ 3\frac{X_3 }{ \widehat\sigma X_2}\frac{ \sin3\beta}{ \sin2\beta}\frac{2\sin3\tau}{3\sin2\tau}})$ \\
\hline
$ \widehat s\in\mho_1$ & $\blacktriangleright\mathfrak{Re}(-i\wp(\frac{\widehat s_1 e^{i\tau_1 } +\widehat s_{1,\ast}}{{\widehat\sigma}},\beta,X))$\\
$\color{black} $&$\color{blue} = \frac{X_3}{\widehat\sigma^3} \sin3\tau_1 \sin3\beta  \widehat s_1^2  (  \widehat s_1+   \frac {\sqrt{1-\Delta}}{3\frac{X_3}{\widehat\sigma X_2}\frac{ \sin3\beta}{ \sin 2\beta}\frac{\sin 3\tau_1}{   3 \sin2\tau_1}}  ) $\\
\hline
$\widehat s\in\mho_2$ & $\blacktriangleright\mathfrak{Re}(-i\wp(\frac{\widehat s_2 e^{i\tau_2 } +\widehat s_{2,\ast}}{{\widehat\sigma}},\beta,X))$\\
$\color{black}  $&$\color{blue} = \frac{X_3}{\widehat\sigma^3} \sin3\tau_2 \sin3\beta  \widehat s_2^2  (  \widehat s_2- \frac {\sqrt{1-\Delta}}{3\frac{X_3}{\widehat\sigma X_2}\frac{ \sin3\beta}{ \sin 2\beta}\frac{\sin 3\tau_2}{   3 \sin2\tau_2}}  ) $\\
\hline\hline
\bf{Case} &\bf{Type $\mathfrak D$, $\mathfrak E$} \cr
\hline\hline  
$\widehat s\in\mho_0$& $\blacktriangleright\mathfrak{Re}(-i\wp(\frac{\widehat s e^{i\tau}}{{\widehat\sigma}},\beta,X))$\\  
& $\color{blue}=\frac{X_3}{\widehat\sigma^3}\sin3\tau \sin3\beta \widehat s   (\widehat s-\frac{    - 1+\sqrt{1-\frac 43 \frac{\sin\tau \sin3\tau  }{\sin^22\tau }\Delta}  }{ 3\frac{X_3 }{ \widehat\sigma X_2}\frac{ \sin3\beta}{ \sin2\beta}\frac{2\sin3\tau}{3\sin2\tau}})    (\widehat s- \frac{    - 1-\sqrt{1-\frac 43 \frac{\sin\tau \sin3\tau }{\sin^22\tau }\Delta}  }{ 3\frac{X_3 }{ \widehat\sigma X_2}\frac{ \sin3\beta}{ \sin2\beta}\frac{2\sin3\tau}{3\sin2\tau}})$ \\
\hline
$\widehat s\in\mho_1$ & $\blacktriangleright\mathfrak{Re}(-i\wp(\frac{\widehat s_1 e^{i\tau_1 } +\widehat s_{1,\ast}}{{\widehat\sigma}},\beta,X))$\\
$\widehat s_{1,\ast}= \widehat s_\pm $ &$\color{blue}= \frac{X_3}{\widehat\sigma^3} \sin3\tau_1 \sin3\beta  \widehat s_1^2  (  \widehat s_1\pm  \frac {\sqrt{1-\Delta}}{3\frac{X_3}{\widehat\sigma X_2}\frac{ \sin3\beta}{ \sin 2\beta}\frac{\sin 3\tau_1}{   3 \sin2\tau_1}}  ) $  
  \cr
  
  \hline
  
\end{tabular}
\end{center} }
\caption{\small $\mathfrak{Re}(-i\wp(\frac{\widehat s}{\widehat\sigma},\beta,X))$ for  deformation defined by Definition \ref{D:deformation}} 
\label{Tb:f3-deformation}
\end{table}
\begin{lemma}\label{L:ideas} 
Define the deformation $\widehat s\mapsto \xi_j=\widehat s_je^{i\tau_{j}}+\widehat s_{j,\ast}$ on $\mho_j$ by Definition \ref{D:deformation}. We have, for $j=0,1,2$,  
\begin{align}
{|\widehat s_j e^{i\tau_{j,\dagger}}-\widehat r_je^{i(\alpha_j-\beta)}| }  \ge & \frac 1C \max\{\widehat r_j,\widehat s_j\},\quad\textit{ if $\,\tau_{j,\dagger}\ne 0$},
 \label{E:f3-denominator}
\end{align} and   
\be\mathfrak{Re}(-i\wp(\frac{\xi_j}{\widehat\sigma},\beta,X))\le   0.
 \label{E:definitive}
 \ee

\end{lemma}
\begin{proof}\hfill \\

\begin{itemize}
\item [$\blacktriangleright$] {\bf Proof of \eqref{E:f3-denominator} :}  According to the definition of $\tau_{j,\dagger}$ in Definition \ref{D:deformation}, if $\tau_{j,\dagger}\ne 0$ then
\be \label{E:angle}
|\alpha_j-\beta-\tau_{j,\dagger}|\ge  \frac{\epsilon_1} {4},\quad j=0,1,2.
\ee As a result,  \eqref{E:f3-denominator} is justified.

\item [$\blacktriangleright$] {\bf Proof of \eqref{E:definitive} :} 
In view of Definition \ref{D:deformation}, Table \ref{Tb:f3-deformation}, and Figure \ref{Fg:signature}, 
\begin{itemize}
\item for Type $\mathfrak A $:
\begin{itemize}
\item On $\mho_1$,   both terms of $\mathfrak{Re}(-i\wp(\frac{\xi_j}{\widehat\sigma},\beta,X))$ are of  the same signatures. Along with Definition \ref{D:deformation}  and Figure \ref{Fg:signature}, yields \eqref{E:definitive}.

\item On $\mho_{0}$, according to the assumption, we need only to consider $Type\, \mathfrak A'$ (i.e., $ \Delta\ge 2$),  $\widehat{\sigma}=X_1$. Along with Definition \ref{D:deformation}  and Figure \ref{Fg:signature},      
\begin{align}
& 3\frac{X_3}{X_1^3}\sin3\beta(\widehat s-\frac32\frac{\sin2\tau}{\sin3\tau } \frac{\widehat s_{+}+\widehat s_{-}}{2} )^2 \label{E:a-1-est-2} \\
+&\frac{X_1}{X_1 }\sin \beta(1- \frac{3\sin^22\tau}{4\sin\tau\sin3\tau }  \frac 1\Delta) \nonumber\\
\le  &-| \frac{X_1}{X_1 }\sin \beta(1- \frac{3\sin^22\tau}{4\sin\tau\sin3\tau }  \frac 1\Delta)|\le - C_0|\sin\beta|.\nonumber
\end{align} Hence we prove   \eqref{E:definitive}.
\end{itemize}

\item For Type $\mathfrak B'',\mathfrak C''$:
\begin{itemize}
\item on $\mho_0$, from $\widehat s\le \frac 1{2\cos\epsilon_1}  \widehat s_{1,\ast}$ and $\epsilon_1\ll 1$, we prove \eqref{E:definitive} by means of
\be\label{E:bc-0}
\hskip1in (\widehat s-\frac{    - 1+\sqrt{1-\frac 43 \frac{\sin\tau \sin3\tau  }{\sin^22\tau }\Delta}  }{ 3\frac{X_3 }{ \widehat\sigma X_2}\frac{ \sin3\beta}{ \sin2\beta}\frac{2\sin3\tau}{3\sin2\tau}})    (\widehat s- \frac{    - 1-\sqrt{1-\frac 43 \frac{\sin\tau \sin3\tau }{\sin^22\tau }\Delta}  }{ 3\frac{X_3 }{ \widehat\sigma X_2}\frac{ \sin3\beta}{ \sin2\beta}\frac{2\sin3\tau}{3\sin2\tau}}) 
\ge   \frac 18\widehat s^2_{1,\ast}. 
\ee  

\item On $\mho_1$, from Figure \ref{Fg:signature}, $\frac{\sin3\beta}{\sin2\beta}\frac{\sin3\tau_1}{\sin2\tau_1} \gtrless 0$ on $\mho_{1,\lessgtr}$. It reduces to proving \eqref{E:definitive} on $\mho_{1,>}$. From the definition of $\mho_{1,>}$ and \eqref{E:roots-distance}, we have 
\be\label{E:bc-1-new}
\widehat s_1\le  \frac {\widehat s_{2,\ast}-\widehat s_{1,\ast}  }{2  \cos\epsilon_1}= \frac{1}{\cos\epsilon_1} |\frac{\sqrt{1-\Delta}}{3{ \frac{X_3}{\widehat\sigma X_2 } \frac{ \sin3\beta  }{  \sin2\beta} }}|.
\ee Therefore, 
\be\label{E:bc-1}
\hskip1.5in \widehat s_1+   \frac {\sqrt{1-\Delta}}{3\frac{X_3}{\widehat\sigma X_2}\frac{ \sin3\beta}{ \sin 2\beta}\frac{\sin 3\tau_1}{   3 \sin2\tau_1}} \le  \frac 12\frac {\sqrt{1-\Delta}}{3\frac{X_3}{\widehat\sigma X_2}\frac{ \sin3\beta}{ \sin 2\beta}\frac{\sin 3\tau_1}{   3 \sin2\tau_1}} \le 0, \quad \widehat s_1\in\mho_{1,>}.
\ee As a result,  \eqref{E:definitive} follows. 

\item On $\mho_2$, from Figure \ref{Fg:signature}, $-\frac{\sin3\beta}{\sin2\beta}\frac{\sin3\tau_2}{\sin2\tau_2} \lessgtr 0$ on $\mho_{2,\lessgtr}$. It reduces to proving \eqref{E:definitive} on $\mho_{2,<}$. From the definition of $\mho_{2,<}$, we have 
\be\label{E:bc-2-new}
\widehat s_2\le  \frac {\widehat s_{2,\ast}-\widehat s_{1,\ast}  }{2  \cos\epsilon_1} = \frac{1}{\cos\epsilon_1}|\frac{\sqrt{1-\Delta}}{3{ \frac{X_3}{\widehat\sigma X_2 } \frac{ \sin3\beta  }{  \sin2\beta} }}|.
\ee Therefore, 
\be\label{E:bc-2}
\hskip1.35in \widehat s_2-   \frac {\sqrt{1-\Delta}}{3\frac{X_3}{\widehat\sigma X_2}\frac{ \sin3\beta}{ \sin 2\beta}\frac{\sin 3\tau_2}{   3 \sin2\tau_2}}\le  -\frac 12\frac {\sqrt{1-\Delta}}{3\frac{X_3}{\widehat\sigma X_2}\frac{ \sin3\beta}{ \sin 2\beta}\frac{\sin 3\tau_2}{   3 \sin2\tau_2}} \le 0  ,\quad \widehat s_2\in \mho_{2,<}. 
\ee Hence  \eqref{E:definitive} is proved. 
\end{itemize} 

\item For Type $ \mathfrak B', \mathfrak C'$: it is sufficient to consider \eqref{E:definitive} for   $\widehat{s}\in \mho_{0,>}$. Hence \eqref{E:definitive} is proved by $\widehat s_\pm\le 0$ and $\widehat s>\frac 12$. 

\item For Type $\mathfrak D,\mathfrak E$: 

\begin{itemize}
\item on $\mho_0$,  we prove \eqref{E:definitive} by means of $\widehat s\le \frac 1{2\cos\epsilon_1}  \widehat s_{1,\ast}$, $\epsilon_1\ll 1$,   and
\be\label{E:de-0}
\begin{split}
\hskip1. in &(\widehat s-\frac{    - 1+\sqrt{1-\frac 43 \frac{\sin\tau \sin3\tau  }{\sin^22\tau }\Delta}  }{ 3\frac{X_3 }{ \widehat\sigma X_2}\frac{ \sin3\beta}{ \sin2\beta}\frac{2\sin3\tau}{3\sin2\tau}})    (\widehat s- \frac{    - 1-\sqrt{1-\frac 43 \frac{\sin\tau \sin3\tau }{\sin^22\tau }\Delta}  }{ 3\frac{X_3 }{ \widehat\sigma X_2}\frac{ \sin3\beta}{ \sin2\beta}\frac{2\sin3\tau}{3\sin2\tau}}) 
\\
\hskip1. in \le &  -\frac 14\widehat s\widehat s_{1,\ast}\le  -\frac 12\widehat s^2.
\end{split}
\ee 

\item on $\mho_1$, if $\widehat s_{1,\ast}=\widehat s_+$, by means of  Table \ref{Tb:dynamic-FNH} and Figure \ref{Fg:signature}, $\frac{\sin3\beta}{\sin2\beta}\frac{\sin3\tau_1}{\sin2\tau_1} \lessgtr 0$ on $\mho_{1,\lessgtr}$. It reduces to proving \eqref{E:definitive} on $\mho_{1,<}$. From the definition of $\mho_{1,<}$, \eqref{E:roots-distance}, and two roots are opposite signs, we have 
\be\label{E:de-1}
\begin{split}
& \widehat s_1+ \frac {\sqrt{1-\Delta}}{3\frac{X_3}{\widehat\sigma X_2}\frac{ \sin3\beta}{ \sin 2\beta}\frac{\sin 3\tau_1}{   3 \sin2\tau_1}} \le +\frac 12  \frac {\sqrt{1-\Delta}}{3\frac{X_3}{\widehat\sigma X_2}\frac{ \sin3\beta}{ \sin 2\beta}\frac{\sin 3\tau_1}{   3 \sin2\tau_1}}  \le -\frac 14\widehat s _1 ,
\end{split}
\ee and  \eqref{E:definitive} is proved.

On $\mho_1$, if $\widehat s_{1,\ast}=\widehat s_-$, by means of  Table \ref{Tb:dynamic-FNH} and Figure \ref{Fg:signature}, $-\frac{\sin3\beta}{\sin2\beta}\frac{\sin3\tau_1}{\sin2\tau_1} \lessgtr 0$ on $\mho_{1,\lessgtr}$. It reduces to proving \eqref{E:definitive} on $\mho_{1,<}$. From the definition of $\mho_{1,<}$, \eqref{E:roots-distance}, and two roots are opposite signs, we have 
\be\label{E:de-1-s-}
\begin{split}
& \widehat s_1- \frac {\sqrt{1-\Delta}}{3\frac{X_3}{\widehat\sigma X_2}\frac{ \sin3\beta}{ \sin 2\beta}\frac{\sin 3\tau_1}{   3 \sin2\tau_1}} \le -\frac 12  \frac {\sqrt{1-\Delta}}{3\frac{X_3}{\widehat\sigma X_2}\frac{ \sin3\beta}{ \sin 2\beta}\frac{\sin 3\tau_1}{   3 \sin2\tau_1}}  \le -\frac 14\widehat s _1 ,
\end{split}
\ee and  \eqref{E:definitive} is proved. 
\end{itemize}

\end{itemize}

\end{itemize} 
\end{proof}

\begin{lemma}\label{L:p-estimate} Let $\tilde\sigma=   \max\{ 1,  X_1 ,  \sqrt{ X_2 },  \sqrt[3]{ X_3 }\}$ and $X_j\ge 0$ defined by \eqref{E:phase}.
\begin{itemize}
\item [(i)] For $\,Type\,\mathfrak B'',\,\mathfrak C'',\,\mathfrak D,\,\mathfrak E$ on $\mho_j$, $j=0,1,2$, and $\,Type\,\mathfrak B',\,\mathfrak C'$ on $\mho_{0,>}$, one has
\begin{align}
\mathfrak{Re}(-i\wp(\frac{ \widehat s_je^{i\tau_{j,\dagger}}+ \widehat  s_{j,\ast}}{\widehat  \sigma},\beta,X)) \le &
     -\frac {| \sin 3\beta|}C \widehat  s_j^{3},\hskip.72in \widehat\sigma=\sqrt[3]{X_3};\label{E:f3-deform-wp}\\
     \mathfrak{Re}(-i\wp(\frac{\widehat   s_je^{i\tau_{j,\dagger}}+ \widehat  s_{j,\ast}}{\widehat  \sigma},\beta,X)) \le& 
     -\frac {| \sin 2\beta|}C \min\{ \widehat s_j^{3},  \widehat s_j^{2}\}, \ \widehat\sigma=\sqrt{X_2}.\label{E:f2-deform-wp}
\end{align}

\item [(ii)] For $\,Type\,\mathfrak A$ on $\mho_1$, 
\begin{align}
\mathfrak{Re}(-i\wp(\frac{ \widehat s_je^{i\tau_{j,\dagger}}+ \widehat  s_{j,\ast}}{\widehat  \sigma},\beta,X)) \le &
     -\frac {| \sin 3\beta|}C \widehat  s_j^{3},\hskip.72in \widehat\sigma=\sqrt[3]{X_3};\label{E:f3-deform-wp-1}\\
     \mathfrak{Re}(-i\wp(\frac{\widehat   s_je^{i\tau_{j,\dagger}}+ \widehat  s_{j,\ast}}{\widehat  \sigma},\beta,X)) \le& 
     -\frac {|  \sin 2\beta|}C   \widehat s_j^{3}, \hskip.72in   \widehat\sigma=\sqrt{X_2}=\tilde\sigma.\label{E:f2-deform-wp-1}
\end{align}
\item [(iii)] For
\begin{itemize}
\item  $Type\,\mathfrak B'',\,\mathfrak C'',\,\mathfrak D,\,\mathfrak E$ on $\mho_{0}$;
\item $Type\,\mathfrak A',\,\mathfrak B',\,\mathfrak C' $  on $\mho_{0,>}$,
\end{itemize}  one has
\begin{align}  
\quad\mathfrak{Re}(-i\wp(\frac{\widehat   s_0e^{i\tau_{0,\dagger}}}{\widehat  \sigma},\beta,X))\le &- \frac 1C| \sin\beta|  \widehat   s,\hskip.8in \widehat\sigma=  {X_1}=\tilde\sigma.     \label{E:all-0}  
\end{align}
\end{itemize}

\end{lemma}
\begin{proof}  \hfill \\
\begin{itemize}
\item [$\blacktriangleright$] {\bf Proof of \eqref{E:f3-deform-wp} and \eqref{E:f3-deform-wp-1}:}
In view of $\widehat\sigma=\sqrt[3]{X_3}$, Definition \ref{D:deformation}, Table \ref{Tb:f3-deformation}, and Figure \ref{Fg:signature}, 
\begin{itemize}
\item for Type $\mathfrak A $ on $\mho_1$, from Figure \ref{Fg:signature},   both terms of $\mathfrak{Re}(-i\wp(\frac{\xi_1}{\sqrt[3]{X_3}},\beta,X))$ are of  the same signatures. Therefore,
\begin{align*}
&{Re}(-i\wp(\frac{\xi_1}{ \sqrt[3]{X_3}},\beta,X))\\
=& \sin3\tau_1\sin3\beta \widehat s_1^3 +\frac{X_1}{\sqrt[3]{X_3}}\sin\tau_1\sin\beta\widehat s_1(1-\frac{1}{\Delta })\\
\le &-|\sin3\tau_1\sin3\beta| \widehat s_1^3
\end{align*} and  \eqref{E:f3-deform-wp-1} is proved.
\item For Type $\mathfrak B'',\mathfrak C''$: 
\begin{itemize}
\item On $\mho_0$, by means of \eqref{E:bc-0},   we prove \eqref{E:f3-deform-wp}. 

\item On $\mho_1$, from Figure \ref{Fg:signature}, $\frac{\sin3\beta}{\sin2\beta}\frac{\sin3\tau_1}{\sin2\tau_1} \gtrless 0$ on $\mho_{1,\lessgtr}$. It reduces to proving \eqref{E:f3-deform-wp} on $\mho_{1,>}$. Then \eqref{E:f3-deform-wp} follows from \eqref{E:bc-1-new}, \eqref{E:bc-1} and the definition  of interval  $\mho_{1,>}$.
\item On $\mho_2$, from Figure \ref{Fg:signature}, $-\frac{\sin3\beta}{\sin2\beta}\frac{\sin3\tau_2}{\sin2\tau_2} \lessgtr 0$ on $\mho_{2,\lessgtr}$. It reduces to proving \eqref{E:f3-deform-wp} on $\mho_{2,<}$. Then \eqref{E:f3-deform-wp} follows from \eqref{E:bc-2-new}, \eqref{E:bc-2} and the definition  of interval  $\mho_{2,<}$.
\end{itemize}

\item For Type $ \mathfrak B', \mathfrak C'$ on $\mho_{0,>}$, \eqref{E:f3-deform-wp} is proved since $\widehat s_\pm\le 0$ and $\widehat s>\frac 12$.

\item For Type $\mathfrak D,\mathfrak E$: 
\begin{itemize}
\item on $\mho_0$, by means of \eqref{E:de-0},   we prove \eqref{E:f3-deform-wp}. 
\item On $\mho_1$, from Figure \ref{Fg:signature}, $\widehat s_{1,\ast}=\widehat s_\pm$, $\pm\frac{\sin3\beta}{\sin2\beta}\frac{\sin3\tau_1}{\sin2\tau_1} \lessgtr 0$ on $\mho_{1,\lessgtr}$. It reduces to proving \eqref{E:f3-deform-wp} on $\mho_{1,<}$. Then \eqref{E:f3-deform-wp} follows from \eqref{E:de-1}, \eqref{E:de-1-s-}, and the definition  of interval  $\mho_{1,<}$.
\end{itemize}

\end{itemize}
\item [$\blacktriangleright$] {\bf Proof of \eqref{E:f2-deform-wp}:} In view of $\widehat\sigma=\sqrt{X_2}$, Definition \ref{D:deformation}, Table \ref{Tb:f3-deformation}, and Figure \ref{Fg:signature}, 
\begin{itemize}
\item For Type $\mathfrak B''$ on $\mho_j$, $j=0,1,2$:  Notice  \eqref{E:determinant} implies 
\be\label{E:f2-coef}
3\frac{X_3}{X_2^{3/2}}\sin3\beta=\frac{X_2^{1/2}}{X_1 } \frac{\sin2\beta}{\sin\beta}\Delta \sin2\beta .
\ee Together with Table \ref{Tb:dynamic-FNH},  $\Delta>\frac 12$,  and similar argument as that for  $\widehat{\sigma}=\sqrt[3]{X_3}$, yields 
\be\label{E:f2-12-a-1}
\begin{gathered}
 \mathfrak{Re}(-i\wp(\frac{\widehat s_je^{i\tau_{j,\dagger}}+\widehat s_{j,\ast}}{ \widehat \sigma },\beta,X)) \le -\frac 1C| \sin 2\beta|\widehat s_j^{3}.
  \end{gathered}
 \ee 
 
\item For Type $   \mathfrak B'$ on $\mho_{0,>}$: Using \eqref{E:f2-coef}, Table \ref{Tb:dynamic-FNH},  $\Delta>\frac 12$,  $\widehat{s}_\pm\le 0$,  and   $\widehat{s}>\frac 12$, yields 
\be\label{E:f2-12-a-2}
\begin{gathered}
 \mathfrak{Re}(-i\wp(\frac{\widehat s_je^{i\tau_{j,\dagger}}+\widehat s_{j,\ast}}{ \widehat \sigma },\beta,X)) \le -\frac 1C| \sin 2\beta|\widehat s_j^{3}.
  \end{gathered}
 \ee

\item For Type $ \mathfrak C'', \mathfrak D, \mathfrak E$ on $\mho_j$, $j=0,1,2$:   
Via \eqref{E:f2-coef}, similar argument as that for   $\widehat{\sigma}=\sqrt[3]{X_3}$, 
\begin{align}
| \widehat s_j+(-1)^{j+1}\frac {\sqrt{1-\Delta}}{3\frac{X_3}{\widehat\sigma X_2}\frac{ \sin3\beta}{ \sin 2\beta}\frac {\sin 3\tau_j}{   2 \sin2\tau_j} }    |\ge\frac 1C|  \frac{ 1}{ \frac{X _3}{X_2^{3/2}}\frac{\sin3\beta}{\sin 2 \beta}  }|,& \  \textit{on $\mho_j$, $j=1,2$, for $Type\ \mathfrak C''$},\nonumber\\
| \widehat s_1\pm\frac {\sqrt{1-\Delta}}{3\frac{X_3}{\widehat\sigma X_2}\frac{ \sin3\beta}{ \sin 2\beta}\frac {\sin 3\tau_1}{   2 \sin2\tau_1} }    |\ge\frac 1C|  \frac{1}{ \frac{X _3}{X_2^{3/2}}\frac{\sin3\beta}{\sin 2 \beta}  }|,&  \ \textit{on $\mho_1, 0<\widehat s_\pm$  for $Type\ \mathfrak D,\mathfrak E $,}\nonumber \\
| \widehat s_{1,\ast}-\widehat s_{2,\ast}|\ge\frac 1C|  \frac{ 1 }{ \frac{X _3}{X_2^{3/2}}\frac{\sin3\beta}{\sin 2 \beta}  }| ,&  \ \textit{for $Type\ \mathfrak C,\mathfrak D, \mathfrak E $}.\label{E:2-0-1}
\end{align}
As a result,
\be\label{E:f2-12-cde}
\begin{gathered}
   \mathfrak{Re}(-i\wp(\frac{\widehat s_je^{i\tau_{j,\dagger}}+\widehat s_{j,\ast}}{ \widehat \sigma },\beta,X))\le-  \frac 1C|\sin  2\beta | \widehat s_j^2.
\end{gathered} 
\ee
\item For Type $ \mathfrak C' $ on $\mho_{0,>}$: 
Thanks to $\widehat s_\pm\le 0$, $\widehat{s}>\frac 12$, $0\le \Delta\le \frac 12$, \eqref{E:f2-coef}, and \eqref{E:2-0-1},  
\be\label{E:f2-12-cde-new}
\begin{gathered}
   \mathfrak{Re}(-i\wp(\frac{\widehat s_je^{i\tau_{j,\dagger}}+\widehat s_{j,\ast}}{ \widehat \sigma },\beta,X))\le-  \frac 1C|\sin  2\beta | \widehat s_j^2.
\end{gathered} 
\ee
\end{itemize} Hence \eqref{E:f2-deform-wp} is justified.

\item [$\blacktriangleright$] {\bf Proof of   \eqref{E:f2-deform-wp-1}:} In this case, 
$\widehat\sigma=\sqrt{X_2}=\max\{X_1,\sqrt{X_2},\sqrt[3]{X_3}\}$ for Type $ \mathfrak A $ on $\mho_1$.  Together with  \eqref{E:f2-coef}, Table \ref{Tb:dynamic-FNH},  $\Delta>\frac 12$,   yields 
\be\label{E:f2-12-a}
\begin{gathered}
 \mathfrak{Re}(-i\wp(\frac{\widehat s_je^{i\tau_{j,\dagger}}+\widehat s_{j,\ast}}{ \widehat \sigma },\beta,X)) \le -\frac 1C| \sin 2\beta|\widehat s_j^{3}.
  \end{gathered}
 \ee  Hence \eqref{E:f2-deform-wp-1} is justified.

 \item[$\blacktriangleright$] {\bf Proof of \eqref{E:all-0} :} In this case, 
$\widehat\sigma= X_1=\max\{X_1,\sqrt{X_2},\sqrt[3]{X_3}\}$,
\begin{itemize}
\item For Type $ \mathfrak A '$ on $\mho_0$, the estimates then follows from \eqref{E:all-0}.
\item For $\,Type\,\mathfrak B'',\,\mathfrak C'',\,\mathfrak D,\,\mathfrak E$  on $\mho_0$, and  for $\,Type\,\mathfrak B',\,\mathfrak C',$   on $\mho_{0,>}$,  we use 
\be\label{E:roots}  
3\frac{X_3}{X_1^3}\sin3\beta =\left(\frac{X_2}{X_1^2}\frac{\sin2\beta}{\sin\beta}\right)^2\sin\beta\Delta,\quad 3\frac{X_3}{X_1X_2 }\frac{\sin3\beta}{\sin2\beta}=\frac{X_2}{X_1^2}\frac{\sin2\beta}{\sin\beta}\Delta.  
\ee Hence  \eqref{E:all-0} follows from
\be\label{E:u-0-f1}
\qquad \left\{
{\ba{ll}   \eqref{E:f3-deform-s}, \eqref{E:roots},  \frac 12\le\Delta\le 1, \textit{ \eqref{E:determinant}, Table \ref{Tb:f3-deformation}}&    \textit{ $Type\,\mathfrak B    $},\\
\textit{\eqref{E:determinant}, Table \ref{Tb:f3-deformation}},  \eqref{E:roots},  |(\widehat s -\widehat s_+')(\widehat s -\widehat s_-')|\ge \frac 1C  \frac{ 1 }{ (\frac{X _2}{X_1^2}\frac{\sin2\beta}{\sin   \beta})^2|\Delta|  },&   \textit{ $Type\,\mathfrak C,\mathfrak D$},\\\textit{\eqref{E:determinant}, Table \ref{Tb:f3-deformation}},  \eqref{E:roots},  |(\widehat s -\widehat s_+')(\widehat s -\widehat s_-')|\ge \frac 1C  \frac{ 1 }{ (\frac{X _2}{X_1^2}\frac{\sin2\beta}{\sin   \beta})^2|\Delta|  },&    \textit{  $Type\,\mathfrak E$}.\ea}\right.   \ee    
\end{itemize}  
\end{itemize}

\end{proof}

\begin{definition}\label{D:dominant} Let $\tilde\sigma=   \max\{ 1,  X_1 ,  \sqrt{ X_2 },  \sqrt[3]{ X_3 }\}$   with $X_j$ defined by \eqref{E:phase}, and scaled coordinates
\[
\zeta =\kappa_1+\frac{\tilde s }{\tilde \sigma}e^{i\beta}\quad\in D_{\kappa_1} 
\]  by replacing $\widehat \sigma$, $\widehat s$, $\widehat s_j$, $\widehat s_{j,\ast}$, $\widehat \lambda$, $\widehat \lambda_{j,\ast}$ by $\tilde \sigma$, $\tilde s$, $\tilde s_j$, $\tilde s_{j,\ast}$, $\tilde \lambda$, $\tilde \lambda_{j,\ast}$ in Definition \ref{D:deformation}. We decompose $X_1,X_2,X_3> 0$ into following three cases
\be\label{E:123-grad-assump}
(F1)   \ \tilde\sigma=       X_1  ,     \qquad\quad
(F2)    \  \tilde\sigma=   \sqrt{ X_2 } ,    \qquad\quad
(F3)    \  \tilde\sigma=  \sqrt[3]{ X_3 }.      
 \ee

\end{definition}
\begin{proposition}\label{P:non-homogeneous-F3}
For Case $(F3)$, $(F2)$, and $f\in L^\infty (D_{\kappa_j})$ is $\tilde s$-holomorphic, 
\begin{align}
   |I_4| _{C^\mu_{\tilde\sigma}(D_{\kappa_j,\frac 1{\tilde\sigma}})}\le &  C\epsilon_0   |  f|_{L^\infty(D_{\kappa_j})},\label{E:32-CIE}\\
|I_5| _{L^\infty(D_{\kappa_j )}}\le &  C\epsilon_0   |  f|_{L^\infty(D_{\kappa_j})}.\label{E:32-CIE-1}   
 \end{align}

\end{proposition}
  
\begin{proof}  \hfill \\

\begin{itemize}
\item {\bf Estimates for $|I_4|_{C^\mu_{\tilde\sigma}(D_{\kappa_1,\frac 1{\tilde\sigma}})}$ :} 
 Using the $\tilde s$-holomorphic property of $f$, and a residue theorem,
\begin{align}
    I_{4} =& - \frac { \theta(1-\tilde r)}{2\pi i}\int_{-\pi}^\pi  d\beta[\partial_ {\beta}   \ln (1-\gamma |\beta| )]   
   \{(  \int_{   S _<}+\int_{  \Gamma  _{40} }) \frac{  e^{-i \wp(\frac{\xi_0 }{{\tilde\sigma}},\beta,X)}f(\frac{ \xi_0 }{\tilde\sigma} ,-\beta,X)}{ \tilde s_0e^{i\tau_0}  -  \tilde r_0 e^{ i (\alpha_0 -\beta) } } d{\xi_0 } \label{E:21-I-4-4-dec-1}\\
   +& \int_{\Gamma  _{41} }    \frac{  e^{-i \wp(\frac{\xi_1 }{{\tilde\sigma}},\beta,X)}f(\frac{ \xi_1 }{\tilde\sigma} ,-\beta,X)}{ \tilde s_1e^{i\tau_1}  -  \tilde r_1 e^{ i (\alpha_1 -\beta) } } d{\xi_1 }+\int_{\Gamma  _{42} }    \frac{  e^{-i \wp(\frac{\xi_2 }{{\tilde\sigma}},\beta,X)}f(\frac{ \xi_2 }{\tilde\sigma} ,-\beta,X)}{ \tilde s_1e^{i\tau_2}  -  \tilde r_1 e^{ i (\alpha_2 -\beta) } } d{\xi_2 }\nonumber\\
+&  \int_{   S  _>  }\frac{  e^{-i \wp(\frac{\xi_h }{{\tilde\sigma}},\beta,X)}f(\frac{ \xi_h }{\tilde\sigma} ,-\beta,X)}{ \tilde s_he^{i\tau_h}  -  \tilde r_h e^{ i (\alpha_h -\beta) } } d{\xi_h }  \}, \nonumber \end{align}with 
\begin{align}
   S    _{<}(\tilde\sigma,\beta,X,\lambda)=& \{\xi_0 : \tilde s=2      \},\label{21-I4-01-contour}  \\
 \Gamma   _{40} (\tilde\sigma,\beta,X,\lambda)= &
  \{\xi_0   : 
  \tilde s\in (2,\tilde\sigma\delta)\cap  \mho _0 ,\  \textit{$\tau_0 =\tau_{0, \dagger}$}    \}, 
  \nonumber\\
  \Gamma   _{41} (\tilde\sigma,\beta,X,\lambda)= &
 \{\xi_1   : 
  \tilde s\in (2,\tilde\sigma\delta)\cap  \mho _1 ,\  \textit{$\tau_1 =\tau_{1, \dagger}$}    \},  
  \nonumber\\
  \Gamma   _{42} (\tilde\sigma,\beta,X,\lambda)= &
 \{\xi_2   : 
  \tilde s\in (2,\tilde\sigma\delta)\cap  \mho _2 ,\  \textit{$\tau_2 =\tau_{2, \dagger}$}    \},  
  \nonumber\\
  S _>(\tilde\sigma,\beta,X,\lambda)  = & \{\xi_h   : h=\sup_{\mho_j\ne\phi}j,\, \tilde s=\tilde \sigma\delta   \},
    \nonumber 
\end{align}
and   $\xi_j  $, $\tau_j$,  $\tau_{j, \dagger}$, $\mho_j=\mho_j(\tilde\sigma,\beta,X)$  defined by Definition \ref{D:deformation}. 

In view of \eqref{E:assumption}, $\tilde r<1$, \eqref{E:f3-denominator}, \eqref{E:definitive}, and \eqref{21-I4-01-contour},%
\begin{multline}\label{E:homo-4}
| \frac { \theta(1-\tilde r)}{2\pi i}  \int_{-\pi}^\pi  d\beta[\partial_ {\beta}   \ln (1-\gamma |\beta| )]  \\
\times(\int_{S _<}+\int_{S  _>  })\frac{  e^{-i \wp(\frac{\tilde s}{\tilde\sigma}e^{i\tau},\beta,X) }f(\frac{\tilde s}{\tilde\sigma}e^{i\tau},-\beta,X)}{\tilde se^{i\tau}-  \tilde r e^{ i (\alpha-\beta) }} d\tilde s e^{i\tau}|_{C^\mu_{\tilde\sigma}(D_{\kappa_1,\frac 1{\tilde\sigma}})}
\le C\epsilon_0|f|_{L^\infty(D_{\kappa_1})}.
\end{multline}

Applying \eqref{E:f3-deform-wp} and \eqref{E:f3-deform-wp-1}  for Case $(F3)$, or \eqref{E:f2-deform-wp} and \eqref{E:f2-deform-wp-1} for $(F2)$,  \eqref{E:f3-denominator}, $\tilde r<1$, \eqref{E:definitive},   \eqref{21-I4-01-contour}, and improper integrals,
\begin{align}
&\sum_{j=0}^2| \frac { \theta(1-\tilde r)}{2\pi i}  \int_{-\pi}^\pi  d\beta[\partial_ {\beta}   \ln (1-\gamma |\beta| )] \label{E:homo-4-gamma}\\ 
 \times& \int_{\Gamma _{4j}}\frac{  e^{-i \wp(\frac{\tilde s_j}{\tilde\sigma}e^{i\tau_j},\beta,X) }f(\frac{\tilde s_j}{\tilde\sigma}e^{i\tau_j},-\beta,X)}{\tilde s_je^{i\tau_j}-  \tilde r_j e^{ i (\alpha_j-\beta) }} d\tilde s e^{i\tau_j}|_{C^\mu_{\tilde\sigma}(D_{\kappa_1,\frac 1{\tilde\sigma}})}\nonumber\\
\le &C \sum_{n= 1 }^2\sum_{j=0}^2|\frac { \theta(1-\tilde r)}{2\pi i}  \int_{-\pi}^\pi  d\beta[\partial_ {\beta}   \ln (1-\gamma |\beta| )]e^{-i(n-1)\beta}\nonumber\\\times&
 \int_{\Gamma _{4j}} \frac{ e^{-i \wp(\frac{\tilde s_j}{\tilde\sigma}e^{i\tau_j},\beta,X)}f(\frac{\tilde s_j}{\tilde\sigma}e^{i\tau_j},-\beta,X)} {( \tilde s_je^{i\tau_j}-  \tilde r_j e^{ i (\alpha_j-\beta) })^n  }d\tilde s_j e^{i\tau_j}|_{L^\infty (D_{\kappa_1})} \nonumber\\
\le &
\left\{
{\ba{ll} 
C \epsilon_0|f|_{L^\infty(D_{\kappa_1})} \int_{-\pi}^\pi  d\beta 
 \int  e^{-\frac 1C\tilde s^3_j|\sin 3\tau_\dagger\sin 3\beta|}d\tilde s _j &\textit{if }\tilde\sigma=\sqrt[3]{X_3},\\
 C\epsilon_0|f|_{L^\infty(D_{\kappa_1})}  \int_{-\pi}^\pi   d\beta  
 \int  e^{-\frac 1C\tilde s^2_j|\sin 2\tau_\dagger\sin 2\beta|}d\tilde s _j &\textit{if }\tilde\sigma=\sqrt {X_2},\ea}\right.   
  \nonumber\\     
 \le& 
 \left\{
{\ba{ll}   C\epsilon_0|f|_{L^\infty(D_{\kappa_1})}  \int_{-\pi}^\pi   d\beta  \frac{1}{\sqrt[3]{|\sin 3\beta|}} 
 \int_0^\infty  e^{-  t^3 |\sin  3\tau_\dagger |   } dt &\textit{if }\tilde\sigma=\sqrt[3]{X_3},\\
 C\epsilon_0|f|_{L^\infty(D_{\kappa_1})}  \int_{-\pi}^\pi   d\beta  \frac{1}{\sqrt {|\sin 2\beta|}} 
 \int_0^\infty  e^{-  t^2 |\sin  2\tau_\dagger |   } dt &\textit{if }\tilde\sigma=\sqrt {X_2},\ea}\right.   
  \nonumber\\ 
  \le &C\epsilon_0|f|_{L^\infty(D_{\kappa_1})}.\nonumber
\end{align} 

Combining \eqref{E:21-I-4-4-dec-1}, \eqref{E:homo-4},   and \eqref{E:homo-4-gamma}, we derive
\be\label{E:i-I_4-k}
\begin{split}
&|I_4|_{C^\mu_{\tilde\sigma}(D_{\kappa_1,\frac 1{\tilde\sigma}})}\le C\epsilon_0 |f|_{L^\infty(D_{\kappa_1})}.
\end{split}
\ee

\item {\bf Estimates for $|I_5|_{L^\infty(D_{\kappa_1})}$
:} Using the $\tilde s$-holomorphic property of $f$, and the residue theorem,
\begin{align}
  I_{5} 
= &- \frac { \theta( \tilde r-1)}{2\pi i}  \int_{-\pi}^\pi  d\beta[\partial_ {\beta}   \ln (1-\gamma |\beta| )]  \{\int_{  \Gamma  _{50} } \frac{ e^{-i \wp(\frac{\xi_0 }{{\tilde\sigma}},\beta,X)}f(\frac{ \xi_0 }{\tilde\sigma} ,-\beta,X)}{\tilde s_0 e^{i\tau_0} -  \tilde r_0 e^{ i (\alpha_0 -\beta) }} d{\xi_0 }  \label{E:21-I-5-dec-1}\\
+& \int_{\Gamma  _{51} }     \frac{  e^{-i \wp(\frac{\xi_1 }{{\tilde\sigma}},\beta,X)}f(\frac{ \xi_1 }{\tilde\sigma} ,-\beta,X)}{\tilde s_1 e^{i\tau_1}-  \tilde r_1 e^{ i (\alpha_1 -\beta) }} d{\xi_1 } + \int_{\Gamma  _{52} }     \frac{  e^{-i \wp(\frac{\xi_2 }{{\tilde\sigma}},\beta,X)}f(\frac{ \xi_2 }{\tilde\sigma} ,-\beta,X)}{\tilde s_2 e^{i\tau_2}-  \tilde r_2e^{ i (\alpha_2 -\beta) }} d{\xi_2 }\nonumber\\
+& \int_{   S  _>  }  \frac{  e^{-i \wp(\frac{\xi_h }{{\tilde\sigma}},\beta,X)}f(\frac{ \xi_h }{\tilde\sigma} ,-\beta,X)}{\tilde s_h e^{i\tau_h}  -  \tilde r_h e^{ i (\alpha_h -\beta) }} d{\xi_h }\}\nonumber\\
-&    \theta(\tilde r-1)\theta(\tilde r_1-\frac 14)\theta(\tilde r_2-\frac 14) \int_{\beta\in\mathfrak\Delta(\lambda)}   d\beta[\partial_ {\beta}   \ln (1-\gamma |\beta| )] \sgn(\beta) \nonumber\\
\times& e^{-i \wp(\frac{\tilde r e^{ i (\alpha-\beta) }}{\tilde \sigma} , \beta,X)  }f(\frac {\tilde r e^{i(\alpha-\beta)}}{ {\tilde\sigma} },-\beta ,X)  , \nonumber 
\end{align}where 
{\be\label{E:alpha-beta}
 \mathfrak\Delta (\lambda) \equiv \{\beta: \ 
\left\{
{\ba{lll}  
  |\alpha_0-\beta|<\frac{\epsilon_1}2, &(\alpha_0-\beta)\beta<0,&\tilde r\in\mho_0,  \\
  ||\alpha_1-\beta|-\pi|<\frac{\epsilon_1}2,&(\alpha_1-\beta)\beta<0, &\tilde r\in\mho_{1<}, \\
  |\alpha_1-\beta|<\frac{\epsilon_1}2,&(\alpha_1-\beta)\beta<0, &\tilde r\in\mho_{1>},  \\
  ||\alpha_2-\beta|-\pi|<\frac{\epsilon_1}2,&(\alpha_2-\beta)\beta<0, &\tilde r\in\mho_{2<}, \\
  |\alpha_2-\beta|<\frac{\epsilon_1}2,&(\alpha_2-\beta)\beta<0, &\tilde r\in\mho_{2>}  
  \ea}\right.  \} ,
\ee} and  $  S _> $ defined by \eqref{21-I4-01-contour},   $  \Gamma  _{5j}= \Gamma  _{5j}(\beta,X,\lambda), \ j=0,1,2$ defined by
\be\label{E:21-I-5-2-new}
\begin{split}
\Gamma_{50}=& 
   \{\xi_0 : 
    \tilde s\in  \mho_0 ,\, \textit{$\tau_0=\tau_{0,\dagger}$} \}\cup S_{50},    
  \\
\Gamma _{51}  =&  \Gamma_{51,out}\cup S_{51}\cup\Gamma_{51,in},    
  \\
\Gamma _{52}  =&  \Gamma_{52,out}\cup S_{52}\cup\Gamma_{52,in},  \end{split}
\ee with
\begin{center}
\begin{tabular}{ll} 
\multirow{2}{*}{$\quad\ S_{50}=  
 \left\{
{\ba{l}
\{\xi_1 : 
      \tilde s_0 =\frac 12^+  \} \\  
   {\phi,}   
  \ea}\right.$} &$Type\,\mathfrak A '\wedge(\widehat{\sigma}=X_1), \, \mathfrak B',\mathfrak C'$,  \\
    &otherwise,\\

    \multirow{3}{*}{$\ \Gamma_{51,in}=  
 \left\{
{\ba{l}  
    {\phi,} \\
  \{\xi_1 : 
    \tilde s\in   \mho_1 ,\, \tau_1=0 \textit{ on }\mho_{1>},   \\
  \hskip.87in\tau_1=\pi \textit{ on }\mho_{1<}, \, \tilde s_1<1/2\},  
  \ea}\right.$} &$ \tilde r_1 >\frac 14$,  \\
    &$ \tilde r_1 <\frac 14$, \\ &\\
\multirow{2}{*}{$\Gamma_{51,out}=  
 \left\{
{\ba{l}  
  {\{\xi_1 : 
    \tilde s\in  \mho_1 ,\, \textit{$\tau_1=\tau_{1,\dagger}$} \},} \\
  {\{\xi_1 : 
    \tilde s\in  \mho_1,\, \textit{$\tau_1=\tau_{1,\dagger}$},\,  \tilde s_1 >1/2 \},}   
  \ea}\right.$}& $  \tilde r_1 >\frac 14$,  \\
    &$ \tilde r_1 <\frac 14$, 
\\ 
    
\multirow{2}{*}{$\quad\ S_{51}=  
 \left\{
{\ba{l}  
   {\phi,} \\
 \{\xi_1 : 
      \tilde s_1 =1/2 \}   
  \ea}\right.$} &$ \tilde r_1 >\frac 14$,  \\
    &$ \tilde r_1 <\frac 14$,
    \end{tabular}
\end{center}
\begin{center}
\begin{tabular}{ll}

    \multirow{3}{*}{$\ \Gamma_{52,in}=  
 \left\{
{\ba{l}  
    {\phi,} \\
  \{\xi_2 : 
    \tilde s\in   \mho_2 ,\, \tau_2=0 \textit{ on }\mho_{2>},   \\
  \hskip.87in\tau_2=\pi \textit{ on }\mho_{2<}, \, \tilde s_2<1/2\},  
  \ea}\right.$} &$ \tilde r_2 >\frac 14$,  \\
    &$ \tilde r_2 <\frac 14$, \\ &\\
 \multirow{2}{*}{$\Gamma_{52,out}=  
 \left\{
{\ba{l}  
  {\{\xi_2 : 
    \tilde s\in  \mho_2 ,\, \textit{$\tau_2=\tau_{2,\dagger}$} \},} \\
  {\{\xi_2 : 
    \tilde s\in  \mho_2,\, \textit{$\tau_2=\tau_{2,\dagger}$},\,  \tilde s_2 >1/2 \},}   
  \ea}\right.$}& $  \tilde r_2 >\frac 14$,  \\
    &$ \tilde r_2 <\frac 14$,\\    
\multirow{2}{*}{$\quad\ S_{52}=  
 \left\{
{\ba{l}  
   {\phi,} \\
 \{\xi_2 : 
      \tilde s_2 =1/2 \}   
  \ea}\right.$} &$ \tilde r_2 >\frac 14$,  \\
    &$ \tilde r_2 <\frac 14$,
 \end{tabular}
\end{center}
and $\alpha_j$,  $\xi_j  $,   $\tau_{ j,\dagger}$, $\mho_j=\mho_j(\tilde\sigma,\beta,X)$  defined by  Definition \ref{D:renormal}.

Using \eqref{E:vekua}, $\tilde r>1$,   and argument as that for $I_4$,
\begin{align}
|I_5|_{L^\infty(D_{\kappa_1})} \le C\epsilon_0|f|_{L^\infty(D_{\kappa_1})}+\sum_{j=1}^2|  \frac { \theta( \tilde r-1)}{2\pi i}  \int_{-\pi}^\pi  d\beta[\partial_ {\beta}   \ln (1-\gamma |\beta| )]\label{E:21-i-5-reduce} \\
\times   \int_{\Gamma  _{5j,in}  }     \frac{  e^{-i \wp(\frac{\xi_j }{{\tilde\sigma}},\beta,X)}f(\frac{ \xi_j }{\tilde\sigma} ,-\beta,X)}{\tilde s_j e^{i\tau_j}-  \tilde r_j e^{ i (\alpha_j -\beta) }} d{\xi_j } |_{L^\infty(D_{\kappa_1})}.\nonumber
\end{align}

Namely, we have to pay extra attention when both $\tilde \zeta$ and $\tilde\lambda$ is close to one of the essential stationary point $\tilde s_{j,\ast}(\tilde\sigma,\beta, X)$, say $\tilde s_{1,\ast}(\tilde\sigma,\beta, X)$ without loss of generality because the other case can be done by analogy.  In this situation,  $\Gamma_{52,in}=\phi$.
  For the estimates on $\Gamma_{51,in}$, in view of 
  \eqref{E:21-I-5-2-new}, we have $|\kappa_1-\tilde\zeta|\ge 1/4$ for  $   \tilde s\in\Gamma_{51,in}  $. Hence  $I_5$ for $\tilde s\in\Gamma_{51,in}$ is no longer a singular integral and we can apply \eqref{E:vekua}. Namely,
\begin{align}
&|\theta(\tilde r-1)\int_{-\pi}^\pi d\beta [\partial_ {\beta}   \ln (1-\gamma |\beta| )] \int_{\Gamma _{51,in }}\frac{  e^{-i \wp(\frac{\xi_1 }{{\tilde\sigma}},\beta,X)}f(\frac{ \xi_1 }{\tilde\sigma} ,-\beta,X)}{\tilde s_1 e^{i\tau_1}-  \tilde r_1 e^{ i (\alpha_1 -\beta) }} d{\xi_1 }|_{L^\infty(D_{\kappa_1})}\label{E:residue-beta} \\
\le &C|\iint_{ \tilde s\in\Gamma_{51,in} }\frac{ \widetilde \gamma_1(\tilde s, \beta) e^{-i\wp(\frac {\tilde s}{{\tilde\sigma}},\beta,X)} f(\frac {\tilde s}{{\tilde\sigma}},-\beta,X)  }{{\tilde\zeta}-\tilde \lambda}d\overline{\tilde\zeta} \wedge d{\tilde\zeta}|_{L^\infty(D_{\kappa_1})}\nonumber\\
\le &    C\epsilon_0 |f|_{L^\infty(D_{\kappa_1})} . \nonumber
\end{align} 

Consequently, 
\be\label{E:I-50-est}
\begin{split}
&|I_{5}|_{L^\infty(D_{\kappa_1})}\le C\epsilon_0|f|_{L^\infty(D_{\kappa_1})}.
\end{split}
\ee

\end{itemize}

\end{proof}

Estimates for Case $(F1)$ are more involved. Let's point out the difficulties and outline our strategy first for clarity.
\begin{itemize}
\item  On $\mho_j$ for $j\ge 1$, in view of Lemma \ref{L:p-estimate}, there are no good uniform estimates (means the coefficients for the polynomial $\tilde s$ only depends on functions of $\beta$)   for $\mathfrak{Re}(-i\wp(\frac{ \tilde s_je^{i\tau_{j,\dagger}}+ \tilde s_{j,\ast}}{X_1}$, $\beta,X)) $. 

We shall take advantage of the scaling invariant properties of the Hilbert transform and estimates    \eqref{E:f2-deform-wp}  and \eqref{E:f3-deform-wp-1}, where $\sqrt{X_2}$ and $\sqrt[3]{X_3}$ are not necessarily equal to $\tilde\sigma=X_1$, to derive estimates on $\Gamma_{4j}$ or $\Gamma_{5j}$, $j\ge 1$.   In view of argument of \eqref{E:residue-beta}, one additional important issue  is   that,  to apply  \eqref{E:vekua}, the Cauchy integral is no longer a singular integral near renormalized essential critical points (see Definition \ref{D:renormal}). This is done in Lemma \ref{L:root-scaling}. 
 
\item On $\mho_0$, the above scaling argument won't work since $0$ is a singular point.

We have to take advantage of estimate \eqref{E:all-0}! If we apply    argument as for Case $(F3)$ or $(F2)$ in Proposition \ref{P:non-homogeneous-F3} directly, then the difficulty encountered is the Jacobian 
 $
\frac 1{|\sin\beta|}d\beta$ instead of $\frac 1{\sqrt[3]{|\sin3\beta|}}d\beta$ or $\frac 1{\sqrt {|\sin2\beta|}}d\beta$.  To overcome this difficulty when $|\beta|\ll 1$, 
 we shall squeeze out extra $|\sin\beta|$-decay of integrals on $\Gamma_{40}$ or $\Gamma_{50}$ via  deformation of a finer decomposition. This is done via $\mathfrak J_1$-$\mathfrak J_5$ decomposition \eqref{E:homo-i-41}-\eqref{E:homo-i-45}.
\end{itemize}

\begin{definition}\label{D:renormal} For Case $(F1)$, introduce new scaled $\sigma_j$-parameters on $\mho_j(\beta,X)$, $j=0,1,2,$
\be\label{E:affine-coord-new}
\begin{split}
&\left\{
{\ba{ll}  
{\sigma_0={\tilde\sigma},\  {  \sigma_1}=  \sigma_2=
 \sqrt[3]{|X_3|}},&  \textit{for }\ Type\ \mathfrak A, \mathfrak B,\mathfrak E,\\
 {\sigma_0={\tilde\sigma},\  {  \sigma_1}=  \sigma_2=
 \sqrt[2]{|X_2|}},&  \textit{for }\ Type\ \mathfrak C,\mathfrak D,
 \ea}\right. 
 \end{split}
 \ee and scaled $\sigma_j$-coordinates on $\mho_j(\tilde\sigma, \beta,X)$
 \be\label{E:12-lambda-sigma}
\begin{gathered}
\lambda=\kappa_1+\frac{\tilde r }{\tilde\sigma}e^{i\alpha}=\kappa_1+\frac{   s_{j,\ast}e^{i\beta}+ r_je^{i\alpha_j}}{  \sigma_j}  ,\\
   s_{j,\ast}\equiv \tilde s_{j,\ast}\frac{\sigma_j}{\tilde\sigma},\ r_j=\tilde r_j \frac{  \sigma_j}{{ \tilde \sigma }} \ge 0,\  \alpha_j = \alpha_j(\beta,X,\lambda), \\  
   \frac{\tilde  s}{ \tilde  \sigma } \mapsto  \frac{ \vartheta_j}{\sigma_j}\equiv \frac{s_{j,\ast}+s_je^{i\tau_j}}{\sigma_j}  ,  \ \ 
\tilde s\equiv (s_{j,\ast}\pm  s_j)\frac{\tilde\sigma}{\sigma_j}  
  \in \mho_j(\tilde\sigma, \beta,X) 
\end{gathered}
\ee  where $\mho_j(\tilde\sigma, \beta,X)$, $\tilde s_{j,\ast}$, and $\tilde r_j$ are defined by Definition \ref{D:dominant}.
 
\end{definition}
\begin{lemma}\label{L:root-scaling}
For Case $(F1)$, introduce the new scaled $\sigma_j$-coordinates defined by Definition \ref{D:renormal}, 
\be\label{E:F1-s-1-inf}
\inf_\beta s_{1,\ast} = c_0,\quad 0<c_0<1. 
\ee 

\end{lemma}

\begin{proof}\hfill \\
\begin{itemize}
\item [$\blacktriangleright$] {\bf Proof  for $ Type\ \mathfrak A'', \mathfrak B'',\mathfrak E$ :} from Table \ref{Tb:dynamic-FNH}, \eqref{E:affine-coord-new}, 
and
\[
\begin{gathered}
|s_+s_-|=|\frac{X_1}{X_3^{1/3}}\frac{\sin\beta}{\sin3\beta}|\ge \frac 13,\quad\
s_\pm=\frac{    - 1\pm\sqrt{1-\Delta}\, }{3\frac{X_3^{2/3}}{ X_2}\frac{\sin3\beta}{\sin2\beta}},
\end{gathered}
\]we derive $|s_+|\sim |s_-| $ and then \eqref{E:F1-s-1-inf} for $ Type\ \mathfrak A'', \mathfrak B'',\mathfrak E$.

\item [$\blacktriangleright$] {\bf Proof  for $ Type\ \mathfrak C'', \mathfrak D$ :} from \eqref{E:determinant}, Table \ref{Tb:dynamic-FNH}, and \eqref{E:affine-coord-new}, 
\begin{align*}
|  s_ +|=&|\frac{-\Delta}{6\frac{X _3}{\tilde \sigma X_2}\frac{\sin 3\beta}{\sin2\beta} } +\mbox{l.o.t.}|=|\frac{-3\frac{ {X_1}{X_ 3}}{X_2^2}\frac{\sin\beta\sin3\beta}{\sin^22\beta}}{6\frac{X _3 }{ X_2^{3/2}}\frac{\sin 3\beta}{\sin2\beta} } +\mbox{l.o.t.}|\ge\frac 1C,\\
|  s_ -|=&|\frac{-2}{3\frac{X _3}{\tilde \sigma X_2}\frac{\sin 3\beta}{\sin2\beta} } +\mbox{l.o.t.}|\ge\frac 1C|  s_ +|\ge\frac 1C.
\end{align*} Hence \eqref{E:F1-s-1-inf} is proved for Type $\mathfrak C''$ and $\mathfrak D$.
\end{itemize}

\end{proof}
  
\begin{proposition}\label{P:non-homogeneous-F1}
For Case $(F1)$, and $f\in L^\infty (D_{\kappa_j})$ is $\tilde s$-holomorphic, 
\begin{align}
   |I_4| _{C^\mu_{\tilde\sigma}(D_{\kappa_j,\frac 1{\tilde\sigma}})}\le &  C\epsilon_0   |   f|_{L^\infty(D_{\kappa_j})},\label{E:f1-CIE}\\
|I_5| _{L^\infty(D_{\kappa_j )}}\le &  C\epsilon_0   |  f|_{L^\infty(D_{\kappa_j})}.\label{E:f1-CIE-1}   
 \end{align}
\end{proposition}  

\begin{proof}

Thanks to Lemma \ref{L:root-scaling}, we can following the same argument \eqref{E:21-I-4-4-dec-1}-\eqref{E:homo-4},  \eqref{E:21-I-5-dec-1}-\eqref{E:21-I-5-2-new} to obtain 
\begin{align}
|I_{4}|_{C^\mu_{\tilde\sigma}(D_{\kappa_1,\frac 1{\tilde\sigma}})} 
\le &C\epsilon_0 |f|_{L^\infty(D_{\kappa_1})} 
     + \sum_{j=0}^2|I_{4j}|_{L^\infty (D_{\kappa_1})} ,\label{E:renormal-1}\\
     |I_5|_{L^\infty(D_{\kappa_1})} \le & C\epsilon_0|f|_{L^\infty(D_{\kappa_1})}+\sum_{j=0}^2|I_{5j}|_{L^\infty (D_{\kappa_1})},\label{E:renormal-5-new}
     \end{align}
 with
\begin{align}
    I_{4j}=&  
   \sum_{h= 1,2}  \frac { \theta(1-\tilde r)}{2\pi i}  \int_{-\pi}^\pi  d\beta[\partial_ {\beta}   \ln (1-\gamma |\beta| )]e^{-i(n-1)\beta}
 \int_{\Gamma _{4j}} \frac{ e^{-i \wp(\frac{\tilde s_j}{\tilde\sigma}e^{i\tau_j},\beta,X)}f(\frac{\tilde s_j}{\tilde\sigma}e^{i\tau_j},-\beta,X)} {( \tilde s_je^{i\tau_j}-  \tilde r_j e^{ i (\alpha_j-\beta) })^h  }d\tilde s_j e^{i\tau_j},   \nonumber \\
 I_{5j}=&\frac { \theta( \tilde r-1)}{2\pi i}  \int_{-\pi}^\pi  d\beta[\partial_ {\beta}   \ln (1-\gamma |\beta| )] \int_{\Gamma  _{5j,out}  }    \frac{ e^{-i \wp(\frac{\tilde s_j}{\tilde \sigma}e^{i\tau_j},\beta,X)}f(\frac{\tilde s_j}{\tilde \sigma}e^{i\tau_j},-\beta,X)} {    \tilde s_je^{i\tau_j}- \tilde r_j e^{ i (\alpha_j-\beta) }   }d \tilde s_j e^{i\tau_j}.\nonumber
 \end{align} 
 
\begin{itemize}
\item [$\blacktriangleright$] {\bf Proof  for $I_{41}, I_{42},I_{51},I_{52}$ :} 
 For $j\ge 1$, using the scaling invariant of the Hilbert transform and
 \begin{itemize}
 \item for $\Gamma _{4j}\cap \{s_j<1\}$: applying Lemma \ref{L:root-scaling},  \eqref{E:vekua}; 
 \item for $\Gamma _{4j}\cap \{s_j>1\}$: thanks to the scaling invariant of the Hilbert transform, applying  \eqref{E:f3-deform-wp-1} on $\mho_1,\,\mho_2 $ for $Type\,\mathfrak A'',\,\mathfrak B''\,\mathfrak E$,  and  \eqref{E:f2-deform-wp-1} on $\mho_1,\,\mho_2 $ for $Type\,\mathfrak C'',\,\mathfrak D$, 
 \end{itemize} we have
\begin{align}
&|I_{4j}|_{C^1(D_{\kappa_1,\frac{1}{\tilde\sigma}})}\label{E:41}\\
\le & \sum_{h=0}^1\{|\frac { \theta(1-\tilde r)}{2\pi i}  \int_{-\pi}^\pi  d\beta[\partial_ {\beta}   \ln (1-\gamma |\beta| )] \nonumber\\\times&
 \int_{\Gamma _{4j}\cap \{s_j<1\}} \frac{ e^{-i \wp(\frac{  s_j}{ \sigma}e^{i\tau_j},\beta,X)}\widetilde \chi(\frac{  s_j}{ \sigma}e^{i\tau_j},-\beta,X)} {   (\tilde s_je^{i\tau_j}-  \tilde  r_j e^{ i (\alpha_j-\beta) }  )^h( s_je^{i\tau_j}-    r_j e^{ i (\alpha_j-\beta) }  ) }d  s_j e^{i\tau_j}|_{L^\infty (D_{\kappa_1})}\nonumber \\
+& \sum_{h=0}^1\{|\frac { \theta(1-\tilde r)}{2\pi i}  \int_{-\pi}^\pi  d\beta[\partial_ {\beta}   \ln (1-\gamma |\beta| )] \nonumber\\
\times&\int_{\Gamma _{4j}\cap \{s_j>1\}} \frac{ e^{-i \wp(\frac{  s_j}{ \sigma}e^{i\tau_j},\beta,X)}\widetilde \chi(\frac{  s_j}{ \sigma}e^{i\tau_j},-\beta,X)} {   (\tilde s_je^{i\tau_j}-  \tilde  r_j e^{ i (\alpha_j-\beta) }  )^h( s_je^{i\tau_j}-    r_j e^{ i (\alpha_j-\beta) }  ) }d  s_j e^{i\tau_j}|_{L^\infty(D_{\kappa_1})} \}\nonumber\\
\le &C\epsilon_0|\widetilde \chi|_{L^\infty(D_{\kappa_1})}.\nonumber
\end{align}

 In an entirely similar way, namely, for $j\ge 1$,
\begin{itemize}
 \item for $\Gamma _{5j}\cap \{s_j<1\}$: applying Lemma \ref{L:root-scaling},  \eqref{E:vekua}; 
 \item for $\Gamma _{5j}\cap \{s_j>1\}$: thanks to the scaling invariant of the Hilbert transform, applying  \eqref{E:f3-deform-wp-1} on $\mho_1,\,\mho_2 $ for $Type\,\mathfrak A'',\,\mathfrak B''\,\mathfrak E$,  and  \eqref{E:f2-deform-wp-1} on $\mho_1,\,\mho_2 $ for $Type\,\mathfrak C'',\,\mathfrak D$, 
 \end{itemize} we have
\be\label{E:51}
\begin{split}
&|I_{5j}|_{L^\infty(D_{\kappa_1})}\le C\epsilon_0|f|_{L^\infty(D_{\kappa_1})}.
\end{split}
\ee 

As a result, proof for $I_{4j}$, $I_{5j}$ for $j\ge 1$ is done.

\item [$\blacktriangleright$] {\bf Proof  for $I_{40 },I_{50}$ :} We shall use the estimate \eqref{E:all-0}. But we have to deal with the singularity $\sin\beta=0$ caused by the change of variables. We shall squeeze out extra $|\sin\beta|$-decay of integrals on $\Gamma_{40}$ or $\Gamma_{50}$ via a finer decomposition. Namely, decompose
\begin{align}
 I_{40}=&-\frac { \theta(1-\tilde r)}{2\pi i} \int_{0}^\pi  d\beta[\partial_ {\beta}   \ln (1-\gamma |\beta| )] \int_{\Gamma_{40}}   \mathfrak J_{<} ,\label{E:scal-4-linear}
\end{align}
\begin{align}
\mathfrak J_<=&
\left\{
{\ba{ll} 
 \mathfrak J_{1}+\mathfrak J_{2}+\mathfrak J_{3}+\mathfrak J_{4}+\mathfrak J_{5} ,  &\textit{if } 0<\beta<\epsilon_1/8,\\
 0,&\textit{if }  -\epsilon_1/8<\beta<0,\\
 \frac{  e^{-i \wp(\frac{\vartheta_0 }{{ \sigma_0}},\beta,X)}f(\frac{ \vartheta_0 }{ \sigma_0} ,-\beta,X)}{s_0 e^{i\tau_0} -    r_0 e^{ i (\alpha_0 -\beta) }} d{\vartheta_0 }, &\textit{if }  |\beta|>\epsilon_1/8; 
 \ea}\right. \label{E:NH-L}
\end{align} and $\mathfrak J_1,\cdots,\mathfrak J_5$ defined by 
 \begin{align}
 \mathfrak J_1= & \theta(\frac 1{|\sin\beta|}-|\tilde s-\tilde r|)   \frac{ {[e^{-i\wp(\frac{\tilde se^{i\tau }}{X_1 } ,\beta,X)  }-1]  f(\frac{\tilde se^{i\tau }}{X_1 } ,-\beta,X) }}{\tilde s e^{i\tau}- \tilde r  e^{ i (\alpha-\beta ) }}d\tilde se^{i\tau },\label{E:homo-i-41} \\
\mathfrak J_2= & \theta(\frac 1{|\sin\beta|}-|\tilde s-\tilde r|)   \frac{ {[ 1-e^{-i\wp(\frac{\tilde se^{-i\tau }}{X_1 } ,-\beta,X)  } ]  f(\frac{\tilde se^{-i\tau }}{X_1 } ,-\beta,X ) }}{\tilde se^{-i\tau } - \tilde r  e^{ i (\alpha-\beta ) }}d\tilde se^{-i\tau },\label{E:homo-i-42} \\
\mathfrak J_3=& \theta(\frac 1{|\sin\beta|}-|\tilde s-\tilde r|) e^{-i\wp(\frac{\tilde se^{-i\tau }}{X_1 } ,-\beta,X)  }  f(\frac{\tilde se^{-i\tau }}{X_1 } ,-\beta,X)  \label{E:homo-i-43} \\
\times&  [\frac{1  }{\tilde se^{-i\tau } - \tilde r  e^{ i (\alpha-\beta ) }}-\frac{ 1  }{\tilde se^{-i\tau } - \tilde r  e^{ i (\alpha+\beta ) }}]d\tilde se^{-i\tau },\nonumber\\
\mathfrak J_4= &\theta(\frac 1{|\sin\beta|}-|\tilde s-\tilde r|)   e^{-i\wp(\frac{\tilde se^{-i\tau }}{X_1 } ,-\beta,X)  }\frac{ f(\frac{\tilde se^{-i\tau }}{X_1 } ,-\beta,X)- f(\frac{\tilde se^{-i\tau }}{X_1 } ,+\beta,X)   }{\tilde se^{-i\tau } - \tilde r  e^{ i (\alpha+\beta ) }} d\tilde se^{-i\tau },\label{E:homo-i-44}\\     
\mathfrak J_5=& \theta(|\tilde s-\tilde r|-\frac 1{|\sin\beta|}) ( e^{-i\wp(\frac{\tilde se^{i\tau }}{X_1 } ,\beta,X)  } \frac{ {f(\frac{\tilde se^{i\tau } }{X_1 } ,-\beta,X) }}{\tilde se^{i\tau } -\tilde r  e^{ i (\alpha-\beta ) }}d\tilde se^{i\tau } \label{E:homo-i-45} \\
  - &e^{-i\wp(\frac{\tilde se^{-i\tau }}{X_1 } ,-\beta,X)  }  \frac{ {f(\frac{\tilde se^{-i\tau } }{X_1 } ,+\beta,X) }}{\tilde se^{-i\tau } - \tilde r  e^{ i (\alpha+\beta ) }}d\tilde se^{-i\tau }),\nonumber
\end{align} with 
 $\tau $  defined by  
Definition \ref{D:deformation} for $\beta\in[0,\pi]$. 

Let's explain our strategy before providing detailed estimates for $\mathfrak J_1,\cdots,\mathfrak J_5$.

For $\mathfrak J_1$-$\mathfrak J_4$, one can squeeze out extra $|\sin\beta|$-decay of nominators of integrands by taking advantage of difference terms.   
Through the change of variables   \be\label{E:second-scale}
\tilde s\mapsto t=\tilde s|\sin\beta|,  
\ee the cut off function $\theta(\frac 1{|\sin\beta|}-|\tilde s-\tilde r|)$ makes the $t$-domain of $\mathfrak J_1$-$\mathfrak J_4$  compact, thus estimates can be derived. For $\mathfrak J_5$,   we use the scaling invariant of the Hilbert transform to cancel the Jacobian $\frac 1{|\sin\beta|}$, the cut off function $\theta(|\tilde s-\tilde r|-\frac 1{|\sin\beta|})$ making the scaled Hilbert transform $\frac{  1 }  {  t-\tilde r|\sin\beta|\,  }$  bounded,    and the nominator is $t$-exponentially decaying. Hence estimates can be derived.

Precisely, 
\begin{itemize}
\item [$\bullet$]   $\mathfrak J_1,\mathfrak J_2$:  From   the mean value theorem  and  \eqref{E:all-0},
{\begin{align} 
  &|e^{-i\wp(\frac{\tilde se^{\pm i\tau }}{X_1 } ,\pm \beta,X)  } -1|\le C  |  \sin\beta|,\quad\textit{ for $\tilde s\in \mho_0(\beta,X)$, $|\beta|<\frac{\epsilon_1}8$}  .\label{E:homo-i-41-42-aux} 
\end{align}}
\item  [$\bullet$] $\mathfrak J_3$:   From $|\beta|<\epsilon_1/8$,   
\begin{align}
 & |\frac1{ \tilde se^{-i\tau_\dagger }-  \tilde r e^{ i (\alpha-\beta) }} -\frac1{ \tilde se^{-i\tau_\dagger }-  \tilde r e^{ i (\alpha+\beta) }}|\label{E:L-H-53-aux}\\
=&  |\frac{\tilde r e^{i\alpha}2\sin\beta}{(\tilde se^{-i\tau_\dagger }-  \tilde r e^{ i (\alpha-\beta) })(\tilde se^{-i\tau_\dagger }-  \tilde r e^{ i (\alpha+\beta) })}| \le C|\sin\beta| .\nonumber
\end{align}

\item [$\bullet$]   $\mathfrak J_4$:  From $\tilde s$-holomorphic properties  of $f$,  
\begin{align} 
    &|\frac{   f(\frac{\tilde s}{\tilde \sigma}e^{-i\tau_\dagger },-\beta,X)-f(\frac{\tilde s}{\tilde \sigma}e^{-i\tau_\dagger },+\beta,X) }{ \tilde se^{-i\tau_\dagger }-  \tilde r e^{ i (\alpha+\beta) }  }|\label{E:homo-i-44-aux}  \\
\le  &  C|f|_{L^\infty(D_{\kappa_1})}\frac{\tilde s| \sin\beta| }{ |\tilde se^{-i\tau_\dagger }-  \tilde r e^{ i (\alpha+\beta) } | } 
\le    C|f|_{L^\infty(D_{\kappa_1})}| \sin\beta|. \nonumber
\end{align}

\end{itemize} 

Applying  \eqref{E:homo-i-41}-\eqref{E:homo-i-44}, \eqref{E:homo-i-41-42-aux}-\eqref{E:homo-i-44-aux}, and the change of variables \eqref{E:second-scale},  for $j=1,\cdots, 4$,
\begin{align}
& |-\frac { \theta(1-\tilde r)}{2\pi i}  \int_{0}^\pi  d\beta[\partial_ {\beta}   \ln (1-\gamma |\beta| )]   +  \int_{  \Gamma _{40}  }  \mathfrak J_j |_{L^\infty(D_{\kappa_1 })} \label{E:H-I-41}\\
\le & C|f|_{L^\infty(D_{\kappa_1})}|\int_{0}^\pi  d\beta[\partial_ {\beta}   \ln (1-\gamma |\beta| )]     \int_{ 2|sin\beta|  } ^{X_1\delta|\sin\beta|} \theta(1-  \left|t-(\tilde r|\sin\beta|)\, \right|)   dt   |_{L^\infty(D_{\kappa_1})}\nonumber\\
 \le &C\epsilon_0 |f|_{L^\infty(D_{\kappa_1})}.\nonumber
\end{align}

On the other hand, using   \eqref{E:homo-i-45}, $\epsilon_1>0$,  the rescaling \eqref{E:second-scale}, the  scaling invariant property of the Hilbert transform,
    one obtains
\begin{align}
 &|-\frac { \theta(1-\tilde r)}{2\pi i}  \int_{0}^\pi  d\beta[\partial_ {\beta}   \ln (1-\gamma |\beta| )]  \int_{  \Gamma _{40}  }  \mathfrak J_5|_{L^\infty(D_{\kappa_1})} \label{E:H-I-45-new} \\
 \le &   C\epsilon_0|f|_{L^\infty(D_{\kappa_1})}\int_{ 2|sin\beta|  } ^{X_1\delta|\sin\beta|}   \theta(| t-\tilde r|\sin\beta|\,|-1)   \frac{  e^{-t\sin\frac{\epsilon_1}4 } }{ |t-\tilde r|\sin\beta|\,| } dt|_{L^\infty(D_{\kappa_1})}  \nonumber\\
   \le &  C\epsilon_0 |f|_{L^\infty(D_{\kappa_1})}. \nonumber
\end{align}

 From \eqref{E:H-I-41} and \eqref{E:H-I-45-new},
$
 |I_{40}|_{L^\infty(D_{\kappa_1})}\le C\epsilon_0 |f|_{L^\infty(D_{\kappa_1})}$. 
In an entirely similar way,
$|I_{50}|_{L^\infty(D_{\kappa_1})}\le C\epsilon_0 |f|_{L^\infty(D_{\kappa_1})}$. 
Combining with  \eqref{E:renormal-1}-\eqref{E:51},     the proof for \eqref{E:f1-CIE}, \eqref{E:f1-CIE-1} is completed.


\end{itemize}

\end{proof}

\section{The IST for perturbed multi line solitons}\label{S:N-line}

 Using the Sato theory \cite{S81,SS82,S89a,S89b,BC07,BK03,KW13} and Boiti et al's  direct scattering theory of KP  equation \cite{P00,BP302,BP309,BP310,BP211,BP212,BP214},  we   adapt our method and extend the approach of the IST for perturbed $1$-solitons  to that for perturbed $   {\mathrm{Gr}(N,M)_{> 0}}$ KP solitons.  We shall state the complete theory for perturbed $   {\mathrm{Gr}(N,M)_{> 0}}$ KP  solitons but only give  explanation of distinct features which  shows  the TP (totally positive) condition is necessary and justifies differences between IST for perturbed $1$-solitons and  perturbed multi line solitons  are mainly algebraic.
 

\subsection{Statement of results}\label{SS:DP} 

\begin{definition}\label{D:N-generic-sd}
Given $0<\epsilon_0\ll 1$,  $d\ge 0$, and a $  {\mathrm{Gr}(N, M)_{> 0}}$ KP soliton $u_s(x)$ defined by $\{\kappa_j\}, A$, a scattering data  $ {\mathcal S} =(\{z_n\},\{\kappa_j\}, \mathcal D ,s _c(\lambda))$ is called    $d$-admissible   if  
\begin{align}
& s_c(\lambda)=
\left\{
{\ba{ll}
 {\frac{ \frac {i}{ 2} \sgn(\lambda_I)}{\overline\lambda-\kappa_j}\frac{\gamma_j}{1-\gamma _j|\alpha|}}+\sgn(\lambda_I)  h_j(\lambda),&\lambda\in   D^ \times_{ \kappa_j },\\
\sgn(\lambda_I) {  \hbar_n}(\lambda),&\lambda\in    D^\times _{ z_n},
\ea}
\right.\label{E:s-c-N},\\
&\mathcal  D=\left( \textit{\tiny${\ba{ccc}\kappa_1^N&\cdots&0\\
\vdots&\ddots&\vdots\\
0&\cdots&\kappa_N^N\\
\mathcal D_{N+1 1}&\cdots&\mathcal D_{N+1 N}\\
\vdots&\ddots&\vdots\\
\mathcal D_{M 1}&\cdots&\mathcal D_{M N}
\ea}$}\right),  \label{E:sym-N-D}
\end{align} and
\begin{align}
&\det  (
\frac 1{\kappa_{k}-z_{h}}
 )_{1\le k, h\le N}\ne 0,\  z_1=0,\{z_n,\kappa_j\}  \textit{ distinct real},\label{E:intro-sym-N-D-flat}\\
&
\epsilon_0\ge     (1-\sum_{j=1}^ME_{{\kappa_j}}  )  \sum_{|l|\le {d+8}}|\left(|\overline\lambda-\lambda|^{l_1} +| \overline\lambda^2-\lambda^2|^{l_2}\right) s_c (\lambda)|   _{    
 L^\infty}\label{E:epsilon-0-BPP-new}
 \\
 &\qquad +  \sum_{j=1}^M(|\gamma_j|+|h_j|_{ L^\infty(D_{\kappa_j})})+\sum_{n=1}^N|\hbar_n|_{ C^1(D_{z_n})}\\
 &\qquad+   | {\corr{\textit{diag}\,(q_1, \cdots, q_M)^{-1}}\times\mathcal D \times \corr{\textit{diag}\,(q_1, \cdots, q_N)}  - { \mathcal D}^\flat  }|_{L^\infty},\nonumber\\
 &s_c(\lambda)=  \overline{s_c( \overline\lambda)},
  h_j(\lambda)=-\overline{h_j( \overline\lambda)}, 
 \hbar_n(\lambda)=-\overline{\hbar_n( \overline\lambda)},\label{E:N-reality-inv}\\
& { \mathcal D}^\flat = \textit{diag}\,(   
\kappa^N_1 ,\cdots,\kappa^N_M )\, A^T,\ \ q_j=\frac{\Pi_{2\le n\le N}(\kappa_j-z_n)}{(\kappa_j-z_1)^{N-1}} \textit{ for }1\le j\le M. \label{E:intro-sym-N-D-flat-new}
\end{align}     
\end{definition} 

Define $T$ as the continuous scattering operator
\be\label{E:cauchy-operator-p-inverse}
T  \phi  (x ,\lambda)
\equiv  {  s}_c(\lambda  )e^{(\overline\lambda-\lambda)x_1+(\overline\lambda^2-\lambda^2)x_2 +(\overline\lambda^3-\lambda^3)x_3 }\phi(x, \overline\lambda).
\ee
 
\begin{definition}\label{D:N-phase} Given  $\{z_n,\kappa_j \}$, $1\le n\le N$, $1\le j\le M$, the {eigenfunction  space} ${ W }=W_x$  consists of $\phi$ satisfying 
\begin{itemize}
\item [$(a)$] $\phi (x, \lambda)=\overline{ \phi (x, \overline\lambda)};$ 
\item [$(b)$] $(1-  \sum_{n=1}^NE_{z_n} )\phi(x, \lambda)\in L^\infty;$ 
\item [$(c)$] for $\lambda \in D_{z_n}^\times$, 
$
\phi(x, \lambda)=\frac{ {\phi_{z_n,\res}(x)}}{\lambda-z_n }  +\phi_{z_n,r}(x, \lambda)$, $ \phi_{z_n,\res}$,   $\phi_{z_n,r}  \in L^\infty( D_{z_n})$; 
\item[$(d)$] for $\lambda =\kappa_j+re^{i\alpha}\in D_{\kappa_j}^\times$, $\phi=\phi^\flat+\phi^\sharp $, 
$\phi^\flat =\sum_{l=0}^\infty \phi_l(X)(-\ln(1-\gamma_j|\alpha|))^l \in L^\infty(D_{\kappa_j}) $, $\phi^\sharp \in C_{\tilde\sigma}^{\mu } (D_{\kappa_j,\frac{1}{\tilde\sigma}}) \cap L^\infty (D_{\kappa_j})$, $\phi^\sharp(x,\kappa_j)=0$. 
\end{itemize}  Here the rescaling parameter $\tilde\sigma$ and rescaled H$\ddot{  o}$lder spaces   $C_{\tilde\sigma}^{\mu } (D_{\kappa_j,\frac{1}{\tilde\sigma}}) $ are defined as in Definition \ref{D:phase}.   Finally, for   $\phi\in W$,   
\begin{align}
|\phi|_W 
\equiv & |(1-\sum_{n=1}^N E_{z_n})\phi|_{L^\infty}  
 +\sum_{n=1}^N (| \phi_{z_n,\res}|_{L^\infty}+ | \phi_{z_n,r}|_{L^\infty(D_{z_n})})\label{E:N-W} \\
  + &\sum_{j=1}^M (| \phi^\flat     |_{ L^\infty(D_{\kappa_j})}+| \phi^\sharp     |_{ C^\mu_{\tilde\sigma}(D_{\kappa_j,\frac{1}{\tilde\sigma}})\cap L^\infty (D_{\kappa_j})})  . \nonumber
  \end{align}

\end{definition}  


 We have shown that for  the IST of perturbed $1$-line solitons, the Sato  eigenfunctions   and   the  Sato   adjoint eigenfunction  play an essential role, for instance, for the construction for the Green function; proving estimates of the Green function; and  estimating of residue of the eigenfunctions at $0$.    For  a $\textrm{Gr}(N,M)_{\ge 0}$ KP soliton $ u_s(x)$   defined by \eqref{E:line-tau} and \eqref{E:line-grassmannian}, we   introduce the Sato  eigenfunctions $\varphi(x,\lambda)$ and   the  Sato   adjoint eigenfunction  $\psi(x,\lambda)$,  defined as follows  %
\begin{align}
 \varphi(x,\lambda) 
=&e^{\lambda x_1+\lambda^2 x_2}\frac{\sum_{1\le j_1< \cdots< j_N\le M}\Delta_{j_1,\cdots,j_N}(A) (1-\frac{\kappa_{j_1}}\lambda)\cdots(1-\frac{\kappa_{j_N}}\lambda)E_{j_1,\cdots,j_N}(x)}{\tau(x)}\nonumber \\
\equiv&e^{\lambda x_1+\lambda^2 x_2}\chi(x,\lambda),\label{E:N-sato}\\
 \psi(x,\lambda) 
=&e^{-(\lambda x_1+\lambda^2 x_2)}\frac{\sum_{1\le j_1< \cdots< j_N\le M}\Delta_{j_1,\cdots,j_N}(A)\frac {E_{j_1,\cdots,j_N}(x)}{(1-\frac{\kappa_{j_1}}\lambda)\cdots(1-\frac{\kappa_{j_N}}\lambda)}}{\tau(x)}\nonumber \\
\equiv & e^{-(\lambda x_1+\lambda^2 x_2)}\xi(x,\lambda)  \label{E:N-sato-1}
\end{align}\cite[(2.12)]{BP214}, \cite[Theorem 6.3.8., (6.3.13) ]{D91}, \cite[Proposition 2.2, (2.21)]{K17}. 
  Here $\chi(x,\lambda)$ and $\xi(x,\lambda)$ are called the Sato normalized eigenfunction and the {Sato normalized adjoint eigenfunction}.   
   They satisfy the Lax equation  for    $ \lambda\ne 0$,
\be\label{E;N-spectral}
\begin{split}
&\mathcal L\varphi(x,\lambda)\equiv \left(-\partial_{x_2}+\partial^2_{x_1}+u_s(x)\right)\varphi(x,\lambda)=0,\\
&L\chi(x,\lambda)\equiv \left(-\partial_{x_2}+\partial^2_{x_1}+2\lambda\partial_{x_1}+u_s(x)\right)\chi(x,\lambda)=0,
\end{split}
\ee and the adjoint Lax equation 
\be\label{E;N-spectral-new}
\begin{split}
&\mathcal L^\dagger\psi(x,\lambda)\equiv\left(\partial_{x_2}+\partial^2_{x_1}+u_s(x)\right)\psi(x,\lambda)=0,\\
&L^\dagger\xi(x,\lambda)\equiv \left(\partial_{x_2}+\partial^2_{x_1}-2\lambda\partial_{x_1}+u_s(x)\right)\xi(x,\lambda)=0 .
\end{split}
\ee  
Proofs for \eqref{E:line-tau}, \eqref{E:line-grassmannian}, \eqref{E:N-sato}-\eqref{E;N-spectral-new} will be provided in $\S$ \ref{SSS:N-solitons} for convenience.

\begin{theorem}  \cite{Wu20,Wu21} Given a perturbed $\textrm{Gr}(N,M)_{\color{black}>0}$ KP soliton $u_0(x_1,x_2)$ satisfying
\be\label{E:intro-ini-data}
  \begin{array}{l}
  u_0(x_1,x_2)=u_s(x_1,x_2,0)+v_0(x_1,x_2),\\
  \textit{$ u_s(x)$  a $\textrm{Gr}(N,M)_{\color{black}>0}$ KP soliton defined by $\kappa_1,\cdots,\kappa_M$ and  $A\in {\mathrm{Gr}(N,M)_{> 0}}$,} \\
  \textit{$   \sum_{|l|\le { d+8}} |{\color{black} (1+|x_1|+|x_2|) } \partial_x^l v _0|_  {L^1\cap L^\infty} \ll 1$,  $d\ge 0$,}\\
    \textit{$z_1=0$, $\{z_n,\kappa_j\}_{1\le n\le N,1\le j\le M}$  distinct reals, $
 \det  (
\frac 1{\kappa_{k}-z_{h}}
 )_{1\le k, h\le N}\ne 0$,}
 \end{array}
 \ee   we have 
 
\begin{itemize}
\item [(1)] the unique solvability of
   \begin{gather}
(-\partial_{x_2}+\partial_{x_1}^2 +2\lambda\partial_{x_1}
+u_0(x_1,x_2))m_0(x_1,x_2 ,\lambda)=0, \ \label{E:Lax}\\
\lim_{|x|\to\infty}m_0(x_1,x_2,\lambda)=  \widetilde\chi(x_1,x_2,0,\lambda)=\frac{(\lambda-z_1)^{N-1}}{\Pi_{2\le n\le N}(\lambda-z_n)}\chi(x_1,x_2,0,\lambda)  \label{E:bdry} 
\end{gather} for $\forall\lambda\in\CC\backslash\{z_n,\kappa_j\}$.

\item [(2)] The {forward scattering transform} is defined as
\be\label{E:SD} 
\mathcal S(u_0,\{ z_n\})=(\{ z_n\},\{\kappa_j\}, \mathcal  D,s _c(\lambda)) 
\ee satisfying  
\begin{gather}   
    {   m}_0(x_1, x_2,\lambda) =1+\sum_{n=1}^N\frac{   m_{0;z_n, \res }(x_1, x_2  )}{\lambda -z_n }  +\mathcal C  T_0
     m_0  \ \in { W_0 }=W_{(x_1,x_2,0)},  \label{E:N-CIE-t0} \\
(e^{\kappa_1x_1+\kappa_1^2x_2 }m_0(x_1,x_2,\kappa^+_1),\cdots,e^{\kappa_Mx_1+\kappa_M^2x_2 }m_0(x_1,x_2,\kappa^+_M))\mathcal D=0 ,\label{E:N-sym-t0}
     \end{gather}where  $\{z_n\}$ and $\{\kappa_j\}$  are blow-up and multi-valued points of  $m_0$ respectively; $\mathcal  D$ are norming constants between values of $m_0$ at $\lambda=\kappa_j^+=\kappa_j+0^+$   and can be computed by
\begin{align}
&  \mathcal  D = \widetilde{\mathcal  D} \times \left(\begin{array}{ccc}\widetilde{\mathcal  D}_{11}&\cdots &\widetilde{\mathcal  D}_{1N}\\\vdots&\cdots &\vdots\\
\widetilde{\mathcal  D}_{N1}&\cdots &\widetilde{\mathcal  D}_{NN}\end{array}\right)^{-1}\textit{diag}\,(\kappa_1^N,\cdots,\kappa_N^N)  ,\label{E:intro-sym-N-D-N}\\
&\widetilde{\mathcal  D}=  
\corr{\textit{diag}\,(\frac{\Pi_{2\le n\le N}(\kappa_1-z_n)}{(\kappa_1-z_1)^{N-1}}, \cdots, \frac{\Pi_{2\le n\le N}(\kappa_M-z_n)}{(\kappa_M-z_1)^{N-1}})}\times\mathcal  D^\sharp,\nonumber\\
& { \mathcal  D^\sharp}= 
  \left({\mathcal  D}_{ji}^\sharp\right)= \left(\mathcal  D^\flat_{ji}+\sum_{l=j}^M\frac{c_{jl}\mathcal  D^\flat_{li}}{ 1-c_{jj}} \right), \nonumber\\
& {\mathcal  D}^\flat= \textit{diag}\,(   
\kappa^N_1 ,\cdots,\kappa^N_M )\, A^T, \nonumber
\end{align}  with $c_{jl}=-\int\Psi _j(x_1,x_2,0 ) v_0(x_1,x_2)\varphi_l(x_1,x_2,0  )dx_1dx_2$, $\Psi_j(x)$, $\varphi_l(x)$ are residues of the adjoint eigenfunction at $\kappa_j$ \cite[(3.17))]{Wu21} and values of the Sato eigenfunction at $\kappa_l$ \eqref{E:residue-eigenfunction}. Moreover, $T_0=T|_{x_3=0}$ and $T$  is the continuous scattering operator defined by \eqref{E:cauchy-operator-p-inverse}, $s_c(\lambda)$ is the continuous scattering data, arising from the $\overline\partial$-characterization
\be \label{E:N-conti-sc-debar}
\begin{gathered}
\partial_{\overline\lambda}m_0(x_1,x_2, \lambda)
=  s_c(\lambda) e^{(\overline\lambda-\lambda)x_1+(\overline\lambda^2-\lambda^2)x_2  }m_0{(x_1,x_2, \overline\lambda)},\ \lambda\notin\RR, \\
  s_c(\lambda) = \frac { \Pi_{2\le n\le N}(\overline\lambda-z_n)}{(  \overline\lambda-z_1)^{N-1}   }\frac {\sgn(\lambda_I)}{2\pi i} \iint e^{-[(\overline\lambda-\lambda)x_1+(\overline\lambda^2-\lambda^2)x_2 ] }  \\
  \times  \xi(x_1,x_2,0, \overline\lambda) v_0(x_1,x_2)m_0(x_1,x_2, \lambda)dx_1dx_2.  
\end{gathered}
\ee 
 
The scattering data is a  {$d$-admissible scattering data}. Namely, it satisfies the algebraic and analytic constraints:
\begin{align}
& s_c(\lambda)=
\left\{
{\ba{ll}
 {\frac{ \frac {i}{ 2} \sgn(\lambda_I)}{\overline\lambda-\kappa_j}\frac{\gamma_j}{1-\gamma _j|\alpha|}}+\sgn(\lambda_I)  h_j(\lambda),&\lambda\in   D^ \times_{ \kappa_j },\\
\sgn(\lambda_I) {  \hbar_n}(\lambda),&\lambda\in    D^\times _{ z_n},
\ea}
\right.\label{E:inverse-s-c-N},\\
&\mathcal  D=\left({\tiny\ba{ccc}\kappa_1^N&\cdots&0\\
\vdots&\ddots&\vdots\\
0&\cdots&\kappa_N^N\\
\mathcal D_{N+1 1}&\cdots&\mathcal D_{N+1 N}\\
\vdots&\ddots&\vdots\\
\mathcal D_{M 1}&\cdots&\mathcal D_{M N}
\ea}\right),  \label{E:inverse-sym-N-D}
\end{align}
 and   
 \begin{align}
 &  {\begin{array}{l} 
 |(1-\sum_{j=1}^ME_{{\kappa_j}}  )  \sum_{|l|\le {d+8}}|\left(|\overline\lambda-\lambda|^{l_1}   +| \overline\lambda^2-\lambda^2|^{l_2}\right) s_c (\lambda)|  _{ 
  L^\infty} \\
   +   \sum_{j=1}^M(|\gamma_j|+|h_j|_{ L^\infty(D_{\kappa_j})})+\sum_{n=1}^N|\hbar_n|_{ C^1(D_{z_n})}  \\
    + | {\corr{\textit{diag}\,(q_1, \cdots, q_M)^{-1}}\times\mathcal D \times \corr{\textit{diag}\,(q_1, \cdots, q_N)}  - { \mathcal D}^\flat  }|_{L^\infty}\\
  \le  {C\sum_{|l|\le  { d+8}} |{\color{black} (1+|x_1|+|x_2|) } \partial_{x }^{l }       v_0|_{L^1\cap L^\infty}} , \end{array}}\label{E:s-c-ana}\\
  &s_c(\lambda)=  \overline{s_c( \overline\lambda)},
  h_j(\lambda)=-\overline{h_j( \overline\lambda)}, 
 \hbar_n(\lambda)=-\overline{\hbar_n( \overline\lambda)},\ \label{E:s-c-reality}\\
 &q_j=\frac{\Pi_{2\le n\le N}(\kappa_j-z_n)}{(\kappa_j-z_1)^{N-1}} \textit{ for }1\le j\le M.\nonumber
 \end{align} 
\end{itemize}

\end{theorem}

 \begin{theorem}   \label{T:N-soliton-evolution}    \cite[Theorem 5]{Wu21} If $\Phi= e^{ \lambda  x_1+ \lambda ^2x_2}    m(x, \lambda)$ satisfies the Lax pair \eqref{E:KPII-lax-1} and  
\begin{gather}
\partial_{\overline\lambda}  m(x, \lambda)=  {  {  s}_c(\lambda,x_3)}e^{(\overline\lambda-\lambda)x_1+(\overline\lambda^2-\lambda^2)x_2}   m(x,\overline\lambda)  ,\label{E:debar-evol-N}\\
  (e^{\kappa_1x_1+\kappa_1^2x_2 }m (x ,\kappa^+_1),\cdots,e^{\kappa_Mx_1+\kappa_M^2x_2 }m (x ,\kappa^+_M)){ \mathcal D(x_3)}=0,\label{E:D-evol-N}
\end{gather} with $\mathcal D(x_3) $ begin in the form of \eqref{E:sym-N-D}, then  
\be\label{E:linearization-D-evol-N}
\begin{gathered}
 {  s}_c(\lambda, x_3)=   {e^{ (\overline\lambda^3-{ \lambda}^3)x_3}}{  s}_c(\lambda ),\quad
   {\mathcal {\mathcal D}}_{mn}(x_3)=  { e^{ (\kappa_m^3-\kappa_n^3)x_3}}  {\mathcal D}_{mn}.  
   \end{gathered}\ee   	
 
\end{theorem}

\begin{theorem}   \cite{Wu22} Given a  {$d$-admissible scattering data} $\mathcal S =(\{z_n\},\{\kappa_j\},\mathcal D,s_c(\lambda))$,  $\epsilon_0\ll 1$, 
\begin{itemize}
\item [(1)] there exists uniquely an eigenfunction  $m\in W$ for   the system  of the Cauchy integral equation   and the {$\mathcal D$-symmetry},
\begin{gather}   
 {  {   m}(x, \lambda) =1+\sum_{n=1}^N\frac{   m_{z_n, \res }(x  )}{\lambda -z_n }  +\mathcal C  T
     m ,\ \lambda\neq z_n,}  \label{E:N-CIE} \\
{ (e^{\kappa_1x_1+\kappa_1^2x_2+\kappa_1^3x_3}m(x,\kappa^+_1),\cdots,e^{\kappa_Mx_1+\kappa_M^2x_2+\kappa_M^3x_3}m(x,\kappa^+_M))\mathcal D=0 },\label{E:N-sym}
     \end{gather}   satisfying
\be\label{E:N-asymp}
  { \sum_{0\le l_1+2l_2+3l_3\le d+5}| \partial^l_{x}\left[m(x ,\lambda)-\widetilde \chi (x ,\lambda)\right]|_{W}\le C\epsilon_0}.
 \ee 
 
\item [(2)]  Moreover,      
\begin{gather}
 \left(-\partial_{x_2}+\partial_{x_1}^2+2 \lambda\partial_{x_1}+ u (x) \right)  m (x ,\lambda)=0 ,\label{E:N-Lax-u}\\\
 u(x )\equiv - 2  \partial_{x_1}\sum _{n=1}^N    m_{z_n,\res}(x  )-\frac 1{\pi i}\partial_{x_1}\iint  T  m  \ d\overline\zeta\wedge d\zeta ,\label{E:N-u-rep}\\
\sum_{0\le l_1+2l_2+3l_3\le d+4}|\partial^l_x\left[u(x )-u_s(x )\right]|_{L^\infty }\le C \epsilon_0,\label{E:N-Lax-u-asymp}
\end{gather}The inverse scattering transform is defined by
\be\label{E:inverse scattering-transform-N}
\mathcal S^{-1}( \{z_n,\kappa_j, \mathcal D,s _c(\lambda)\})= -\frac 1{\pi i}\partial_{x_1}\iint  T  m \ d\overline\zeta\wedge d\zeta -2  \partial_{x_1}\sum_{n=1}^N m_{z_n,\res}(x) ;
\ee

\item [(3)]   $u :\mathbb R\times\mathbb R\times \mathbb R^+\to \mathbb R$ solves the KPII equation.
\end{itemize}

\end{theorem}

It can be seen that when discrete scattering data or continuous scattering data vanish, the forward and inverse scattering transform constructed for perturbed line solitons degenerate into those transforms  for rapidly decaying potentials or for $N$-line solitons. 

\subsection{Comments on distinct features  }\label{SS:N-solitons} 

\subsubsection{The Lax-Sato formulation of the KP equation} \label{SSS:N-solitons}\hfill\\
Our approach  to establish an IST of perturbed $\textrm{Gr}(N,M)_{ >0}$ KP solitons is based on \eqref{E:line-tau}, \eqref{E:line-grassmannian}, \eqref{E:N-sato}-\eqref{E;N-spectral-new}. To verify these formula, we summarize the Lax-Sato formulation of the KP equation \cite[\S 2.1-2.4]{K17} in this subsection.

\begin{itemize}

\item [$\blacktriangleright$]{\bf (The KP hierarchy, the KP equation, and the Lax pair) :} Suppose the Lax operator $\mathfrak L$ can be gauge transformed into the trivial operator $\partial=\partial_{x_1}$, i.e.
\be\label{E:2-5}
\partial=W^{-1}\mathfrak  LW,
\ee
where 
\be\label{2-6}
\begin{split}
W=&1-w_1\partial^{-1}-w_2\partial^{-2}-w_3\partial^{-3}+\cdots,\\
w_j=&w_j(x_1,x_2,x_3,\cdots).
\end{split}
\ee
If the Sato equation
\be\label{E:2-7}
\partial_{x_n}W=B_nW-W\partial^n\quad\textit{for $n=1,2,\cdots,$}
\ee holds where  $B_n=\left(W\partial^nW^{-1}\right)_{\ge 0}$, the polynomial part of $W\partial^nW^{-1}$ in $\partial$, then 
\begin{gather}
\partial_{x_n}\mathfrak  L=[B_n, \mathfrak L],\label{E:2-2}\\
\partial_{x_m} B_n-\partial_{x_n}  B_m+[ B_n, B_m]=0,\label{E:2-3}
\end{gather}
 and the gauge transform \eqref{E:2-5} transforms the linear system of the vacuum wave function 
\be\label{E:2-8}
\left\{
{\ba{ll}
 {\partial \phi_0}=\lambda\phi_0,&\\
\partial_{x_n}\phi_0=\partial^n\phi_0=k^n\phi_0,&n=1,2,\cdots,
\ea}
\right.\qquad \phi_0(x,\lambda)=\exp(\sum_{n=1}^\infty \lambda^nx_n),
\ee to the KP linear system
\be\label{E:2-4}
\left\{
{\ba{ll}
 {\mathfrak L \phi }=\lambda\phi ,&\\
\partial_{x_n}\phi =B_n\phi ,&n=1,2,\cdots,
\ea}
\right.\qquad \phi (x,\lambda)=W\phi_0.
\ee

Note that given a pair $(n,m)$ with $n>m$, from \eqref{E:2-3} (called the Zakharov-Shabat equations), we obtain a system of $n-1$ equations for $u_2,u_3,\cdots, u_n$, 
\be\label{E:2-1}
\begin{gathered}
\mathfrak L=\partial+u_2\partial^{-1}+u_3\partial^{-2}+\cdots,\\
u_2=w_{1,x_1},\ u_3=w_{2,x_1}+w_1w_{1,x_1}, \cdots,
\end{gathered}
\ee in the variables $x_1,x_m,x_n$. For $(n,m)=(3,2)$, the Zakharov-Shabat equations yield
the Kadomtsev-Petviashvili equation \eqref{E:KPII-intro}  for $u=2u_2=2w_{1,x_1}$, and, from the second equation of \eqref{E:2-4}, we derive the Lax pair \eqref{E:KPII-lax-1}.

\item [$\blacktriangleright$]{\bf (The tau function  and wave eigenfunctions) :} In particular, let
\[
\begin{split}
W=& 1-w_1\partial^{-1}-w_2\partial^{-2}-\cdots-w_N\partial^{-N},\\
W_N\equiv &W\partial^N= \partial^{N}-w_1\partial^{N-1}-w_2\partial^{N-2}-\cdots-w_N.
\end{split}
\]Hence the Sato equation \eqref{E:2-7} turns into
\be\label{E:Sato-reduction}
\partial_{x_n}W_N=B_nW_N-W_N\partial^n\quad\textit{for $n=1,2,\cdots $}
\ee
which yield 
\[
\partial_{x_n}(W_Nf)=B_n(W_Nf)+W_N(\partial_{x_n}f-\partial_{x_1}^nf).
\]We conclude, the $N$-th order differential equation $W_Nf=0$ is invariant under any flow of the linear heat hierarchy, $\{\partial_{x_n}f=\partial_{x_1}^nf:n=1,2\cdots\}$. Conversely, if $\partial_{x_n}f_j=\partial_{x_1}^nf_j$, $W_Nf_j=0$, $  n=1,2\cdots,j=1,2,\cdots, N$, then the Sato equation \eqref{E:Sato-reduction} holds and an explicit KP solution can be found via the $\tau$-function
\be\label{E:explicit-tau}
u(x)=2w_{1,x_1}= 2\partial^2_{x_1}\ln\tau(x)\equiv   2\partial^2_{x_1}\ln\textrm{Wr}(f_1,\cdots,f_N)  
\ee by writing $W_Nf_j=0$ as
\[
\left[
\ba{cccc}
f_1&  f_1^{(1)} &\cdots &  f_1^{(N-1)}\\
\vdots&\vdots&\ddots&\vdots\\
f_N& f_N^{(1)} &\cdots &  f_N^{(N-1)}
\ea
\right]  
\left[
\ba{c}
w_N\\
\vdots\\
w_1
\ea
\right]
=
\left[
\ba{c}
 f_1^{(N)}\\
\vdots\\
 f_N^{(N)}
\ea
\right].
\]

For the wave function $\phi=W_N\phi_0$ of the KP  linear system \eqref{E:2-4}, one has \cite[Proposition 2.2]{K17}
\begin{align*}
\phi=&W_N\phi_0=(1-\frac{w_1}{\lambda}-\frac{w_2}{\lambda^2}-\cdots-\frac{w_N}{\lambda^N})\phi_0=\frac{1}{\tau}\left|    
\ba{cccc}
f_1 &f_1^{(1)}&\cdots &f_1^{(N)}\\
\vdots&\vdots&\ddots &\vdots\\
f_N &f_N^{(1)}&\cdots &f_N^{(N)}\\
\lambda^{-N} &\lambda^{-N+1} &\cdots &1
\ea
\right|\phi_0.
\end{align*}

Using elementary column operatorations, the determinant  of the above expression can be rewritten as
\[
\frac{(-1)^N}{\lambda^N}\left|
\left(f_i^{(j)}-\lambda f_i^{(j-1)}
\right)_{1\le i,j\le N}
\right|.
\]Besides, let
 $
f_i(x)=\int_Ce^{\sum\zeta^nx_n}\rho_i(\zeta)d\zeta$.
 Hence
\begin{align*}
&f_i^{(j)}(x)-\lambda f_i^{(j-1)}(x)\\
=&-\lambda\int_C\zeta^{j-1}(1-\frac{\zeta}{\lambda})e^{\sum\zeta^nx_n}\rho_i(\zeta)d\zeta\\
=&-\lambda\int_C\zeta^{j-1}e^{-\sum\frac{\zeta^n}{n\lambda^n}} e^{\sum\zeta^nx_n}\rho_i(\zeta)d\zeta\\
=&-\lambda f_i^{(j-1)}(x_1-\frac{1}{\lambda},x_2-\frac{1}{2\lambda^2},x_3-\frac{1}{3\lambda^3},\cdots).
\end{align*}
As a result,
\be\label{E:2-12}
\phi(x,\lambda)=\frac{\tau(x_1-\frac{1}{\lambda},x_2-\frac{1}{2\lambda^2},x_3-\frac{1}{3\lambda^3},\cdots)}{\tau(x)}\phi_0(x).
\ee

For the eigenfunction $\phi^\dagger$ of the adjoint system \eqref{E:2-4}, one has \cite[\S 2.4]{K17}
\be\label{E:2-12-ast}
\phi^\dagger(x,\lambda)=\frac{\tau(x_1+\frac{1}{\lambda},x_2+\frac{1}{2\lambda^2},x_3+\frac{1}{3\lambda^3},\cdots)}{\tau(x)}\phi^{-1}_0(x).
\ee

\item [$\blacktriangleright$] {\bf (The   $N$-solitons  and Sato eigenfunctions) :} The $N$-solitons \eqref{E:line-tau} is defined by letting 
\be\label{E:N-soliton}
\begin{gathered}
\left(
\ba{c}
f_1(x)\\
\vdots\\
f_N(x)
\ea
\right)=
\left(
\begin{array}{cccc}
a_{11} &a_{12} & \cdots & a_{1M}\\
\vdots & \vdots &\ddots &\vdots\\
a_{N1} &a_{N2} & \cdots & a_{NM}
\end{array}
\right)
\left(
\begin{array}{c }
E_{1}  \\
E_{2}  \\
\vdots  \\
E_{M}  \\
\end{array}
\right),\\
A=(a_{ij})\in {\mathrm{Gr}(N, M)_{\ge 0}},\\
E_j(x)=\exp\theta_j(x)=\exp( \kappa_j x_1+\kappa_j^2 x_2+\kappa_j^3 x_3),\ \ \kappa_1<\cdots<\kappa_M,
\end{gathered}
\ee in \eqref{E:explicit-tau}; and formula   \eqref{E:N-sato} of the Sato eigenfunction  $
 \varphi(x,\lambda)$ (\eqref{E:N-sato-1} of the Sato adjoint eigenfunction  $
 \psi(x,\lambda)$) is derived by \eqref{E:2-12} (\eqref{E:2-12-ast}), \eqref{E:N-soliton},  and noting
\begin{align*}
&e^{\theta_i(x_1-\frac{1}{\lambda},x_2-\frac{1}{2\lambda^2},x_3-\frac{1}{3\lambda^3},\cdots)}=e^{\sum_{n=1}^\infty \kappa_i^n(x_n-\frac{1}{n\lambda^n})}=e^{\left(\sum_{n=1}^\infty \kappa_i^n x_n\right)-\ln (1-\frac{\kappa_i}{ \lambda })}=\frac{\lambda-\kappa_i}{\lambda }e^{\theta_i(x)}.
\end{align*}

\end{itemize}  

\subsubsection{The Lax equation}  It  can be proved by replacing the Sato eigenfunction and Sato adjoint eigenfunction by \eqref{E:N-sato} and adapting the  procedure \eqref{E:c-integral-eq-p}-\eqref{E:z-mn} for the proof of perturbed $1$-solitons.  Major difficulties and differences occur in proving the orthogonality relation (for construction of Green's function $G$) and boundedness of $G_d$. More precisely, 

\begin{itemize}

\item [$\blacktriangleright$] {\bf (The orthogonality relation) :}  Let
\be\label{E:residue-eigenfunction}
\begin{split}
\varphi_j(x)=\varphi(x,\kappa _j),&\quad
\psi_j(x)=\textit{res}_{\lambda=\kappa_j}\psi(x,\lambda),\\
\varphi(x,\kappa)= (\varphi_1(x), \cdots, \varphi_M  (x)),&\quad
 \psi(x,\kappa)= (\psi_1(x), \cdots, \psi_M  (x)).
\end{split}
\ee   
Define
\be\label{E:D'}
\begin{split}
&\mathcal D^\flat= \textrm{diag}\,(   
\kappa^N_1 ,\cdots,\kappa^N_M )\, A^T ,\\
&\mathcal D^{\flat,\dagger} = \left(\ba{c}-d^T,\ I_{M-N}\ea\right)\,\pi\,\textrm{diag}\,(   
\kappa^{-N}_1 ,\cdots,\kappa^{-N}_M  ),
\end{split}
\ee where  $\pi$ is an $M\times M$ permutation matrix and $d$ is an $N\times (M-N)$ matrix satisfying
\be\label{E:D}
 A = \left(\ba{c} 
I_N,\ d \ea \right)\pi.
\ee

We can implement various Pl$\ddot{\textrm u}$cker relations to   \eqref{E:N-sato} , and \eqref{E:2-12-ast} to prove 
\[
\mathcal D^{\flat,\dagger}\mathcal D^\flat=0,\ \ 
\varphi(x,\kappa)\mathcal D^\flat=0,\ \ 
\mathcal D^{\flat,\dagger}\psi(x,\kappa)^T=0
\]and {orthogonality relation}   
(see \cite[Lemma 2.1, 2.2]{Wu21} for detailed proofs)
between Sato  eigenfunctions $\varphi(x,\lambda)$ and   Sato   adjoint eigenfunctions $\psi(x,\lambda)$
\be\label{E:ortho-N}
\sum_{j=1}^M\varphi_j (x)\psi_j(x')=0.
\ee

\item [$\blacktriangleright$] {\bf (Boundedness of $G_d$) :} 

\begin{itemize}

\item 
following argument to  {permute and exchange} cells,  
one obtains the decomposition \cite[(3.16),(3.17)]{BP211}, \cite{BP212},   
\be\label{E:N-discon-G-decomposition}
G_d(x,x',\lambda)=G_{d}^1(x,x',\lambda)+G_{d}^2(x,x',\lambda),
\ee
with
\begin{align}
&G_{d}^1(x,x',\lambda)\label{E:N-discon-G-decomposition-1}\\
= &-\frac{\theta(x_2-x_2')}{[(N-1)!]^2(N+1)}\sum_{\{m_i\},\{n_i\}} \sgn(z_{m_Nn_N}-z'_{m_Nn_N})\nonumber\\
\times& e^{-k_{m_Nn_N}(x_2-x_2')} V(\{m_i\},n_N) V(n_1,\cdots,n_{N-1})    \nonumber\\
\times &\theta((\lambda_R-\kappa_{m_N})(z_{m_Nn_N}-z'_{m_Nn_N})) e^{-(\lambda_R-\kappa_{m_N})(z_{m_Nn_N}-z'_{m_Nn_N})}\nonumber\\
 \times&\dfrac{\mathcal D^\flat(\{m_i\}) \exp(\sum_{l=1}^{N-1} E_{m_l}(x)+E_{n_N}(x))}{\tau(x) } \nonumber\\
\times&\dfrac{\mathcal D^\flat(\{n_i\}) \exp(\sum_{l=1}^{N-1} E_{n_l}( x')+E_{m_N}(x'))}{\tau(x') } \nonumber ,
\end{align}and
\begin{align}
&G_{d}^2(x,x',\lambda)\label{E:N-discon-G-decomposition-2}\\
= &+\frac{\theta(x_2-x_2')}{[(N-1)!]^2(N+1)}\sum_{\{m_i\},\{n_i\}} \sgn(z_{m_Nn_N}-z'_{m_Nn_N})\nonumber\\
\times& e^{-k_{m_Nn_N}(x_2-x_2')} V(\{m_i\},n_N) V(n_1,\cdots,n_{N-1})    \nonumber\\
\times &\theta((\lambda_R-\kappa_{n_N})(z_{m_Nn_N}-z'_{m_Nn_N})) e^{-(\lambda_R-\kappa_{n_N})(z_{m_Nn_N}-z'_{m_Nn_N})}\nonumber\\
 \times&\dfrac{\mathcal D^\flat(\{m_i\}) \exp(\sum_{l=1}^{N } E_{m_l}(x) )}{\tau(x) } \dfrac{\mathcal D^\flat(\{n_i\}) \exp(\sum_{l=1}^{N } E_{n_l}(x '))}{\tau(x') } \nonumber ,
\end{align}where
\be\label{E:N-z-mn}
\begin{gathered}
z_{mn}=x_1+(\kappa_m+\kappa_n)x_2, \quad z'_{mn}=x'_1+(\kappa_m+\kappa_n)x'_2,\\
k_{mn}=\lambda_I^2-(\lambda_R-\kappa_m)(\lambda_R-\kappa_n),
\end{gathered}
\ee and 
\begin{gather*}
V(\{n_i\})= \det\left(   
\ba{cccc}
1&1&\cdots &1\\
\kappa_{n_1}&\kappa_{n_2}&\cdots &\kappa_{n_N}\\
\vdots&\vdots&\ddots&\vdots\\
\kappa_{n_1}^{N -1}&\kappa_{n_2}^{N -1}&\cdots &\kappa_{n_N}^{N -1}
\ea
\right),\\
\mathcal D^\flat(\{n_i\})= \det\left(   
\ba{ccc}
\mathcal D^\flat_{n_1,1}&\cdots & \mathcal D^\flat_{n_1,N}\\
\vdots&\ddots&\vdots\\
\mathcal D^\flat_{n_N,1}&\cdots & \mathcal D^\flat_{n_N,N}
\ea
\right),
\end{gather*} for $m,\,n\in\{1,\,2\}$ and $\{m_i\}=\{m_1,\cdots,m_N\},\{n_1,\cdots, n_N\}$ denote unordered set of $N$ indices from $\{1,\cdots,M\}$. Now, in the $x$-variable, all of the exponentials in the nominators of \eqref{E:N-discon-G-decomposition-1}, \eqref{E:N-discon-G-decomposition-2} are either bounded or dominated by the tau functions   in the denominators by totally  positive condition of $\textrm{Gr}(N, M)_{ >0}$. So 
\begin{align}
|G_d(x,x',\lambda)|<&C  (C_{x'}+1 ),\label{E:bdd-+}
\end{align}and then
\be 
|G(x,x',\lambda)|< C  (C_{x'}+\frac1{\sqrt{|x_2-x_2'|}} ).\label{E:bdd--}
\ee

\item We will use the duality to prove boundedness of  Green's function. Namely, suppose that  $u(x)$ satisfies the KPII equation
then $u(-x)\equiv w(x)$ also satisfies the KPII equation. In particular, if 
\be\label{E:Duality-ini-data}
  \begin{split}
  &u_0(x_1,x_2)=u_s(x_1,x_2,0)+v_0(x_1,x_2),\\
  &u(x_1,x_2,0)=u_0(x_1,x_2),\ \
   \textit{$ u_s(x)$  a ${\mathrm{Gr}(N,M)_{> 0}}$ KP soliton,  } 
 \end{split}
 \ee then
\be\label{E:Duality-ini-data-w}
 \begin{split}
  &\qquad\qquad w_0(x_1,x_2)=u_s(-x_1,-x_2,0)+v_0(-x_1,-x_2),\\
  &\qquad\qquad w(x_1,x_2,0)=w_0(x_1,x_2),\ \
   \textit{$ u_s(-x)$  a ${\mathrm{Gr}(M-N,M)_{> 0}}$ KP soliton.  } 
 \end{split}
 \ee Here we have used the duality of between $\mathrm{Gr}(N,M)_{> 0}$ and $\mathrm{Gr}(M-N, M)_{> 0}$ \cite[Section 4.4]{K18}.

Consider the Lax operators
\begin{equation}\label{E:Duality-KPII-lax-1}
(-\partial_{x_2}+\partial_{x_1}^2+u_0 (x_1,x_2))\Phi_+(x_1,x_2,\lambda) =0,
\end{equation}and
\begin{equation}\label{E:KPII-lax-2}
(-\partial_{x_2}+\partial_{x_1}^2+w_0(x_1,x_2) )\Phi_-(x_1,x_2,\lambda)= 0. 
\end{equation}Notice that \eqref{E:KPII-lax-2} is equivalent to
\begin{equation}\label{E:KPII-lax-3}
(+\partial_{x_2}+\partial_{x_1}^2+u_0 (x_1,x_2))\Phi_-(-x_1,-x_2,\lambda)=0.
\end{equation}

Denote the Lax operators
\begin{align}
\mathcal L_\pm =&-\partial_{x_2}+\partial_{x_1}^2+u_s(\pm x_1,\pm x_2,0),\label{E:sym-0}
 \\
\mathcal L_1=& +\partial_{x_2}+\partial_{x_1}^2+u_s(+ x_1,+ x_2,0),\label{E:sym-0-1} 
\end{align}
and the associated Green functions by $\mathcal G_\pm(x,x', \lambda)$, $\mathcal G_1(x,x', \lambda)$ respectively,  namely,
\be
\mathcal L_\pm\mathcal G_\pm(x,x', \lambda) = \mathcal L_1\mathcal G_1 (x,x', \lambda) =\delta(x-x')  ,\label{E:sym-1}
\ee and define
\begin{align} 
 \mathcal G_\pm(x,x', \lambda) =&e^{ \lambda (x_1-x_1')+\lambda^2 (x_2-x_2') } G_\pm (x,x', \lambda),\label{E:sym-1-1}\\
\mathcal G_1(x,x', \lambda) =&e^{ \lambda (x_1-x_1')+\lambda^2 (x_2-x_2') } G_1 (x,x', \lambda),\label{E:sym-1-2}\\
G_\pm (x,x', \lambda)=&G_{\pm ,d}(x,x', \lambda)+G_{\pm,c} (x,x', \lambda),\label{E:sym-1-3}\\ 
G_1(x,x', \lambda)=&G_{1,d} (x,x', \lambda)+G_{1,c} (x,x', \lambda).\label{E:sym-1-4}
\end{align}

Thanks to {\cite[(1.6)]{BP212} 
\be\label{E:Duality-adjoint} 
\mathcal L_1=\mathcal L^d_+, \quad
G_1(x,x',\lambda)= G_+(x',x,\lambda).
\ee}
From
$
\mathcal L_1(x_1,x_2)=\mathcal L _-(-x_1,-x_2)$,
  one has
\be\label{E:adjoint-1-new}
\begin{split}
G_1(x,x',\lambda)= G_-(-x,-x',\lambda).
\end{split}
\ee

Applying \eqref{E:Duality-ini-data}, \eqref{E:Duality-ini-data-w},   \eqref{E:bdd--},   one has
\begin{align}
|G_+(x,x',\lambda)|<&C  (C_{x'}+\frac1{\sqrt{|x_2-x_2'|}} ),\label{E:Duality-bdd-+}\\
|G_-(x,x',\lambda)|<&C  (C_{x'}+\frac1{\sqrt{|x_2-x_2'|}} ).\label{E:Duality-bdd--}
\end{align}
Combining with \eqref{E:Duality-adjoint} and \eqref{E:adjoint-1-new}, 
\be\label{E:M-1}
|G_1(x,x',\lambda)|< C  (1+\frac1{\sqrt{|x_2-x_2'|}} ).
\ee
Hence we conclude
\begin{align}
|G_+(x,x',\lambda)|<&C  (1+\frac1{\sqrt{|x_2-x_2'|}} ) \label{E:bdd}
\end{align}and improve the estimate \eqref{E:bdd--}.
\end{itemize} 
\end{itemize}

\subsubsection{The inverse problem} Via the iteration sequence  
\begin{align}
& \phi^{(k)}(x ,\lambda)  
=   1+\sum_{n=1}^N \frac{ \phi^{(k)}_{z_n,\res} (x   )}{\lambda -z_n }  +\mathcal CT \phi^{(k-1)}(x, \lambda)  ,\ \ k>0,\label{E:recursion-iteration-N}\\
&(e^{\kappa_1x_1+\kappa_1^2x_2+\kappa_1^3x_3}\phi^{(k)}(x,\kappa^+_1),\cdots,  e^{\kappa_Mx_1+\kappa_M^2x_2+\kappa_M^3x_3}\phi^{(k)}(x,\kappa^+_M))\mathcal D=0,\label{E:recursion-iteration-D-N}\\
&\phi^{(0)}(x ,\lambda)= \widetilde \chi(x,\lambda) ,\label{E:recursion-iteration-bdry-N}
 \end{align} one can adapt the  procedure   of perturbed $1$-solitons to solve the CIE and $\mathcal D$-symmetry. The only issue we have to address more is that, for perturbed $1$-line solitons,  by evaluating the CIE \eqref{E:recursion-iteration-introduction} at $\kappa_1^+$, $\kappa_2^+$, and using  the   $\mathcal D$-symmetry  \eqref{E:recursion-iteration-D-introduction}, one obtains 
 \begin{align}
 \phi^{(k)}_{0,\res} (x   )=&-\frac{\kappa_1 e^{\kappa_1x_1+\kappa_1^2x_2+\kappa_1^3x_3}+a \kappa_2e^{\kappa_2x_1+\kappa_2^2x_2+\kappa_2^3x_3}}{e^{\kappa_1x_1+\kappa_1^2x_2+\kappa_1^3x_3}+a e^{\kappa_2x_1+\kappa_2^2x_2+\kappa_2^3x_3}},\label{E:recursion-iteration-D-N-new} 
 \end{align} and successfully  reduces solvability to deriving uniform estimates of $CT\phi^{(k)}$ on $D_\infty$, $D_{\kappa_j}$, $D_{0}$.

In the following proposition, we show the same argument works for perturbed $N$-line solitons for totally positive types. 
\begin{proposition}\label{P:N-alg-sym}
Suppose $ {\mathcal S}=(\{z_n\},\{\kappa_j\}, \mathcal D,s _c)$ is   $d$-admissible and $\phi^{(k)}$, $\phi^{(k)}_{z_n,\res}$ satisfy \eqref{E:recursion-iteration-N}, \eqref{E:recursion-iteration-D-N}. Then for $k>0$, 
{\small \be \label{E:N-alg-sym}
 {\left(
\ba{c}
\phi^{(k)}_{z_1,\res}  \\
\vdots\\
\phi^{(k)}_{z_N,\res}  
\ea
\right)= -B^{-1}  \widetilde A
\left(
\ba{c}
1+\mathcal C_{{ \kappa_1^+}}T\phi^{(k-1)}\\
\vdots\\
\vdots\\
\vdots\\
\vdots\\
1+\mathcal C_{{ \kappa_M^+}}T\phi^{(k-1)}
\ea
\right),} \ee} where
{\tiny\be\label{E:sym-BA}
\begin{split}
\widetilde A= &  \left(
\ba{cccccc}
\kappa_1^Ne^{\theta_1}&\cdots&0&\mathcal D_{N+1,1}e^{\theta_{N+1}}&\cdots&\mathcal D_{M,1}e^{\theta_M} \\
\vdots&\ddots&\vdots&\vdots&\ddots&\vdots \\
0&\cdots&\kappa_N^Ne^{\theta_N}&\mathcal D_{N+1,N}e^{\theta_{N+1}}&\cdots&\mathcal D_{M,N}e^{\theta_M} \ea 
\right)  ,\ \    
B=     \widetilde A 
   \left(
\ba{ccc} 
\frac 1{\kappa_1-z_1}&\cdots&\frac 1{\kappa_1-z_N}\\
\vdots&\ddots&\vdots\\
\vdots&\ddots&\vdots\\
\vdots&\ddots&\vdots\\
\vdots&\ddots&\vdots\\
\frac 1{\kappa_M-z_1}&\cdots&\frac 1{\kappa_M-z_N} 
\ea
\right), 
\end{split}
\ee} and $ e^{\theta_j}= e^{\kappa_j x_1+\kappa_j^2 x_2+\kappa_j^3 x_3}$.  

Moreover,    for $k>0$,
 \begin{align}
 \sum_{0\le l_1+2l_2+3l_3\le d+5}\left|\partial_x^l\phi^{(k)}_{z_n,\res}\right|_{L^\infty}\le   C(1+\epsilon_0\sum_{0\le l_1+2l_2+3l_3\le d+5}&\left|\partial_x^l\phi^{(k-1)} \right|_W)\label{E:k-est-new},\\
 \sum_{0\le l_1+2l_2+3l_3\le d+5}\left|\partial_x^l\left[\phi^{(k)}_{z_n,\res}-\phi^{(k-1)}_{z_n,\res}\right]\right|_{L^\infty}\le & (C\epsilon_0)^{k },  \label{E:k-difference-new}\\
 \sum_{0\le l_1+2l_2+3l_3\le d+5}\left|\partial_x^l\left[\phi^{(k)}_{z_n,\res}-\widetilde\chi_{z_n,\res}\right]\right|_{L^\infty}\le&  C\epsilon_0 .  \label{E:k-difference-0}
 \end{align}

\end{proposition} 
\begin{proof} Write the $\mathcal D$-symmetry and the evaluation at $ \kappa_j ^+ $ of $\phi^{(k)}$ as   a  linear system  for $M+N$ variables $\{\phi^{(k)}(x,\kappa_j^+),\phi^{(k)}_{z_n,\res}(x)\}$, 
{\tiny\begin{align}
&\left(
\ba{ccccccccc}
\kappa_1^Ne^{\theta_1}&\cdots&0&\mathcal D_{N+1,1}e^{\theta_{N+1}}&\cdots&\mathcal D_{M,1}e^{\theta_M}&0&\cdots&0\\
\vdots&\ddots&\vdots&\vdots&\ddots&\vdots&0&\ddots&0\\
0&\cdots&\kappa_N^Ne^{\theta_N}&\mathcal D_{N+1,N}e^{\theta_{N+1}}&\cdots&\mathcal D_{M,N}e^{\theta_M}&0&\cdots&0\\
-1&\cdots&0&0&\cdots&0&\frac 1{\kappa_1-z_1}&\cdots&\frac 1{\kappa_1-z_N}\\
\vdots&\ddots&\vdots&\vdots&\ddots&\vdots&\vdots&\ddots&\vdots\\
\vdots&\ddots&\vdots&\vdots&\ddots&\vdots&\vdots&\ddots&\vdots\\
\vdots&\ddots&\vdots&\vdots&\ddots&\vdots&\vdots&\ddots&\vdots\\
\vdots&\ddots&\vdots&\vdots&\ddots&\vdots&\vdots&\ddots&\vdots\\
0&\cdots&0&0&\cdots&-1&\frac 1{\kappa_M-z_1}&\cdots&\frac 1{\kappa_M-z_N}
\ea
\right) 
\left(
\ba{c}
\phi^{(k)}(x,\kappa_1^+)\\
\vdots\\
\vdots\\
\vdots\\
\vdots\\
\phi^{(k)}(x,\kappa_M^+)\\
\phi^{(k)}_{z_1,\res} (x)\\
\vdots\\
\phi^{(k)}_{z_N,\res} (x)
\ea
\right)\nonumber\\
&=
\left(
\ba{c}
0\\
\vdots\\
0\\
-1-\mathcal C_{\kappa_1^+}T\phi^{(k-1)}\\
\vdots\\
\vdots\\
\vdots\\
\vdots\\
-1-\mathcal C_{\kappa_M^+}T\phi^{(k-1)}
\ea
\right)  .\label{E:N-system}
 \end{align}} Solving $\phi^{(k)}(x,\kappa_j^+)$ in terms of $\phi^{(k)}_{z_n,\res}(x)$ and plugging the outcomes into \eqref{E:N-system} again yields 
{\small\be\label{E:bdd-res} 
 \begin{split}
&B\left(
\ba{c}
\phi^{(k)}_{z_1,\res}(x)\\
\vdots\\
\phi^{(k)}_{z_N,\res}(x)
\ea
\right)=-\widetilde A
\left(
\ba{c}
 1+\mathcal C_{\kappa_1^+}T\phi^{(k-1)}\\
\vdots\\
\vdots\\
\vdots\\
\vdots\\
 1+\mathcal C_{\kappa_M^+}T\phi^{(k-1)}
\ea
\right) ,
\end{split}\ee} with  $B$ and $\widetilde A $ defined by \eqref{E:sym-BA}.   By the $d$-admissible condition,    the   system \eqref{E:N-system} is just determined and is equivalent to \eqref{E:N-alg-sym}.
 
Proofs of \eqref{E:k-est-new}-\eqref{E:k-difference-0} rely heavily on Sato theory.    The key observation is that  the leading term of the right hand side of \eqref{E:N-alg-sym} can be realized by residue of some normalized Sato eigenfunction $\widetilde\chi'$. Through the direct computation of $\widetilde\chi'$ and matching with \eqref{E:line-grassmannian}, \eqref{E:N-sato}, we shall prove boundedness of $B^{-1}A$ and then verify \eqref{E:k-est-new}-\eqref{E:k-difference-0} by multi linearity and the totally positive (TP) condition. 

Precisely, 
the $d$-admissible condition implies that 
{\small\[
\begin{gathered}
A''= \textit{diag}\,(q_1, \cdots, q_M)^{-1} \times\mathcal D \times  \textit{diag}\,(q_1, \cdots, q_N) =\left(
\ba{ccc}
a''_{11}&\cdots &a''_{1N}\\
\vdots &\ddots &\vdots\\
a''_{N1}&\cdots &a''_{NN}\\
\vdots &\ddots &\vdots\\
a''_{M1}&\cdots &a''_{MN}
\ea
\right),\\
A'=A''\times 
\left(
\ba{ccc}
a''_{11}&\cdots &a''_{1N}\\
\vdots &\ddots &\vdots\\
a''_{N1}&\cdots &a''_{NN}
\ea
\right)^{-1},\qquad
A'\in  \mathrm {Gr}(N,M)_{\color{black}> 0},
\end{gathered}
\]}   such that 
setting 
\be\label{E:sato-prime}
\widetilde \chi'(x,\lambda)=\widetilde \chi_{\{z_n\},\{\kappa_j\},A'}(x,\lambda) 
\ee as the normalized Sato eigenfunction with data $(\{z_n\},\{\kappa_j\},A',0)$, one has
\begin{gather} 
{  {  \widetilde\chi'}(x, \lambda) =1+\sum_{n=1}^N\frac{   \widetilde\chi'_{z_n, \res }(x   )}{\lambda -z_n }  ,} \label{E:sato-CIE} \\
(e^{\kappa_1x_1+\kappa_1^2x_2+\kappa_1^3x_3}\widetilde \chi'(x,\kappa _1),\cdots,e^{\kappa_Mx_1+\kappa_M^2x_2+\kappa_M^3x_3}\widetilde \chi'(x,\kappa _M))\mathcal D=0,\label{E:sato-D}
\end{gather}and, from the $d$-admissible condition,   $\forall k$,
\be\label{E:asy-chi'}
|\widetilde\chi'_{z_n, \res }(x   )-\widetilde\chi _{z_n, \res }(x   )|_{C^k}
\le C_k\epsilon_0.
\ee 

Moreover, using $\mathcal D$-symmetry and evaluating $\widetilde\chi'$ at $\kappa_j$ and   the above argument, yields 
{\small\be\label{E:k-diff}
\quad{ \left(
\ba{c}
\widetilde\chi'_{z_1,\res} (x) \\
\vdots\\
\widetilde\chi'_{z_N,\res}(x)  
\ea
\right)= -  B^{-1}  \widetilde {  A}
\left(
\ba{c}
1 \\
\vdots\\  \vdots\\ \vdots\\ \vdots\\
1 
\ea
\right),}
\ee} with $B$ and $\widetilde A $ defined by \eqref{E:sym-BA}. Let $E_j =e^{\theta_j}  = e^{\kappa_j x_1+\kappa_j^2 x_2+\kappa_j^3 x_3} 
 $ and write
\begin{align}
\widetilde {  A}= &  
\mathcal D^T \textit{diag}\,( E_1, \cdots,  E_M),\label{E:tilde-B-chi}\\
B
=&   
\mathcal D^T   \textit{diag}\,( E_1, \cdots,  E_M)\left(
\ba{ccc} 
 \frac 1{\kappa_1-z_1 }&\cdots&\frac{1}{\kappa_1-z_N }\\
\vdots&\ddots&\vdots\\
\vdots&\ddots&\vdots\\
\frac 1{\kappa_M-z_1 }&\cdots&\frac{1}{\kappa_M-z_N }
\ea
\right).\nonumber
\end{align} 

 From Sato theory,  \eqref{E:N-sato},  \eqref{E:intro-sym-N-D-flat}, \eqref{E:k-diff}, \eqref{E:tilde-B-chi}, elementary row and column operations, and matching the   coefficients of $E_1\times\cdots\times E_N$,  
\begin{align}
  &B^{-1} = \frac{1}{\tau'(x)}
\left(
\ba{ccc} 
  b_{11}&\cdots &  b_{1N}\\
\vdots&\ddots&\vdots\\
  b_{N1}&\cdots &  b_{NN} 
\ea
\right),\nonumber
\\
&  { b_{kl}= \sum_{ J(kl) =( j _{(kl),1},\cdots,j _{(kl),N-1})}\Delta _{J(kl)} E_{J(kl)}(x),\ \ 1\le j _{(kl),1}<\cdots<j _{(kl),N-1}\le M,}\label{E:determinant-B}\\
& \textit{$\tau'(x)$ is the tau function with data $\kappa_j$, $A'$,}\nonumber\\
 & |\Delta _{J(kl)}|=|\Delta _{J(kl)} (\{z_n\},\{\kappa_j\},A')|<C. \nonumber
\end{align}   
Moreover,   
 \begin{align}
   &\tau'(x) \widetilde\chi'_{z_h,\res}(x)\label{E:mathfrak-BA-0}\\
=&\textit{the $h$-row of}\ {\tiny \left(
\ba{ccc} 
 b_{11}&\cdots &  b_{1N}\\
\vdots&\ddots&\vdots\\
  b_{N1}&\cdots &  b_{NN} 
\ea
\right)
\left(
\ba{c}
\kappa_1^NE_1+\cdots+\mathcal D_{N+1,1}E_{N+1}+\cdots+\mathcal D_{M,1}E_M\\
\vdots\\  \vdots\\ \vdots\\ \vdots\\
\kappa_N^NE_{  1}+\cdots+\mathcal D_{N+1,N}E_{N+1}+\cdots+\mathcal D_{M,N}E_M 
\ea
\right)} \nonumber 
\\
=&(\kappa_1^NE_1+\cdots+\mathcal D_{N+1,1}E_{N+1}+\cdots+\mathcal D_{M,1}E_M)\sum_{|J(h1)|=N-1}\Delta _{J(h1)}  E_{J(h1)}(x)\nonumber\\
+&\cdots
+(\kappa_N^NE_{  1}+\cdots+\mathcal D_{N+1,N}E_{N+1}+\cdots+\mathcal D_{M,N}E_M) \sum_{|J(hN)|=N-1}\Delta _{J(hN)}  E_{J(hN)}(x)    
\nonumber\\
\equiv&(\widetilde{  a}_{11}E_1+\cdots+\widetilde{  a}_{1M}E_M)\sum_{|J(h1)|=N-1}\Delta _{J(h1)}  E_{J(h1)}(x)\nonumber\\
+&\cdots
+(\widetilde{ a}_{N1}E_1+\cdots+\widetilde{  a}_{NM}E_M) \sum_{|J(hN)|=N-1}\Delta _{J(hN)}  E_{J(hN)}(x),   
\nonumber
\end{align}and 
\be \label{E:chi-plucker}
\begin{split}
0=& \widetilde {  a}_{1k}E_k \ \sum_{k\in J(h1),\ |J(h1)|=N-1}\Delta _{J(h1)}  E_{J(h1)}(x) 
\ +\ \cdots \\
+&\widetilde {  a}_{Nk}E_k \sum_{k\in J(hN), |J(hN)|=N-1}\Delta _{J(hN)}  E_{J(hN)}(x)  .
\end{split} \ee  

Using \eqref{E:N-alg-sym}, \eqref{E:tilde-B-chi}-\eqref{E:chi-plucker}, and  multi linearity, 
\begin{align*}
&\tau'(x) \phi^{(k)}_{z_h,\res}(x) 
= \tau'(x) \widetilde\chi'_{z_h,\res}(x)\\
+&\textit{the $h$-row of } {\tiny \left(
\ba{ccc} 
 b_{11}&\cdots &  b_{1N}\\
\vdots&\ddots&\vdots\\
  b_{N1}&\cdots &  b_{NN} 
\ea
\right)
\left(
\ba{c}
\widetilde a_{11}E_1\mathcal C_{\kappa_1^+}T\phi^{(k-1)}+\cdots+ \widetilde a_{1M}E_M\mathcal C_{\kappa_M^+}T\phi^{(k-1)}\\
\vdots\\  \vdots\\ \vdots\\ \vdots\\
\widetilde a_{N1}E_1\mathcal C_{\kappa_1^+}T\phi^{(k-1)}+\cdots+ \widetilde a_{NM}E_M \mathcal C_{\kappa_M^+}T\phi^{(k-1)}
\ea
\right)}\\
=&\sum_{|J(h)|=N}\Delta _{J(h)} E_{J(h)}(x),\qquad 
\end{align*} with 
\begin{gather*}
 \sum_{0\le l_1+2l_2+3l_3\le d+5}|\partial_x^l \Delta _{J(h)}|
 <C(1+\sum_{j=1}^M \sum_{0\le l_1+2l_2+3l_3\le d+5}|\partial_x^l\mathcal C_{\kappa_j^+}T\phi^{(k-1)}|).
\end{gather*}
 Along with  the totally positive (TP) condition of $A' $,  yield 
\[
 \sum_{0\le l_1+2l_2+3l_3\le d+5}|\partial_x^l \phi^{(k)}_{z_n,\res}(x)|\le C(1+\sum_{j=1}^M \sum_{0\le l_1+2l_2+3l_3\le d+5}|\partial_x^l\mathcal C_{\kappa_j^+}T\phi^{(k-1)}|), 
\] and \eqref{E:k-est-new}-\eqref{E:k-difference-0} follow  from     and \eqref{E:asy-chi'}.

\end{proof}

\end{document}